\documentclass[aps,preprint,unsortedaddress]{revtex4-1}

\usepackage[english]{babel}

\usepackage{amsmath}%
\usepackage{graphicx}%
\usepackage{xcolor}%
\usepackage{dcolumn}%
\usepackage{bm}%
\usepackage{ulem}
\usepackage{setspace}
\usepackage{siunitx}
\usepackage{lineno,hyperref}
\usepackage[utf8]{inputenc}

\DeclareUnicodeCharacter{0301}{\'{e}}

\begin{document}

\title{Insight into the inclusion of heteroatom impurities in Silicon structures}

\author{Rita Maji}
\email{rita.maji@unimore.it}
\address{Dipartimento di Scienze e Metodi dell'Ingegneria, Universit{\`a} di Modena e Reggio Emilia,Via Amendola 2 Padiglione Tamburini , I-42122 Reggio Emilia, Italy}

\author{Eleonora Luppi}
\address{Laboratoire de Chimie Th\'eorique, Sorbonne Universit\'e and CNRS  F-75005 Paris, France}

\author{Elena Degoli}
\email{elena.degoli@unimore.it}
\address{Dipartimento di Scienze e Metodi dell'Ingegneria, Universit{\`a} di Modena e Reggio Emilia and Centro Interdipartimentale En$\&$Tech, Via Amendola 2 Padiglione Morselli, I-42122 Reggio Emilia, Italy \\
Centro S3, Istituto Nanoscienze-Consiglio Nazionale delle Ricerche (CNR-NANO),Via Campi 213/A, 41125 Modena, Italy\\
Centro Interdipartimentale di Ricerca e per i Servizi nel settore della produzione, stoccaggio ed utilizzo dell'Idrogeno H$2$–MO.RE., Via Universit{\`a} 4, 41121 Modena, Italy}

\author{Julia Contreras-Garc\'{\i}a}
\email{contrera@lct.jussieu.fr}
\address{Laboratoire de Chimie Th\'eorique, Sorbonne Universit\'e and CNRS  F-75005 Paris, France}

\date{\today}


\begin{abstract}

The bonding properties of tilt boundary in poly-silicon and the effect of interstitial impurities are investigated by first-principles. In order to obtain thorough information on the nature of chemical bondings in these solid systems, an accurate topological analysis is performed,  through partitioning of the electron localization function.
Although the mechanism of segregation of single light impurities, such as carbon, nitrogen and oxygen in Si-based systems is known,  it is only in the presence of multiple segregation that the distinctive structures of the various interstitial impurities emerge.
The structural analysis of the modified Si systems and the comparison with the corresponding molecular structure within these solid phases provide an adequate description of interesting properties, for which bond charges provides more insight than bond length. 
It is shown that, in the presence of isovalent carbon, all systems try to preserve the tetrahedral coordination, on the contrary, trivalent nitrogen induces a strong local distortion to fit in the tetrahedral Si matrix while oxygen is the impurity that segregates more easily and more regularly.
This work shows that impurities lead to local distortions and how the electron distribution rearranges to smooth it. Overall, it shows how the analysis of bonds and their correlation with energetics and electronic structure is of fundamental importance for the understanding of the defects induced properties and of the basic mechanisms that influence them.

\end{abstract}

\maketitle

\section{Introduction}

The existence of defects is inherent in materials, despite the more and more efficient growth techniques. \cite{bookdefects} Defects may cost energy to be created, however, in some cases the configurational entropy can make favorable the incorporation of a certain concentration of defects in order to lower the free energy of the system. \cite{AM76} Moreover, impurities are also often intentionally introduced to modify material properties, such as doping of semiconductors with acceptors and donors is essential for electronic and optoelectronic applications. \cite{Defect2016,Defect2020,PhysRevB.72.113303}

In silicon-based devices, defects such as impurities, vacancies, interstitial and grain boundaries (GBs) modify the functionality and the performance of the devices. \cite{HM2018,bookdefects,JCP_Rmaji2021,MAJI2021116477} In this context, polycrystalline silicon (poly-Si) is very attractive for several technological applications \cite{PV2011,MAJI2021116477}. However, its efficiency can be strongly modified because poly-Si is characterized by a high density of GBs which are the primary hindrance to its functionality. Furthermore, the electrical properties of poly-Si and GBs are additionally altered by the complex interaction with intrinsic point defects (vacancies and/or self-interstitials) and/or impurity atoms, such as carbon, oxygen, nitrogen, and various types of metals. \cite{SinnoJAP2013,JCP_Rmaji2021,Majipssp21} Moreover, the interaction of these light impurities (C, O, N) with poly-Si under certain circumstances can facilitate the formation of microstructures of SiC, SiO$_2$, and Si$_3$N$_4$\cite{SinnoJAP2013}.

Oxygen defects in Si have been reported focusing on several aspects, such as doping concentration, different complexes, modification of electronic and mechanical properties. \cite{Kissinger2019,PhysRevB1999_Odefect} Carbon defects studies 
focused on self-interstitial trapping mechanism\cite{PhysRevB.66.195214}, C-C complexes\cite{CC2018}, and transient enhanced diffusion of other impurities\cite{PhysRevB.66.195214}. Nitrogen defects
have been considered as one of the key dopants for exceptional enhancement of mechanical stability\cite{N_Si2002,Yuan2019}. Moreover, recently electronic and optical properties in crystalline Si hyperdoped with N is reported as a source of sub-bandgap absorptance\cite{SR2015}.

Most of the previous first-principles works focused on the  understanding of the complex mechanisms in polycristalline Si following experimental observations. Even if this characterization provides information about the local performance of such areas, a correlation to the light impurity distribution remains ambiguous. 

Another approach, increasingly popular for crystalline systems in the past decade, is the analysis of the chemical bonding based on the topological analysis obtained from density distribution. In particular, electron localization function (ELF) \cite{ijms16048151,Koumpouras_2020,D0CP06073A} has emerged as a powerful tool to understand the behavior of the electrons in atoms, molecules and solids. A great variety of bonding situations can be explained and therefore interpreted in relation to physical properties. \cite{ContrerasGarca2011ElectronDA,Gatti+2005+399+457,FUENTEALBA200757,ContrerasGarcia2009,Feng14,D0CP06073A,molecules26144227,JCC2020}

In this work, we highlight on chemical bonding of the C, N and O local structures in Silicon  bulk and in Silicon GB $\Sigma$3\{111\} with and without vacancies. In order to understand the organization and the underlying electronic structure of the inclusion of C, O and N atoms in Si structures, we focus on topological analysis obtained from density distribution. We present the characteristics of electron localization function (ELF) obtained from first-principles calculations. \cite{ijms16048151,Koumpouras_2020,D0CP06073A} ELF analysis contribute to the understanding of the behavior and properties of C, N and O atoms in Silicon structures by the characterization of the chemical bonding.

\section{Methodology}

\subsection{Systems}
The $\Sigma$3\{111\} Si GB consists of two Si grains, misoriented by an angle $\Omega = 60^{\circ}$ which form an interface along the crystallographic plane \{111\} (coincidence site lattice). To describe the grain, we used an orthorhombic supercell ($a$ $\ne$ $b$ $\ne$ $c$ and $\alpha$ = $\beta$ = $\gamma$= $90^{\circ}$) composed of 96 Si atoms generated with {\it GB Studio program}. \cite{GBStudio2006}. The lattice parameters are $a$=13.30~\AA, $b$=7.68~\AA~and $c$=18.81~\AA.  \cite{MAJI2021116477} The formation energy of the $\Sigma$3\{111\} Si GB is $E^{\text{f}}_{\text{GB}} = 0.002$ eV/\AA$^{2}$ ($E^{\text{f}}_{\text{GB}} = 0.05$ J/m$^{2}$) \cite{MAJI2021116477} which indicates that this GB has a very regular structure, i.e. bond lengths and angles are close to bulk silicon. \cite{ZHAO2017599,PhysRevB.91.035309} This is also the most stable GB since there are no dangling bonds and little bond distortions. \cite{SARAU20112264,MAJI2021116477}

We consider the structure of Si bulk as the reference to compare our results. In this case, we used a cubic supercell ($a=b=c$ and $\alpha$ = $\beta$ = $\gamma$= $90^{\circ}$) of 64 atoms with $a$=10.86~\AA. This value for $a$ was deduced from the experimental lattice constant 5.431\AA, for a face-centered cubic unit cell of two Si atoms. \cite{PhysRevB.32.3792}.

Then, we create a vacancy in the $\Sigma$3\{111\} Si GB. There exists only two distinct positions of Si where vacancies can be created. \cite{MAJI2021116477} These vacancies are indicated throughout the paper as V1 and V2 and are shown respectively in panels (b) and (c) of Fig. \ref{fig:silicon-V}.
The formation energy of the vacancy V1 in the GB is $E^{\text{f}}_{\text{VGB}}$ = 2.91 eV while for the vacancy V2 we found $E^{\text{f}}_{\text{VGB}}$ = 3.04 eV. \cite{MAJI2021116477} The presence of the vacancy induces coordination defects. Most specifically, both in V1 and V2 we have two Si atoms that are threefold coordinated while all the other Si atoms are fourfold coordinated.\cite{MAJI2021116477} These atoms are easily identified in Fig. \ref{fig:silicon-V} as those having a lone pair (brown sphere).

To study the role of C, N and O impurities, we have sequentially inserted from one to four of these atoms as interstitials in Si-bulk and in the Si GB with and without vacancies. Different configurations were found but only lowest total energy ones are reported in each case. More details can be found in Ref.\cite{JCP_Rmaji2021}.

\subsection{Computational method}

\subsubsection{Energetics}

The calculations were performed using density functional theory (DFT) as implemented in the plane-wave based Vienna Ab initio Simulation Package (VASP)\cite{Hafner,Kresse} We employed the generalised gradient approximation PBE (Perdew–Burke–Ernzerhof) for the exchange-correlation functional \cite{PhysRevLett.77.3865} and projector augmented-wave (PAW) pseudopotentials with a cutoff of 400 eV. K-points sampling within the Monkhorst Pack scheme \cite{Monkhorst} was used for integration of Brillouin-zone together with the linear tetrahedron method including Bl\"ochl corrections. \cite{PhysRevB.49.16223} We used a k-mesh of 3$\times$3$\times$3. For the structural optimisation, we used a force threshold value of 10$^{-2}$ eV/\AA{} per atom. For all the systems the total energy is calculated as the total energy of the unit cell where all the atoms are interacting. Moreover, the pressure of the systems is calculated directly from VASP and it is the hydrostatic pressure of a cell of a given volume. Finally, in these calculations we did not include the spin polarization. \cite{SR2015,PhysRevB.80.144112} 

Formation energy ($E^{\text{XB}}_\text{f}$) in bulk Si is calculated as
\begin{equation}
E_{\text{XB}}^\text{f} = E_{n_{\text{X}}+\text{B}} - E_{\text{B}}  - n_\text{X} \mu_{\text{X}}.
\label{Bulkimp} 
\end{equation}
Here X: C/N/O refer to different impurity elements, 
$E_{n_\text{X}+\text{B}}$ and $E_{\text{B}}$ are the total energies of Si bulk containing $n_\text{X}$ number of impurity atoms and of pristine Si bulk, respectively. 
$\mu_{\text{X}}$ is the chemical potential of impurity elements (X: O/N/C), calculated as the energy per atom of an O$_2$ molecule in vacuum for oxygen, as the energy per atom of an N$_2$ molecule in vacuum for nitrogen and for C, it is calculated from diamond phase.

Formation energy ($E^{\text{XGB}}_\text{f}$) in Si GB is calculated as
\begin{equation}
E_{\text{XGB}}^{\text{f}} = E_{n_{\text{X}}+\text{GB}} - E_{\text{GB}}  - n_\text{X} \mu_{\text{X}},
\label{BulkimpGB} 
\end{equation}
where $E_{n_\text{X}+\text{GB}}$ and $E_{\text{GB}}$ are the total energies of Si GB containing $n_\text{X}$ number of impurity atoms and of pristine Si GB, respectively. 

Formation energy ($E^{\text{XVGB}}_\text{f}$) in Si GB with vacancy is calculated as 
\begin{equation}
E_{\text{XVGB}}^{\text{f}} = E_{n_{\text{X}}+\text{VGB}} - E_{\text{VGB}}  - n_\text{X} \mu_{\text{X}}
\label{BulkimpV} 
\end{equation}
$E_{n_\text{X}+\text{VGB}}$ is the total energies of Si GB containing $n_\text{X}$ number of impurity atoms in presence of a vacancy and $E_{\text{VGB}}$ is the energy of Si GB with a vacancy. \cite{JCP_Rmaji2021}

\subsubsection{Chemical bond}

To study the structural changes induced by the impurities in the Si-bulk, and in the Si GB with and without vacancies we have analysed bond lengths and the bond charges \cite{SJ92}.

In order to dissect bond charges we have resorted to the Electron Localization Function (ELF). 
Becke and Edgecombe \cite{BE90} introduced
the ELF as a local measure of electron localization. Its core, $\chi$,
reflects the Pauli principle as a local measure of electron kinetic energy density. More specifically, it depends on the Pauli kinetic energy density, $t_P$ (where the bosonic contribution has been taken away), renormalized by the kinetic energy density in the homogeneous electron gas, $t_{TF}$:

\begin{equation}
\chi=\frac{t_P}{t_{TF}}.
\end{equation}

It is desirable to delimit electron localization and to obtain a
direct relationship between the localization and the function. For
those reasons, it was mapped following the lorentzian function
$\eta=\frac{1}{1+\chi^2}$, so that when $\eta$ tends to 1 electrons are  highly localized, whereas $\eta$ tends to 0.5 in those regions where the electrons follow the homogeneous electron gas distribution (i.e. delocalized).

Following Bader's approach \cite{BA90}, this function can be analyzed
within the dynamical system theory context \cite{AM94} to induce a
partition into non overlapping basins. The gradient of the ELF is the key function to recover chemical entities in real space. In general,
the ELF presents a variable topology with mountains, valleys, plateau
zones, and different kinds of critical points (maxima, saddle, ring,
and cage points) where the ELF gradient vanishes. Zero flux surfaces
of the ELF gradient enclose 3D regions or basins that can be associated
with electron shells and sub-shells, bonds, and lone pairs. The power of
the ELF partition in the case of crystalline systems lies in the fact
that these basins are finite, disjoint, and space filling, which means
that adding all of them up, we recover the macroscopic volume. Integration of the electron density within those basins leads to bond and lone pair charges that also recover the total number of electrons when added together.

We have resorted to the  {\tt CRITIC} code \cite{OT08} in order to compute 
basin  charges
($q$). In all cases, we
have checked the performance of the partition by adding up all the
volumes of the basins in which the unit cell is split, since the total
unit cell volume should be recovered. 

\section{Results}
\label{sec:res}

\subsection{Pure silicon}
\begin{figure}[t]
    \centering
     \includegraphics[scale=0.2]{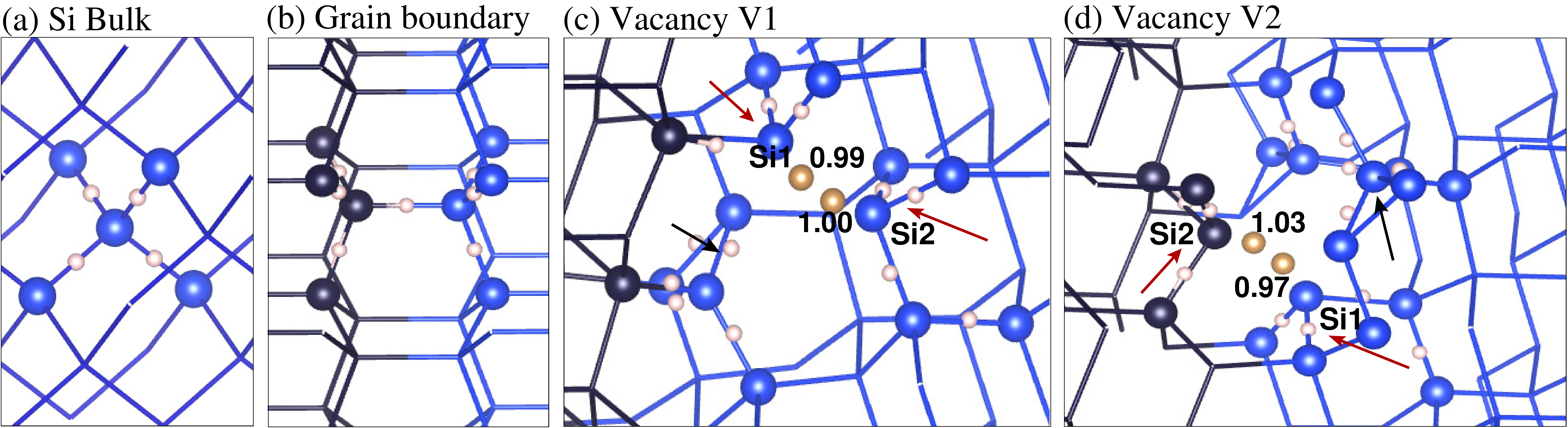}
    \caption{Structures of a) Si bulk b) Grain boundary (GB) and vacancy sites c) V1 d) V2. Silicon atoms in blue. Bonds and lone pairs are highlighted with small white and brown spheres respectively. Amplitude for lone pairs have been mentioned, 3-fold coordinated sites in (c, d) marked as Si1 and Si2. Light and dark blue have been used in order to highlight the grain boundary. Only Si atoms, for which connected bonds are shown here, are represented with "balls". Triangular and pyramidal conformation marked with black and red arrows respectively.}
    \label{fig:silicon-V}
\end{figure}
Results for pure silicon are shown in Fig. \ref{fig:silicon-V}(a).
Bulk silicon has a regular tetrahedral coordination with a Si-Si covalent bond (ELF maxima shown as white spheres) holding 2 electrons.
When the grain boundary is introduced (Fig. \ref{fig:silicon-V}b) the bonding is very much preserved in the four-folded atoms: we observe the appearance of an elongated Si-Si bond (2.37 \AA{}) with a slightly smaller charge (1.98 electrons) that joins the two domains (different colors have been used in order to highlight the grain boundary).

The introduction of vacancies  leads to important bonding changes (Fig. \ref{fig:silicon-V}c-d). 
3-fold coordinated Si atoms appear which  lead in both V1 and V2 to an overall increase of the charge in the neighboring Si-Si bonds (2.01-2.07e). 
The 3-folded atoms appear in most cases in a pyramidal structure, with a lone pair on each Si taking the direction of the ``missing'' Silicon atom. These lone pairs (in brown) hold roughly 1 electron  each (0.99-1.0 electrons) in V1. A greater charge asymmetry is found in V2, where lone pairs hold 1.03 and 0.975 electrons each. Note that in both cases, the two lone pairs add up to the 2 electrons expected for the missing Silicon bond.
On top of the pyramidal conformations, a planar triangular conformation is also observed in V1 where instead of the lone pairs, two confronted Si atoms form a new bond with two ELF maxima. This new bond (highlighted with an arrow in Fig. \ref{fig:silicon-V}c, \ref{fig:silicon-V}d), although weaker than the others (1.54 electrons) allows the atoms involved to recover the 4-fold coordination. 

Finally, it is worth noting that in non-strained cases, as is the case of the grain boundary bond in Fig. \ref{fig:silicon-V}b, the bond appears on the internuclear line. Instead, strained bonds are characterized by ELF maxima outside of the bonding line. This is the case of both V1 and V2, where many of the bonds of the 3-folded atoms are offset. Note that hydrostatic pressure of Si bulk super cell and  Si GB are respectively 1.973 GPa and 1.89 GPa. Whereas in presence of vacancy (V1, V2) hydrostatic pressure of Si GB is P(V1GB)=1.42 GPa and P(V2GB)=1.402 GPa, respectively.\cite{MAJI2021116477} 
The different pressure in the vacancies can be understood from the position of the vacancies with
respect to the grain boundary: whereas V1 lies within one of the domains, V2 lies at the border. This will lead (see below) to important differences when impurities will be adsorbed.

\subsection{Carbon}
\begin{figure}[ht]    
    \centering
   \includegraphics[scale=0.2]{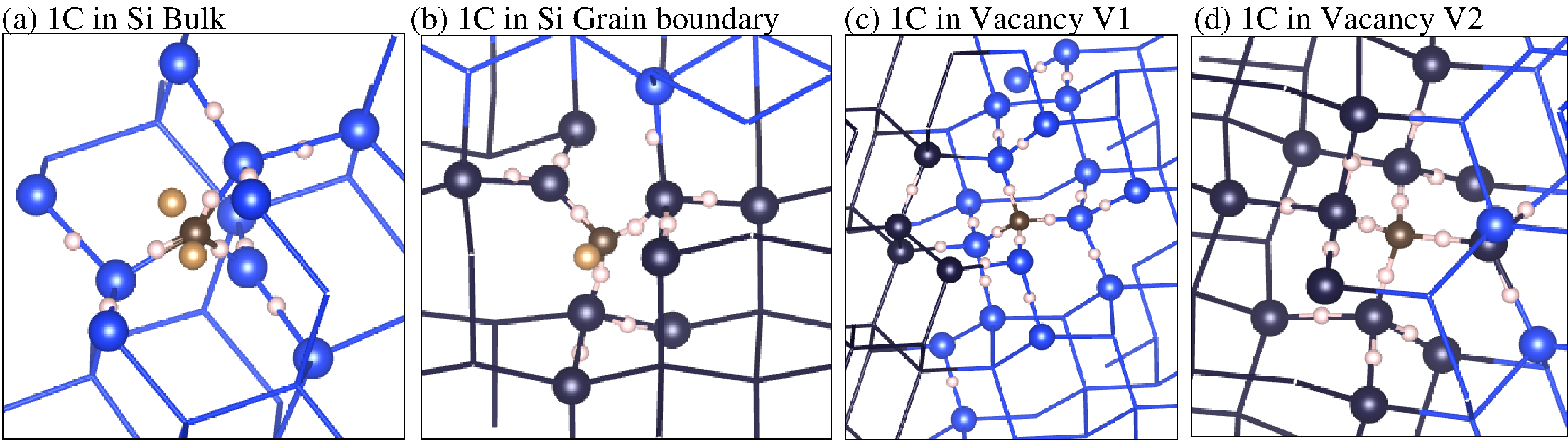}
   \includegraphics[scale=0.178]{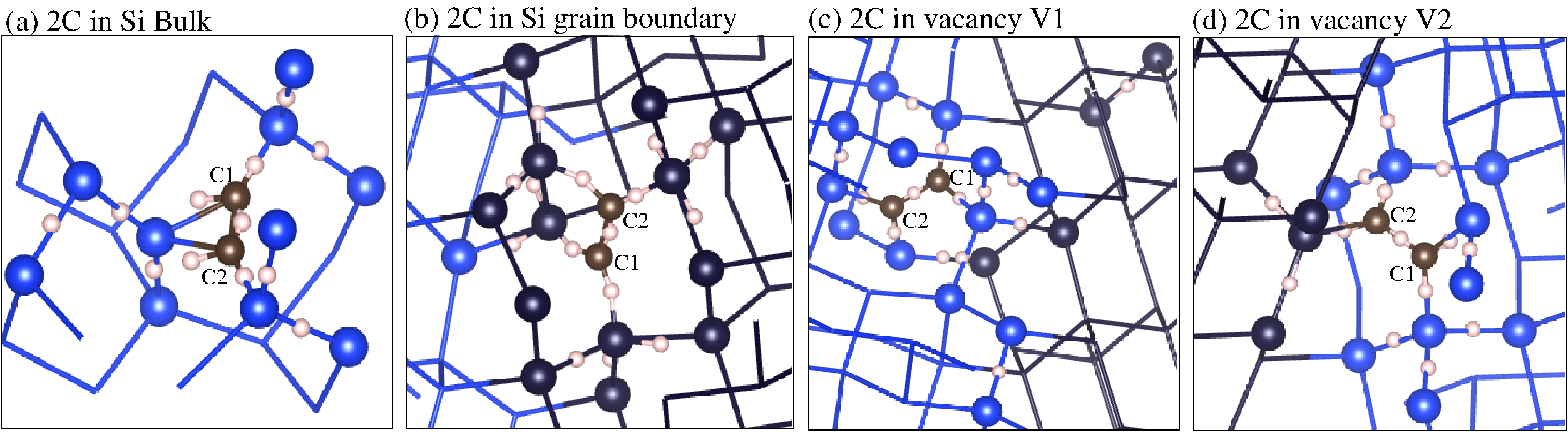} 
    \caption{Structures with the inclusion of 1 (top) and 2 (bottom) carbon atom(s) in  a) bulk Si b) GB b) V1 c) V2. Silicon atoms in blue and carbon atoms in dark brown. Bonds and lone pairs are highlighted with small white and brown spheres respectively.}
    \label{fig:1Celf}
\end{figure}

Given the stable Si bulk and GB structures, it is unfavorable to introduce an extra C atom (see Formation energy in Fig. \ref{fig:en-n}, discussed later). The C atom is 3-fold coordinated in both cases, and places the rest of its valence charge in lone pairs (Figs. \ref{fig:1Celf}a-b): two lone pairs in bulk Si with 0.13+0.012=0.142 electrons in total and one lone pair in the GB with 0.6 electrons. Very low population lone pairs are usually only observed in delocalized and/or unstable configurations. Another signature of instability are the strained C-Si bonds shown both in bulk (see SI: Table \ref{C@Si-Bulk} in comparison to bulk SiC value of Table \ref{Bulk_ref}) and GB (see SI: Table \ref{C@Si-GB} in comparison to bulk SiC value of Table \ref{Bulk_ref}).
The image in the case of vacancies, where there is space for the inclusion of the atom, is very different (Figs. \ref{fig:1Celf}c-d).
Carbon belonging to the same group as Si, it is able to take up Si missing positions in the case of vacancies. 
This is what we observe in the case of V1 and V2: symmetric tetrahedral arrangements are obtained around carbon. Moreover, C and Si have similar electronegativities, leading to 2 electron bonds (see SI: Table \ref{C@Si-V1GB}, \ref{C@Si-V2GB}). 
Note that in the case of V1 bonds are strained, whereas they are not in V2 (i.e. bond charges are offset). Noteworthy, V1 leads to a greater bond length scattering (1.99-2.03 \AA) than V2 (2.01-2.03\AA).

\begin{figure}[t]    
    \centering
    \includegraphics[scale=0.32]{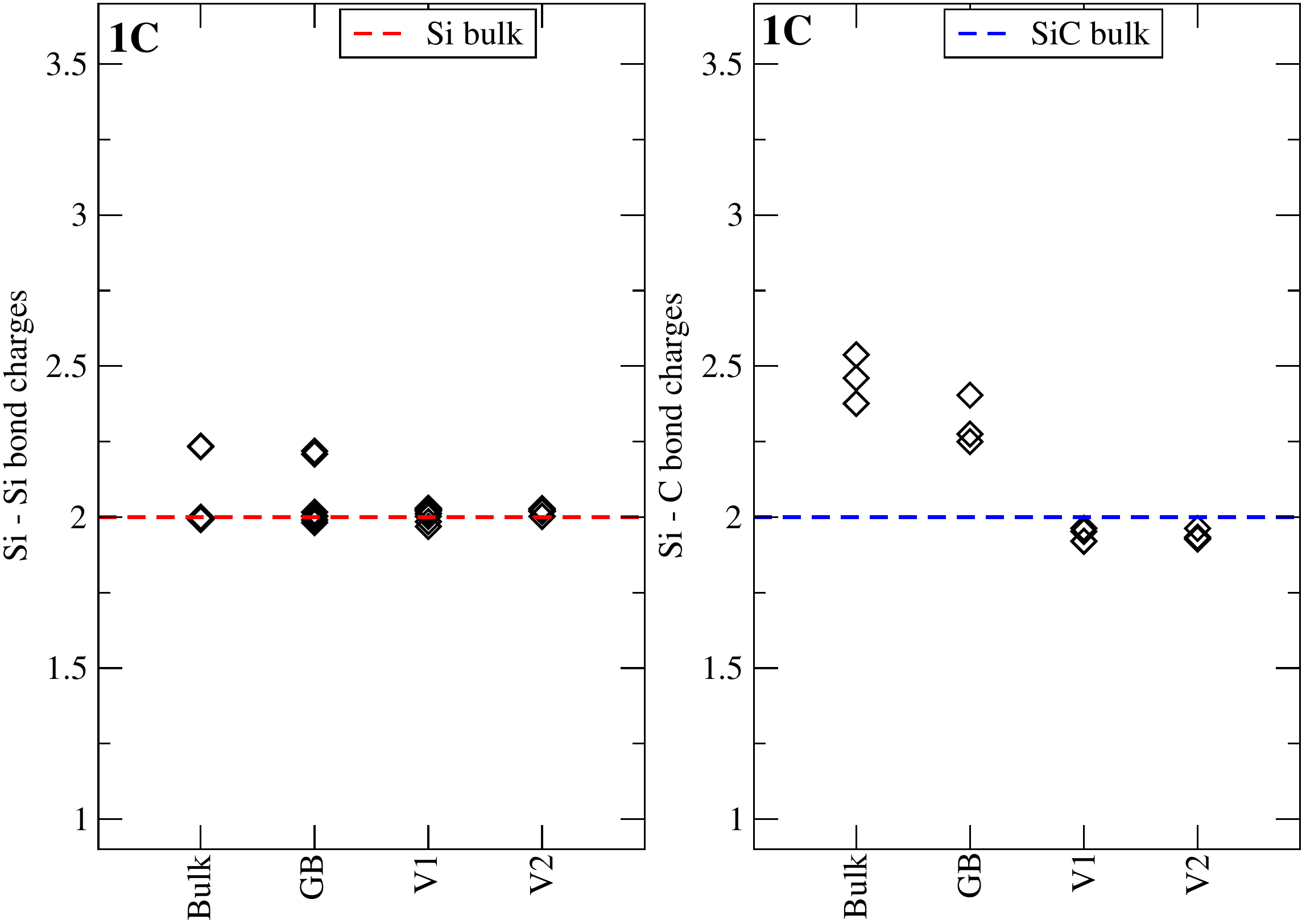}
    \includegraphics[scale=0.32]{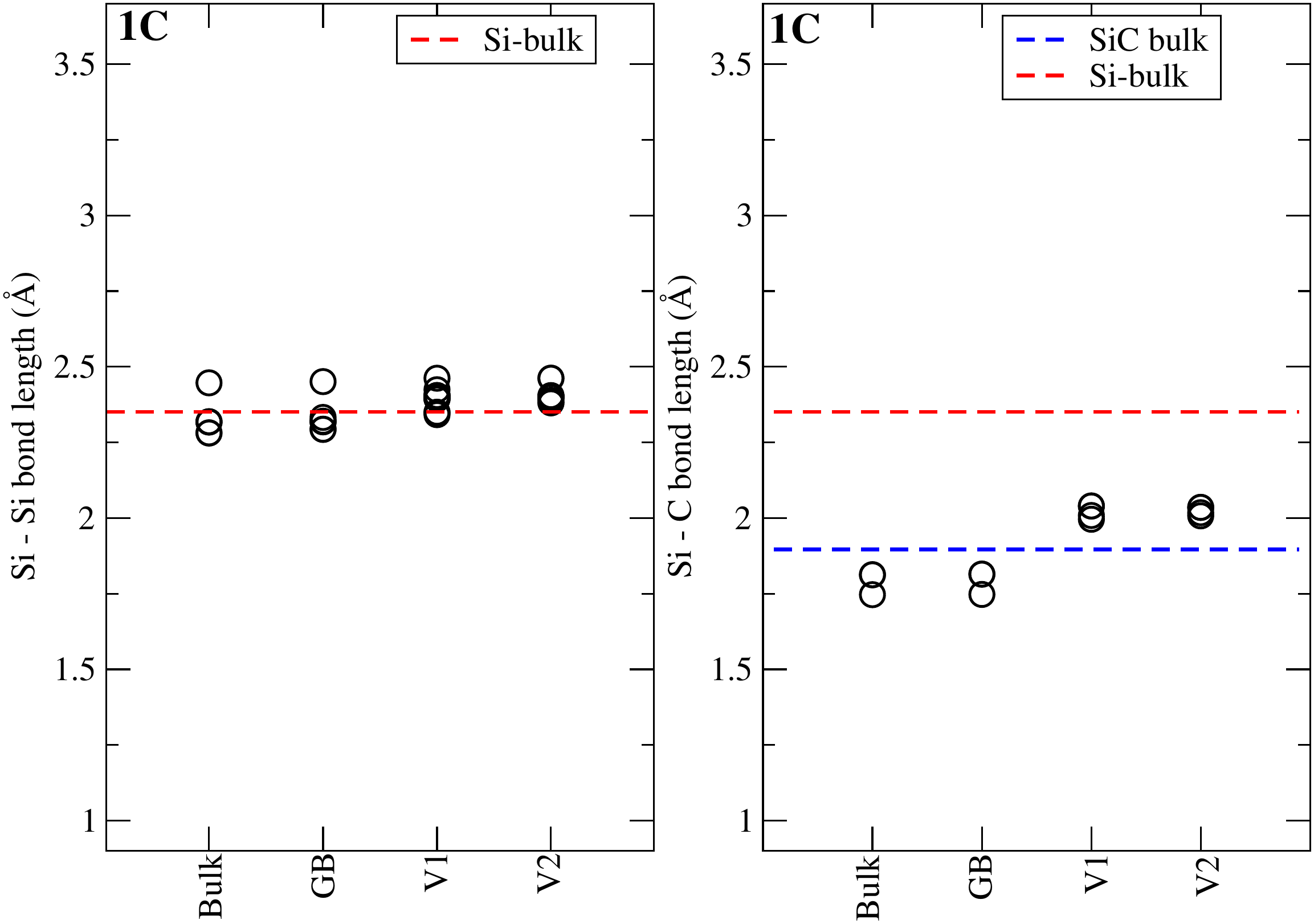}
    \caption{Bond charge and bond length variation of Si-Si and Si-C bonds for the inclusion of one carbon atom in Si-bulk, Si-GB and in presence of vacancy V1 and V2. As a reference Si-Si/Si-C bond charges/lengths from cubic bulk phases of Si and SiC are marked by dotted lines in respective plots. }
    \label{fig:1cbonds}
\end{figure}

Fig. \ref{fig:1cbonds} 
shows Si-Si and Si-C bond charges and lengths in the different systems. The red line provides the bulk Si reference value, while the blue line provides the bulk SiC reference value. We can see that charges (Fig. \ref{fig:1cbonds} top) shed more light on the electronic structure of these strained systems than distances (Fig. \ref{fig:1cbonds} bottom).
Indeed, both Si-Si and Si-C distances are spread around the reference values indicating a strained structure. Instead, bond charges in the V1 and V2 systems are very close to the nominal values of the two (whereas distances are not). In other words, the vacancy systems are able to relax to lead to a stable electronic conformation, even if the structure is locally distorted. We will thus in the following mainly focus on the charges. 

As we have seen, bulk and GB systems with 1C impurity are not able to easily accommodate the carbon atom and lead to offset bond charges, and even to lone pairs. In general, a charge flow towards the impurity is observed in bulk and GB, with greater Si-Si and Si-C charges than in the undoped case.

\begin{figure}[t]    
    \centering
    \includegraphics[scale=0.4]{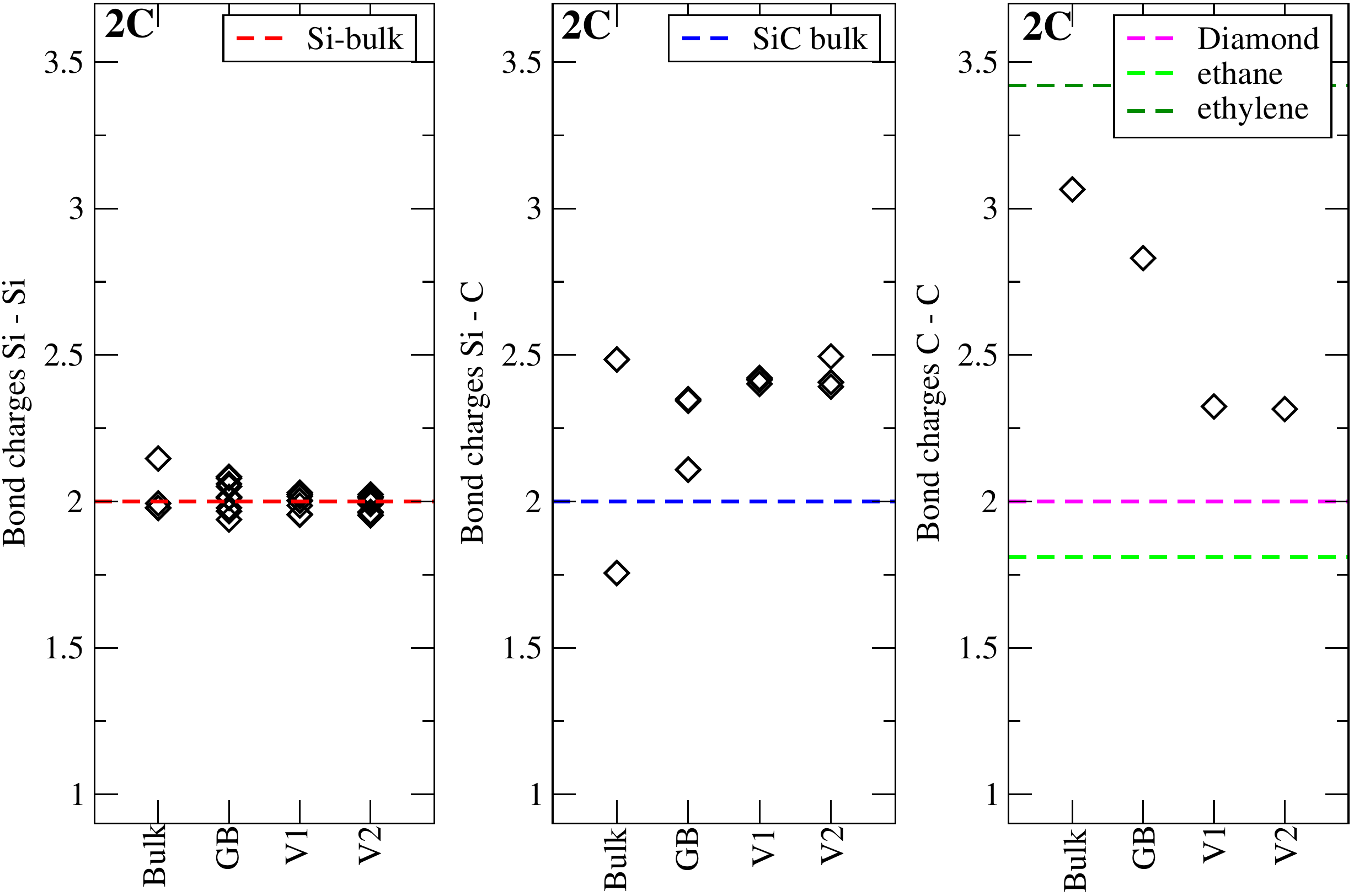}
    \caption{Bond charge variation of Si-Si, Si-C, and C-C bonds for the inclusion of two carbon atoms in Si-bulk, Si-GB and in presence of vacancy V1 and V2. As a reference, Si-Si/Si-C bond charges from cubic bulk phases of Si and SiC, and C-C bond charges from diamond, ethane and ethylene are marked as mentioned in the inset of respective plots.}
    \label{fig:2cbonds}
\end{figure}

When two carbon atoms are introduced in the silicon systems (Fig. \ref{fig:1Celf} bottom), the two C atoms are always found forming a C-C bond in an arrangement similar to ethylene: two Si atoms are attached to each C (i.e. 3-fold coordinated). 
Among the four systems, a slight difference appears in the coordination in the case of bulk silicon. 
The little place available in the case of the bulk leads to a Si atom coordinated to the C=C unit in a similar fashion to the Dewar–Chatt–Duncanson model.  The charge involved in each of these C-Si bonds is of 1.76 e. 
In the case of GB, a unit rather close to ethylene is found, with a C-C charge slightly lower than in C$_2$H$_2$ (ELF calculation on molecular ethylene leads to a C-C bond charge of 3.4 e - blue line in Fig. \ref{fig:2cbonds}).
In the case of the vacancy systems, and in spite of the 3-fold coordinated C atoms, the C-C bond holds 2.3 e, much closer to that of ethane than ethylene.
Overall, a charge deficiency is observed in the C-C bond, which is balanced by a higher charge in the C-Si bonds of ca. 2.4 e. Exceptions are the coordination Si-C bonds in the bulk and some of the bonds in the GB.
The charge balance thus happens locally, so that a  little Si-Si bond charge dispersion is found (Fig. \ref{fig:2cbonds}). 

\begin{figure}[b]    
    \centering
   \includegraphics[scale=0.18]{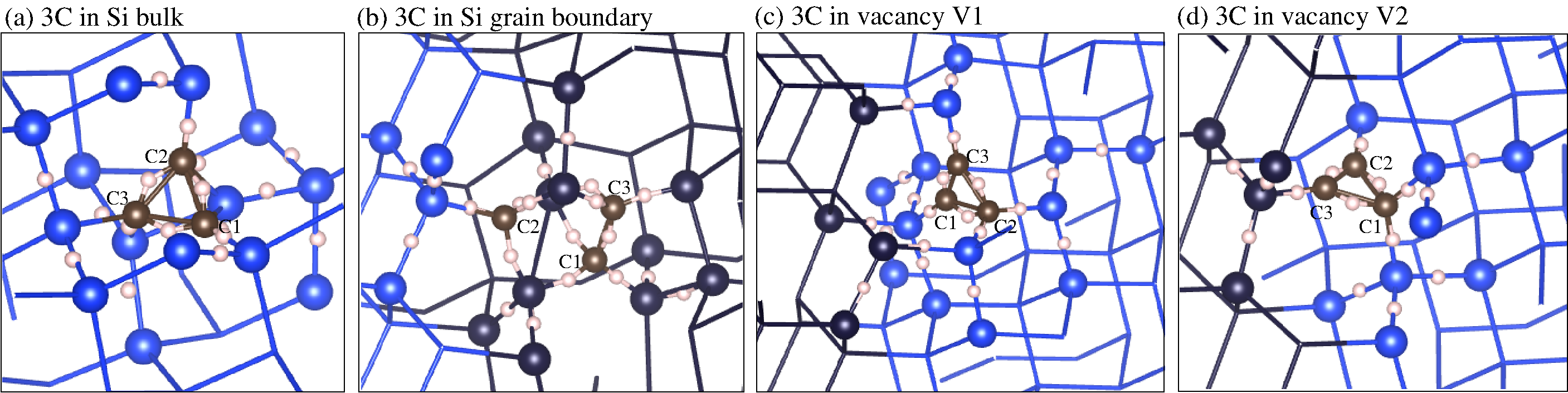}
   \includegraphics[scale=0.145]{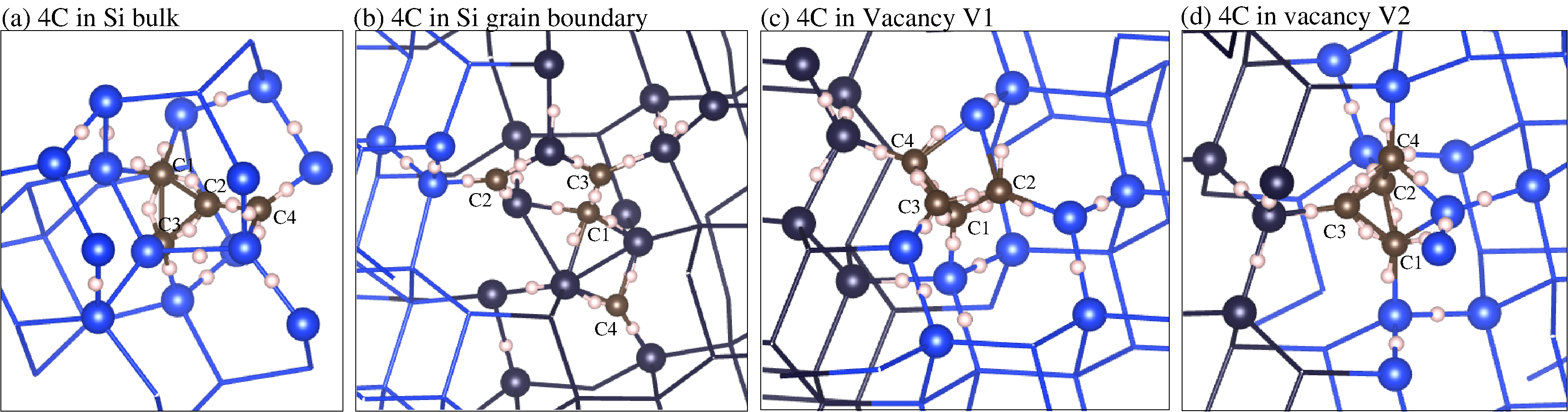}
    \caption{Structures with the inclusion of three (top) and four (bottom) carbon atoms in  a) bulk Si b) GB b) V1 c) V2. Silicon atoms in blue. Bond charges are highlighted with small white spheres. Lone pairs with values $\leq$ 0.01 not shown here are mentioned in SI: Table \ref{C@Si-Bulk}, \ref{C@Si-GB}, \ref{C@Si-V1GB}.}
    \label{fig:3Celf}
\end{figure}
Three carbon atoms (Fig. \ref{fig:3Celf} top) show different patterns depending on the Si structure. Bulk and vacancies lead to the formation of a triangular structure, whereas in the grain boundary the 3C atoms are distributed separately.

A difference similar to the 2C inclusion appears between the bulk and the vacancies: a coordinated Si bridge appears in the bulk. In this case, the bridge in the bulk leads to all the 3C atoms being four fold coordinated with all C having 2 C-C bonds and 2 Si-C bonds. C-C bonds hold 1.5-1.6 e.
C-C bonds are hence close to the single bonds in cyclopropane (1.7 e). The C-Si bridge shows the same pattern as for the 2C atoms: Si-C populations are high (2.4 e) for the terminal bonds and lower (1.9 e) for the bridge bonds.

\begin{figure}[t]    
    \centering
    \includegraphics[scale=0.4]{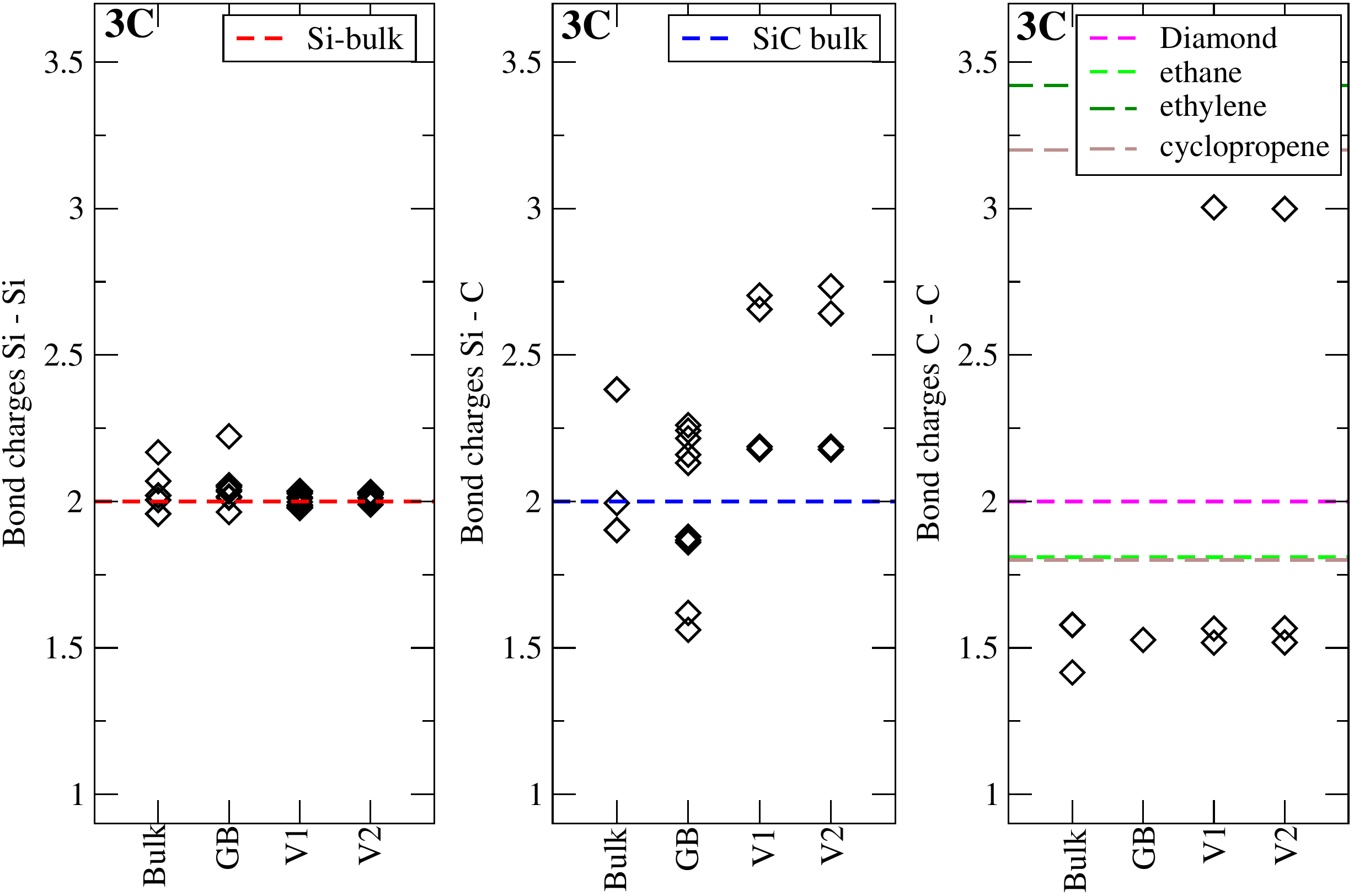}
    \caption{Bond charge variation of Si-Si, Si-C, and C-C bonds for the inclusion of three carbon atoms in Si-bulk, Si-GB and in presence of vacancy V1 and V2. As a reference, Si-Si/Si-C bond charges from cubic bulk phases of Si and SiC, and C-C bond charges from diamond, ethane, ehtylene and cyclopropene are marked in the inset of respective plots.}
    \label{fig:3cbonds}
\end{figure}

Instead, in the vacancies
two carbon atoms are 3-fold coordinated, leading to a unit similar to cyclopropene: they show two single 1.5 e bonds and one double 3.0 e bond (Fig. \ref{fig:3cbonds}). This is to be compared with cyclopropene which has 1.8 e in the C-C bond and 3.2 e in the C=C one. 

Just like in molecular cycles in general, bonds are very strained, which is reflected in the electron pair localized outside of the C-C bonding line. Note that a smaller strain is found in V2 than in V1 (maxima are closer to the internuclear lines), which explains the easier formation of the inclusion of 3C in V2 than in V1 (0.9758 eV and 0.8144 eV in V1 and V2, respectively - see SI: Table \ref{Ef_C}).

Si-C bonds next to the bridge in the bulk and next to the double bond in the vacancies are highly populated, releasing again the charge from the ``strained'' areas.
This leads again to Si-Si bonds close to 2 e except for slight deviations close to the bridging Si atom.

If we now look at the organization in the grain boundary we see that the 3C atoms are distributed separately: one C-C bond inside a domain and one separate C atom at the boundary. The separate C atom, just like for the 1C inclusion, shows a small lone pair (LP) with negligible charge.
In this case, the C-C bond corresponds to an ethane configuration, with both carbon atoms being four folded. The C-C bond charge is consequently smaller than in the 2C case, with only 1.57 e. The Si-C charges are very variable. This is again linked to the appearance of a bridge atom, where the Si-C charge is 1.5e, whereas the terminal C-Si bonds hold 2.1-2.2e. This also leads to the appearance of a higher Si-Si bond charge (2.2 e) at the $\beta$ bridge position (bond next to the Si-C bridge). 


Four carbon atoms (Fig. \ref{fig:3Celf} bottom) are difficult to intercalate all together, so they lead to very different conformations in the various matrices.
In the bulk they adopt a methylenecyclopropane arrangement. The methylene bond holds 3 e. The bonds within the cyclopropane hold ca. 2 e, except for the bond opposite to the methylene, which holds 1 electron only. This weak bond is accompanied by the appearance of a small lone pair (0.02 electron) on one of its atoms. Overall this represents a rather unstable doping.
The C-Si bonds show a charge of 1.8-1.86 e and the Si-Si bonds are in most cases nominal (2.0 e) except one $\beta$-Si-Si bond with respect to the C atoms holding the lone pair. This shows how this atom perturbs the structure in a longer range and highlights a general pattern in the bulk adsorption and the formation of bridges.

When a grain boundary exists, defaults and voids are present allowing separate carbon conformations: a C-C bond and two independent C atoms are adsorbed, all of them with uncommon coordinations: the independent atoms are 3-fold coordinated (C-Si bonds with 2.5 e in one of the atoms and 2.1-2.7 for the other one). The C-C bond also shows one of its ends with a 3-fold coordination and a C-C bond of 2.5 electrons. The different coordinations are balanced out to a good extent by the populations. Whereas the 4-folded carbon has two deficient bonds (C1: 1.1. and 1.7), the 3-folded atom has two overpopulated C-Si bonds (C3: 2.2 and 2.3) (SI: Table \ref{C@Si-GB}).
The Si structure shows in general little perturbations except again for the Si-Si bond $\beta$ with respect to the 3-folded atom in the C-C bond.

\begin{figure}[t]  
    \centering
    \includegraphics[scale=0.4]{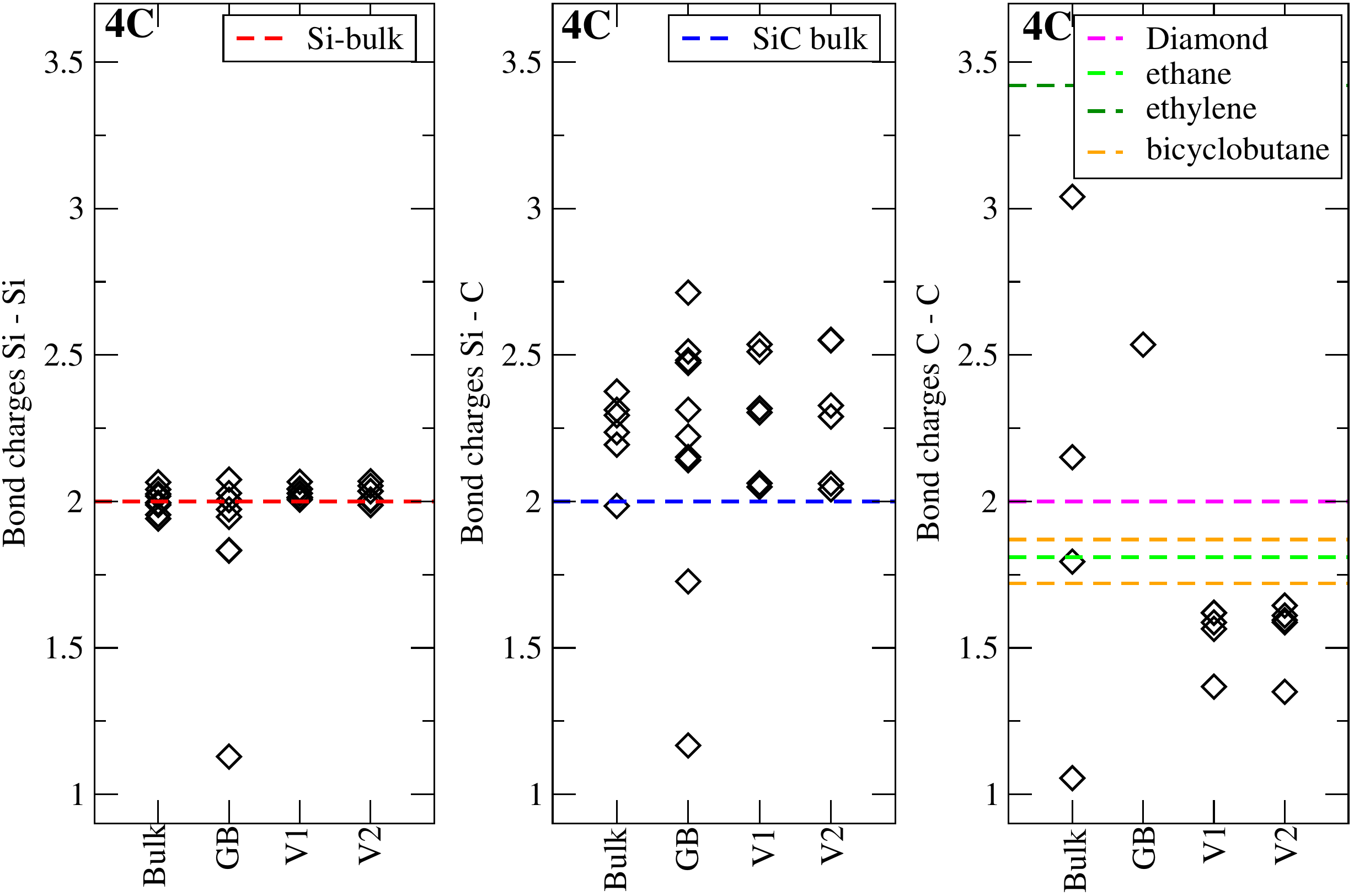}
    \caption{Bond charge variation of Si-Si, Si-C, and C-C bonds for the inclusion of four carbon atoms in Si-bulk, Si-GB and in presence of vacancy V1 and V2. As a reference, Si-Si/Si-C bond charges from cubic bulk phases of Si and SiC, and C-C bond charges from diamond, ethane, ethylene and bicyclobutane are marked in the inset of respective plots}
    \label{fig:4cbonds}
\end{figure}

\begin{figure}[t]   
    \centering
  \includegraphics[width=1.0\textwidth]{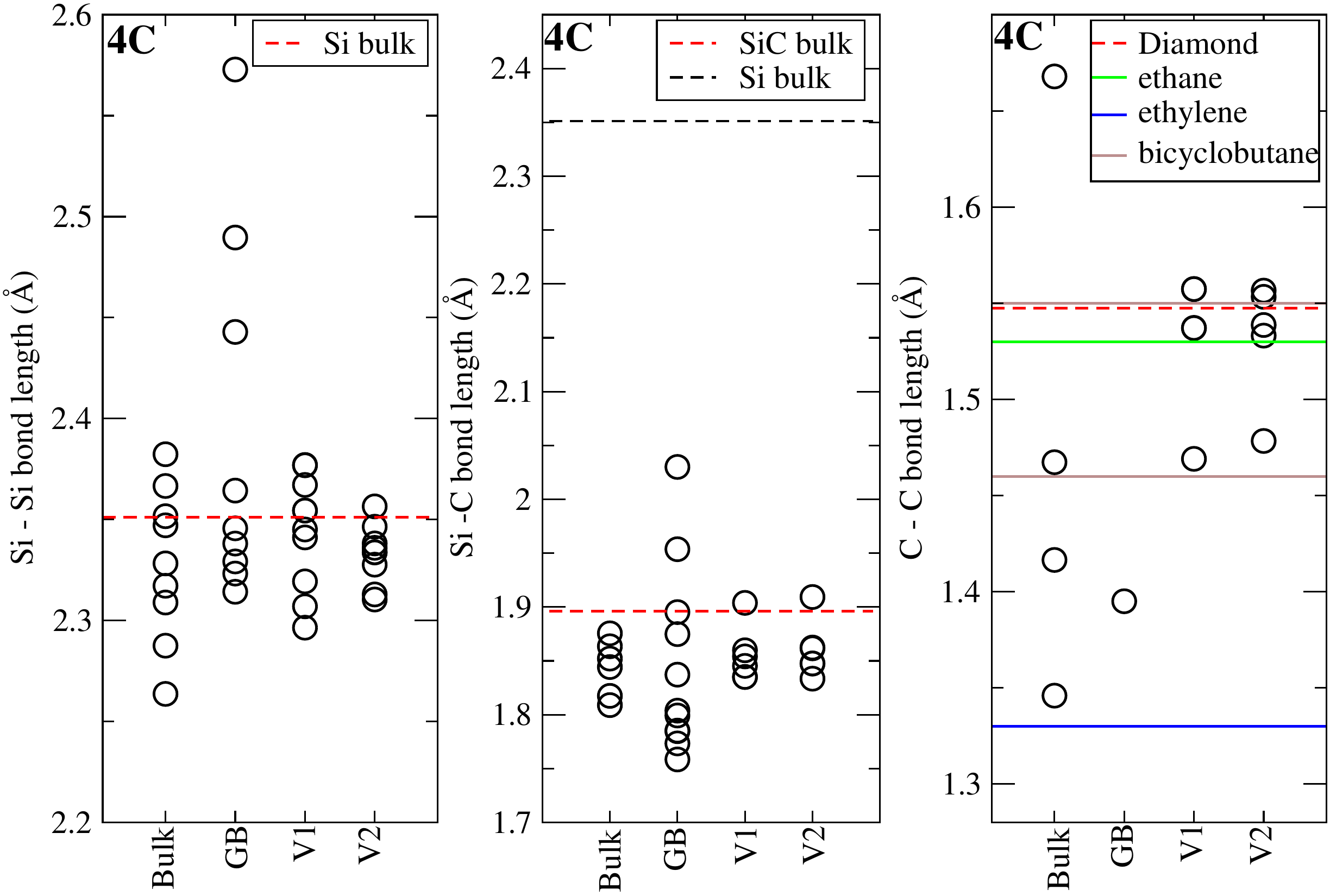}
    \caption{Bond length variation of Si-Si, Si-C and C-C for the inclusion of four carbon atoms in Si-bulk, Si-GB and in presence of vacancy V1 and V2. As a reference, Si-Si/Si-C/C-C bond length from bulk phases of Si, SiC, diamond and molecular systems ethylene, ethane, and bicyclobutane are marked in the inset of respective plots}
    \label{fig:4cbondsl}
\end{figure}

In the presence of vacancies, the bicyclobutane is the most stable organization for both V1 and V2. The C atoms form a bicycle where a weak C-C bond is observed in the bridge (1.6 e and 1.36 e in V1 and V2, respectively). The other C-C bonds are single and strained (1.55-1.6 e). The head C-Si bonds are again strained due to the tension induced by the bicycle. This leads to smaller charges (2e) with respect to the other C-Si bonds (2.3 for the head C-Si bonds, even 2.0 in V1, and 2.5 for the bridge atoms). Although unstable, the bicycle structures lead to only local perturbations, with Si-Si bonds holding charges very close to the nominal 2.0 electrons. This is easily seen in Fig. \ref{fig:4cbonds} top left, where the dispersion in V1 and V2 of Si-Si values is very small. Note that once again, this does not correspond to the Si-Si distances image (see SI: Fig.\ref{fig:4cbondsl}), where for example in V1 a greater distance is observed. Nevertheless, the structure rearrangement manages to preserve the Lewis structure.
Note that the formation energy of these structures (SI: Table \ref{Ef_C})
is given by 2.2861 eV and 2.1306 eV for 
V1 and V2, respectively. Once again, the local distortion of V2 requires a slightly smaller energy than V1.

\begin{figure}[t]
    \centering
    \includegraphics[scale=0.5]{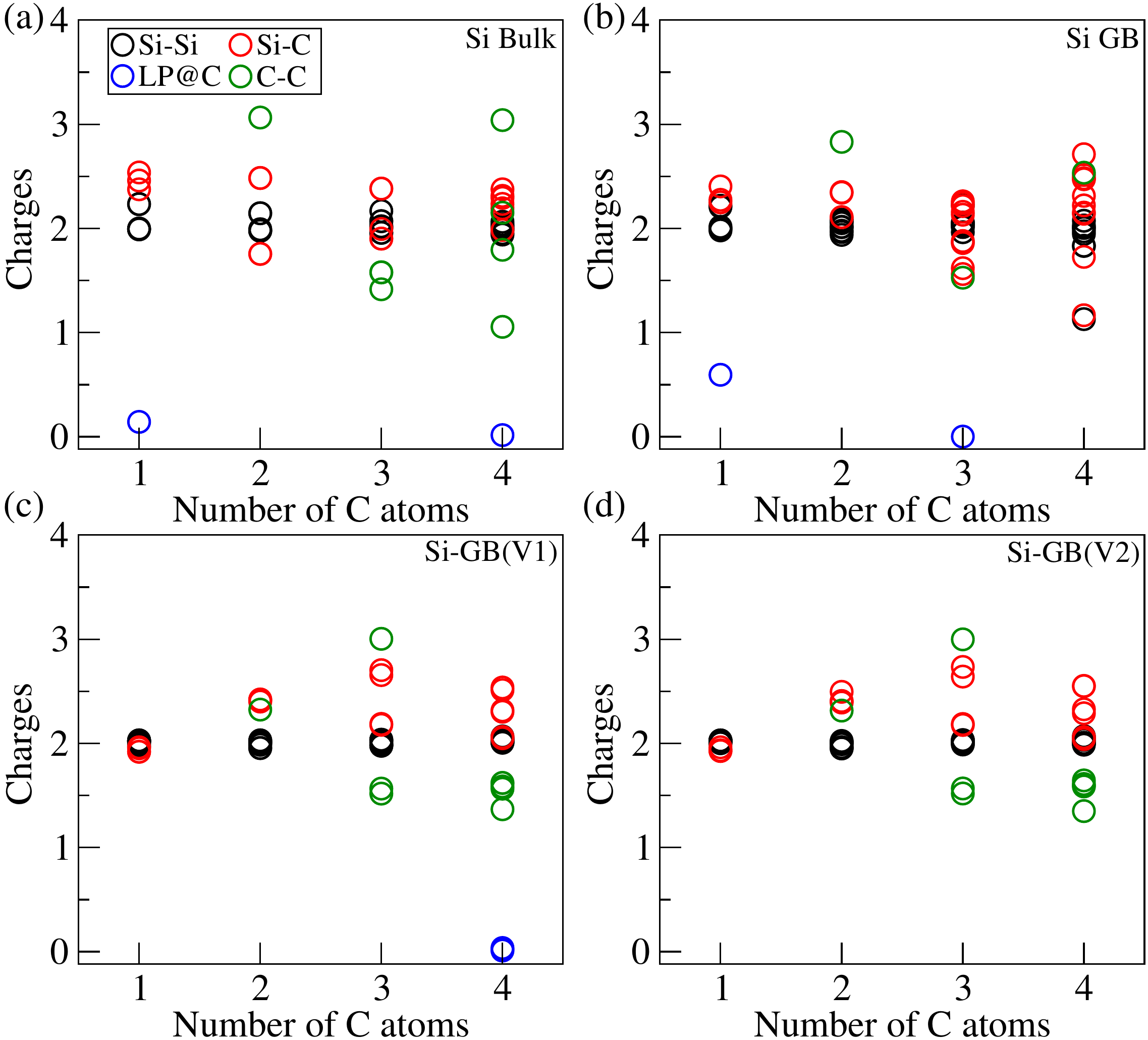}
    \caption{Bond charges (Si-Si, Si-C and C-C) and lone pairs at C (LP@C) variation  for the inclusion of different numbers of carbon atoms in a)bulk Si b) GB c) V1 d) V2.}
    \label{CLP}
\end{figure}

Fig. \ref{CLP} summarizes the trends with the number of atoms. We can see that the Si matrix is more affected in the bulk and GB structures than V1 and V2. In all cases, the charge dispersion increases as we increase the number of dopants, reflecting the difficulty to fit them into the matrix. One exception is the inclusion of 4C in V1 and V2, which leads to a smaller dispersion than 3C. Lone pair charges have also been included in blue. This allows to identify the cases in which charges cannot be accommodated in bonds. Looking at their respective charges, we can see that even in the case they appear, their charge is negligible with one exception only, the lone pair for 1C in the GB. If we now check the general trends in formation energy (Fig. \ref{fig:en-n}), we can see that bulk and GB on the one side and V1 and V2 on the other follow parallel trends with the only exception of 3C, where the separate inclusion of 2C and 1C in the GB seems more favorable than the 3 atoms together in the bulk.

\subsection{Nitrogen}

In this case it is worth starting by analyzing the stable Si-N mixture, Si$_3$N$_4$ (SI: Fig. \ref{fig:references}). In this compound, where valences are naturally compensated, Si is 4-fold coordinated and N is 3-fold coordinated. The Si-N distance is 1.75\AA~and Si-N bond charge is 2.3 e. Each N atom holds two lone pairs of 0.528 e each (see SI: Table \ref{Bulk_ref}). Taking into account the coordination, total charge leads to 7.96 e around the nitrogen atom.  

\begin{figure}[h]    
    \centering
    \includegraphics[scale=0.22]{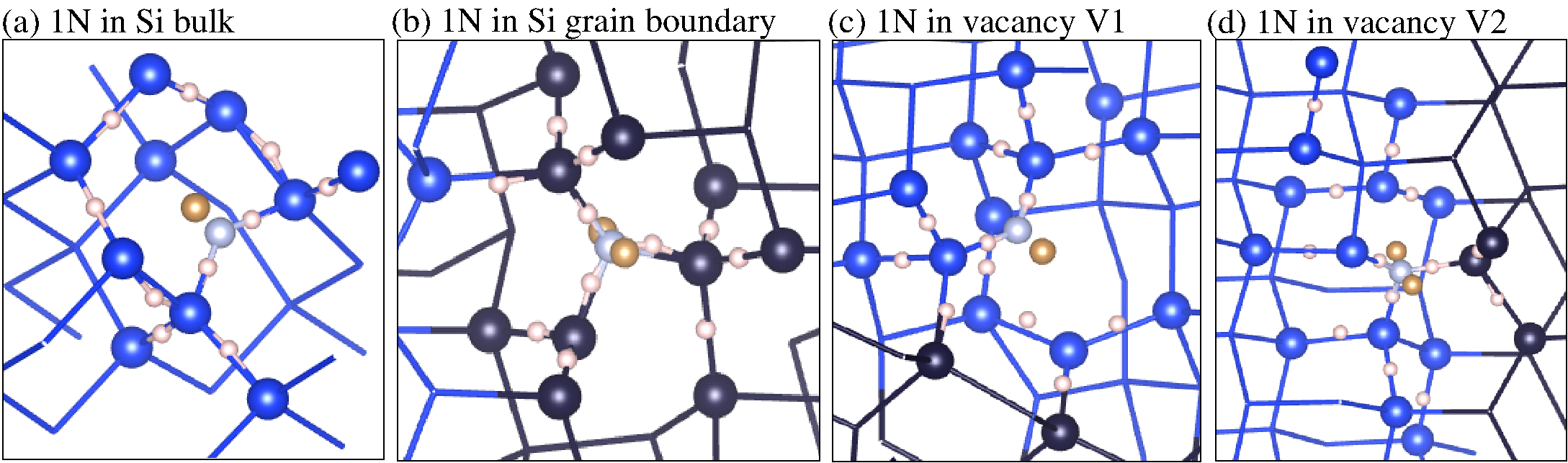}
    \caption{Structures with the inclusion of one nitrogen atom in a) bulk Si b) GB c) V1 d) V2. Silicon atoms in blue. Bonds and lone pairs are highlighted with small white spheres and brown spheres respectively.}
    \label{fig:1Nbond}
\end{figure}
When we introduce one N atom in the Si matrices, two different arrangements appear: 2-fold coordination with one LP, and 3-fold coordination with 2 LPs.
The first case is observed in the bulk and in V1. In the bulk, the lone pair  on N leads to a planar electronic symmetric distribution (ca. 2.4 e on each bond and on the LP). This structure leads to dispersion in the Si-Si bond charges, which show 1.9 e in the $\beta$ bonds with respect to the inclusion. 
The configuration for the interstitial in V1 is similar to the one in the bulk: two fold coordinated and 1 LP. However, there is a bigger charge polarization. The LP holds 2.9 e and Si-N bonds have 1.7 e. A smaller charge Si-Si dispersion is observed in this case, with Si-Si bonds holding 1.95-2.0 e.

The GB enables N to be inserted in a 3-fold coordination. The bonds are  lowly populated (1.75-2.0 e). Two LPs appear (1.9 e in total) to complete the Lewis structure.
The arrangement for V2 is conceptually similar to the one in the GB for 1 Nitrogen atom: it is 3-folded with a slightly deficient population (1.95 e) and two LPs (1.69 e in total). However, a difference arises at the Si level: the 3-folded Si shows a LP (0.4 e) in the direction of the missing bond (SI: Table \ref{N@Si-V2GB}).

\begin{figure}[h]    
    \centering
   \includegraphics[scale=0.2]{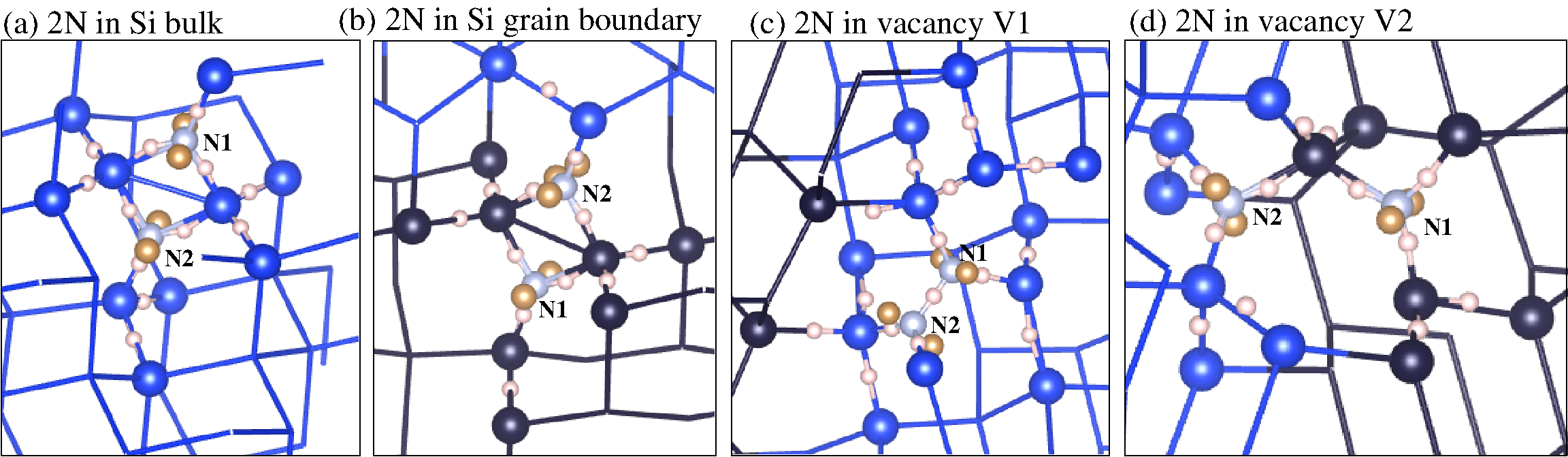}
    \caption{Structures with the inclusion of two nitrogen atoms in a) bulk Si b) GB c) V1 d) V2. Silicon atoms in blue. Bonds and lone pairs are highlighted with small white spheres and brown spheres respectively.}
    \label{fig:2Nbond}
\end{figure}
When we introduce 2N atoms in the bulk they alternate with Si atoms, instead of forming N-N bonds. This leads to a local configuration similar to Si$_3$N$_4$ (see SI: Table \ref{Bulk_ref}): N is 3-fold coordinated and has 2 LPs. However, electrons are distributed differently: N-Si bonds hold 2-2.3 e and the lone pairs hold 0.7 e each. The total charge around N remains nevertheless similar, with 7.7 e. 
The GB leads to a similar structure, though N-Si bonds are slightly more populated (2-2.5 e). One of the N atoms remains with a similar LP distribution whereas in the other one, there is a LP split majorly in two (N2: 0.4 e + 0.5 e). 

\begin{figure}[h]    
    \centering
    \includegraphics[width=1.0\textwidth]{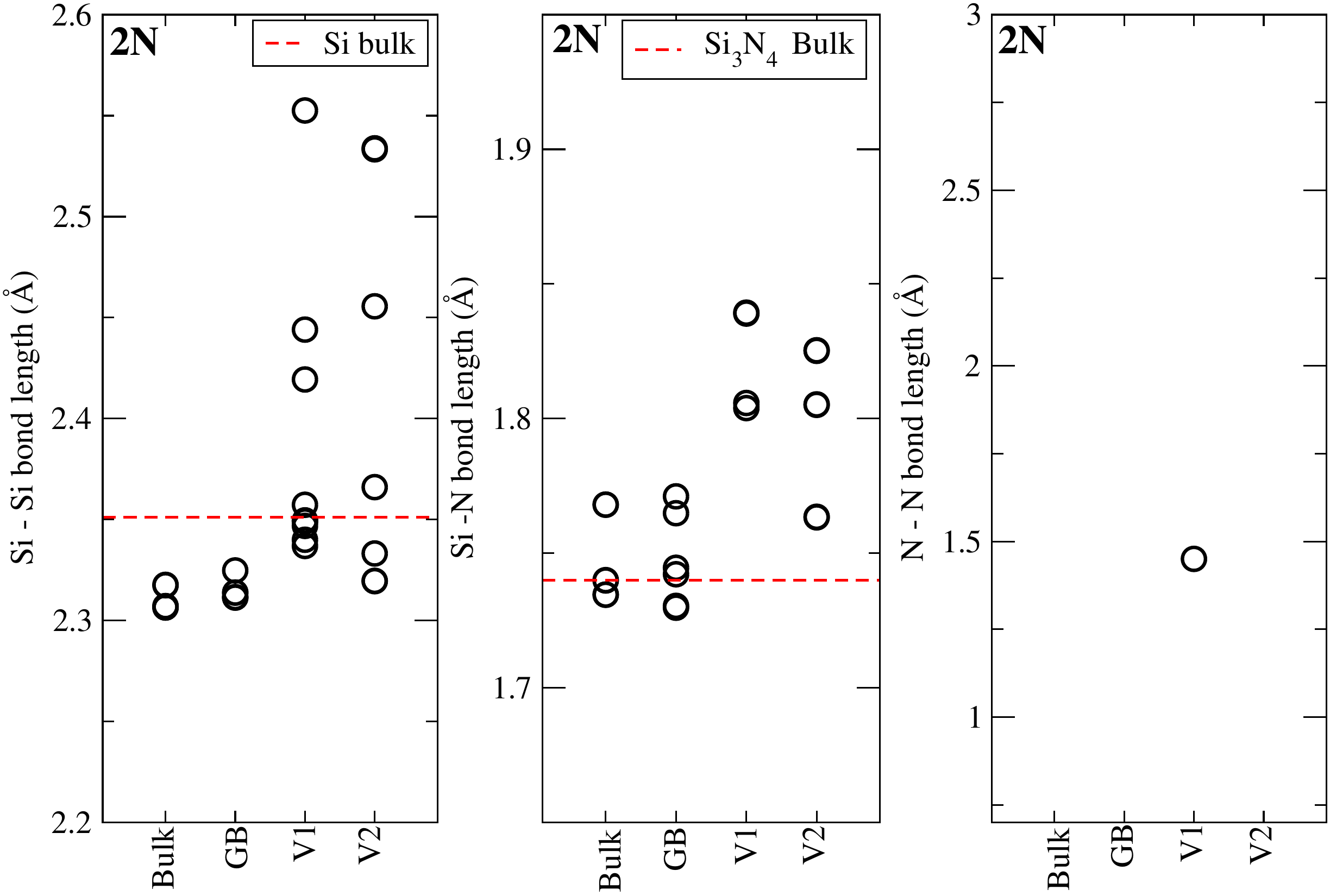}
    \caption{Bond length variation of Si-Si, Si-N and N-N for the inclusion of two nitrogen atoms in Si-bulk, Si-GB and in presence of vacancy V1 and V2. As a reference, Si-Si/Si-N bond length from bulk phases of Si, Si$_3$N$_4$ are marked in the inset of respective plots.}
    \label{fig:2nbondsl}
\end{figure}

V1 is the only case we have observed where a N-N bond is formed (Fig. \ref{fig:2Nbond} c). This N-N bond is single, keeping the 3-fold coordination around each N. The N-N bond holds 1.45 e and each N shows two LPs whose added charge is in the range 2.8-3.1 e.
V2 shows an structure similar to 1N, with the two N adsorbed around the same Si atom, lowly populated bonds (1.8-1.9 e) and two LPs (1.8 e in total).
In all cases Si-Si bonds are overpopulated, even though a great dispersion of distances appears with Si-Si bonds both longer and shorter than in bulk Si (SI: Fig. \ref{fig:2nbondsl}).

\begin{figure}[t]    
    \centering
    \includegraphics[scale=0.173]{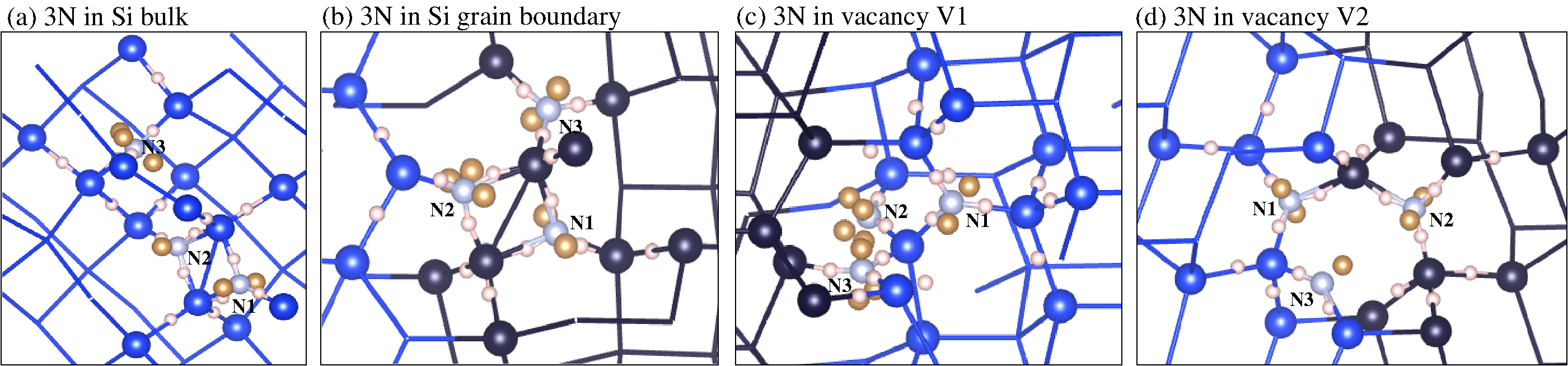}
    \includegraphics[scale=0.18]{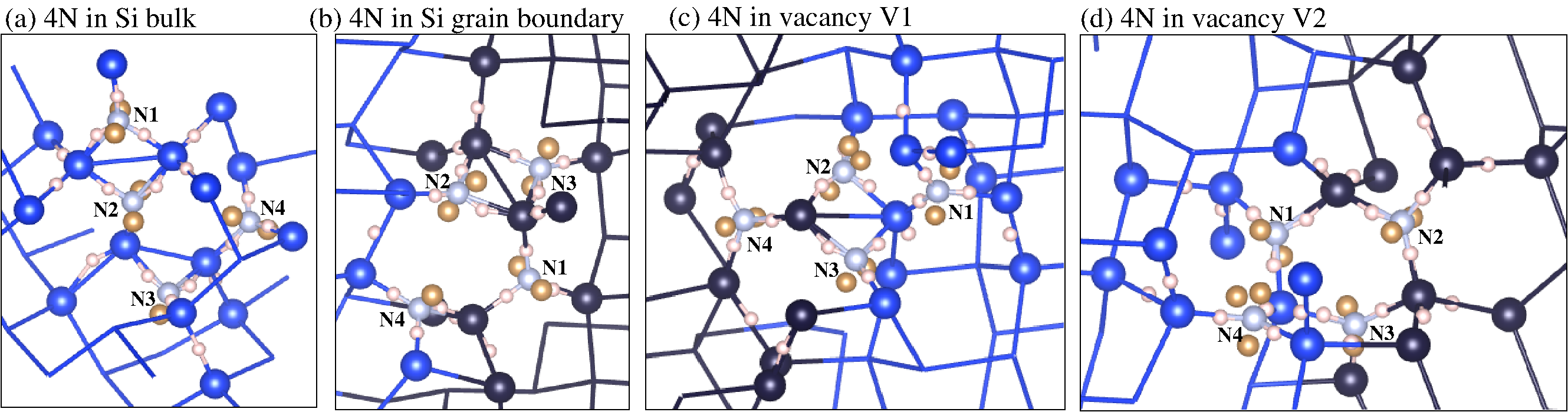}
    \caption{Structures with the inclusion of three (top) and four (bottom) nitrogen atoms in a) bulk Si b) GB c) V1 d) V2. Silicon atoms in blue. Bonds and lone pairs are highlighted with small white spheres and brown spheres respectively}
    \label{fig:3Nbond}
\end{figure}

The adsorption of three N atoms in bulk Si leads to two separate interstitials (Fig. \ref{fig:3Nbond}a): a structure similar to 2N and an isolated N atom. This is similar to what was already observed for 3C in the bulk. The bond charges for the 2N complex are similar to the 2N results, though N LPs have smaller total charges (0.7-0.9 e). The remaining N atom is 3-fold coordinated by e-deficient bonds (1.8-1.9e) and shows 3 LPs which add up to 1.97 e.
Bonds hold ca. 2 e except for the Si-N bonds which are not shared by the two nitrogen atoms, which hold ca. 2.3 e.

For V2 (Fig. \ref{fig:3Nbond}d top), the adsorption takes place in a Si-N chain, with one of the terminal atoms only two-fold coordinated (aka N3). This atom shows highly populated Si-N bonds (2.5-2.7e) and two LPs which add up to 2.1 e. The N atom in the middle of the chain (aka N1) shows a high bond population in the direction of N3 (2.4e) and two LPs with a low charge of 1.25 e. The terminal 3-fold coordinated N atom (aka N2) has a 1.95 e bond with toward N1 and lowly populated bonds with the rest of the matrix (1.7-1.8 e). Its LPs add up to 2.1 e. We have thus different coordinations for the terminal atoms but both holding ca. 2 e in their LPs, whereas the intercalated N atom is surrounded by highly populated bonds, which leads to depleted LPs.

Both GB and V1 (Figs. \ref{fig:3Nbond}b-c top) show a dendritic arrangement (all N atoms linked to the same Si atom).
The globularity in the GB interstice is rather symmetric, with Si-N bonds holding 1.7-1.8 e and LPs 1.1-1.2 e.
V1 structure shows instead a bigger asymmetry since one of the N atoms (aka N2) is only 2-fold coordinated. This atom has Si-N bonds of ca. 2.3 e bonds. As seen before, N2 compensates the 2-fold with a highly populated LP (2.7 e). N1 shows a regular distribution with ca. 2 e bonds and LP. Finally, N3, although 3-fold coordinated leads to a highly populated bond (2.3 e) and a lowly populated LP (1.5 e). This reflects an uncomfortable geometry around this atom. 

For 4N atoms, the 2N bridge arrangement shows up both in the bulk and in V1. The two extra atoms appear alternated with Si in a chain. 
In the bulk, Si-N bonds are in the 2-2.2e range for if further away from the bridge and smaller populations in bond closer to the 2N unit (1.7-2.0e). LPs are distributed like in the previous interstitials: N atoms in the bridge hold two small LPs (1.2-1.5 e) whereas the isolated atoms hold greater charges (up to 1.95 e).
V1 leads to a highly symmetric charge distribution, with bond charges ranging 1.7-1.8 e, whereas LPs themselves are asymmetric (1.2-2.2 e).
The GB structure accommodates the 4N in a 3N dendritic distribution plus 1N. Charges are rather high with the dendritic bridge structure (up to 2.3 e) and low for the terminal bonds (1.7-1.8 e). The LP distribution is regular, with 1.1-1.2 e on each N atom.  
For V2, nitrogen atoms are distributed forming a Si-N cycle plus one terminal N (aka N4). Charges are again rather homogeneous, with some 2.3 e bonds inside the cycle and a 1.8 e bond in N4. The nitrogen atom (N1) not showing large bond charges compensates once again with a higherly populated LP (1.87 e) whereas the other atoms in the cycle hold 1.3 e in their LPs. The low bond charges in N4 are also compensated by 2.02 e in the lone pair (see SI: Table \ref{N@Si-V2GB}).

\begin{figure}[t] 
    \centering
    \includegraphics[scale=0.5]{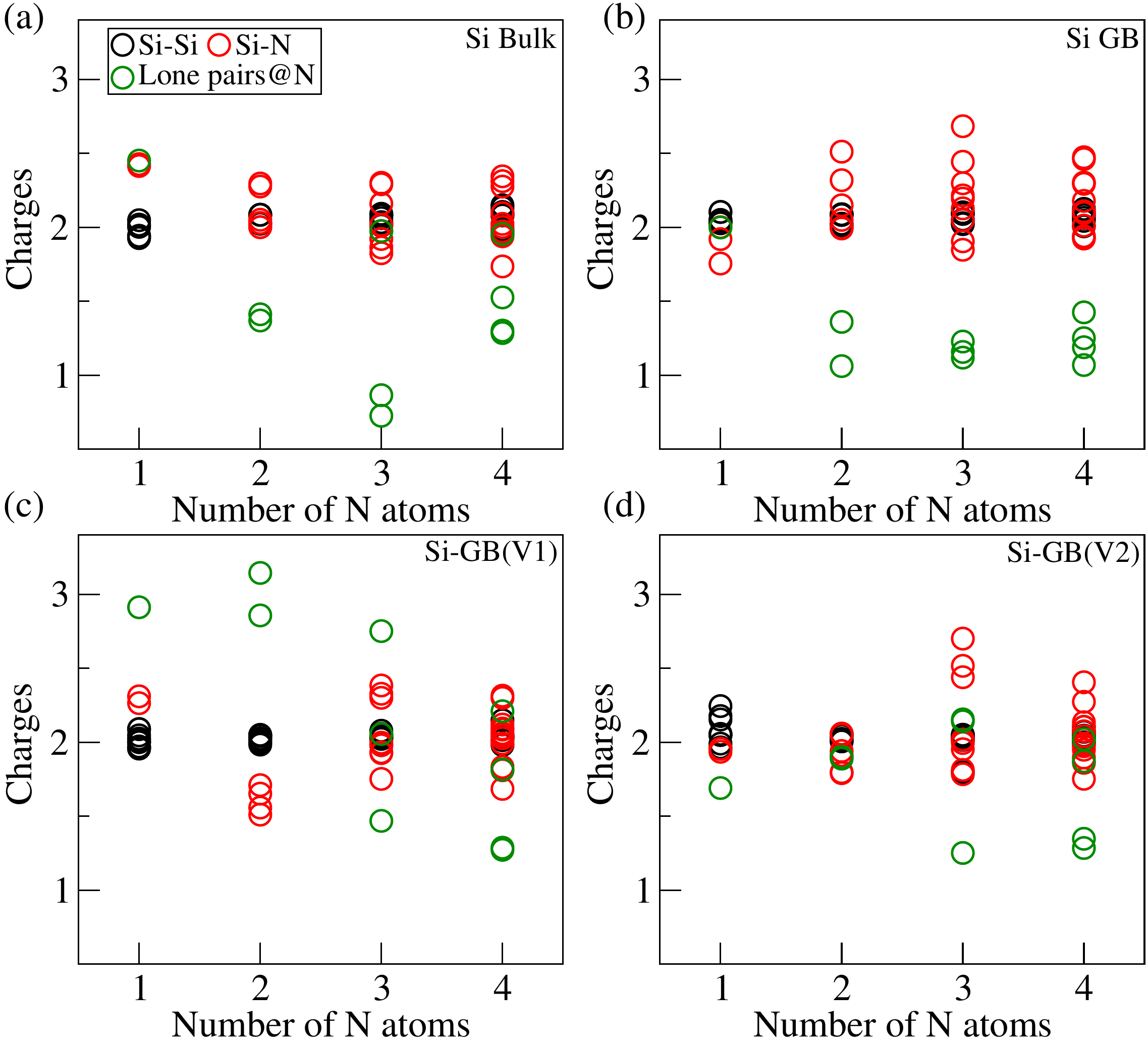}
    \caption{Bond charges (Si-Si and Si-N) and lone pairs at N (LP@N) variation  for the inclusion of different numbers of nitrogen atoms in a)bulk Si b) GB c) V1 d) V2.}
    \label{NLP}
\end{figure}

In general, we see that the natural coordination of N being 3, its insertion in the 4 fold structure leads to delocalized uncommon conformations (e.g. flat N coordinations). This is linked to the fact that some of these structures are gapless (see SI: Table \ref{Eg_N}) \cite{JCP_Rmaji2021}. Bond charges get far away from nominal and are compensated by LPs.
This can be seen in  
Fig. \ref{NLP}, which shows the evolution of charges for different number of atoms on each structure. Lone pairs are shown in green. It can be seen that lone pairs compensate the Si-N charges, with Si-Si charges finally being little affected by the N atoms. It is worth noting that in general, a higher Si-N population is promoted, with small LP charges. There is one exception to this trend, that is V1. For V1 highly populated LPs are found for 1-3 N atoms.

\subsection{Oxygen}

\begin{figure}[h!]
    \centering
        \includegraphics[width=1.0\textwidth]{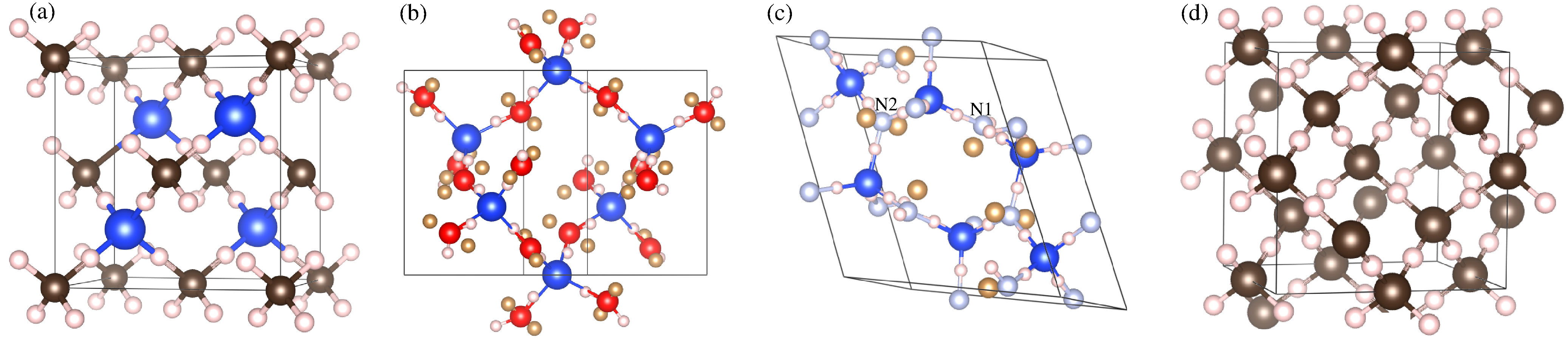}
    \caption{a) SiC b) SiO$_2$ c) Si$_3$N$_4$ d)Diamond}
    \label{fig:references}
\end{figure}

In this case, the reference system is SiO$_2$. This structure shows bond lengths of ca. 1.62~\AA, leading to bond charges in the range 1.91-1.96 e. Each oxygen atom shows two lone pairs with a total charge of 4.13 e (SI: Fig. \ref{fig:references}). This leads to 7.98 e around each oxygen atom.

\begin{figure}[t]
    \centering
    \includegraphics[scale=0.5]{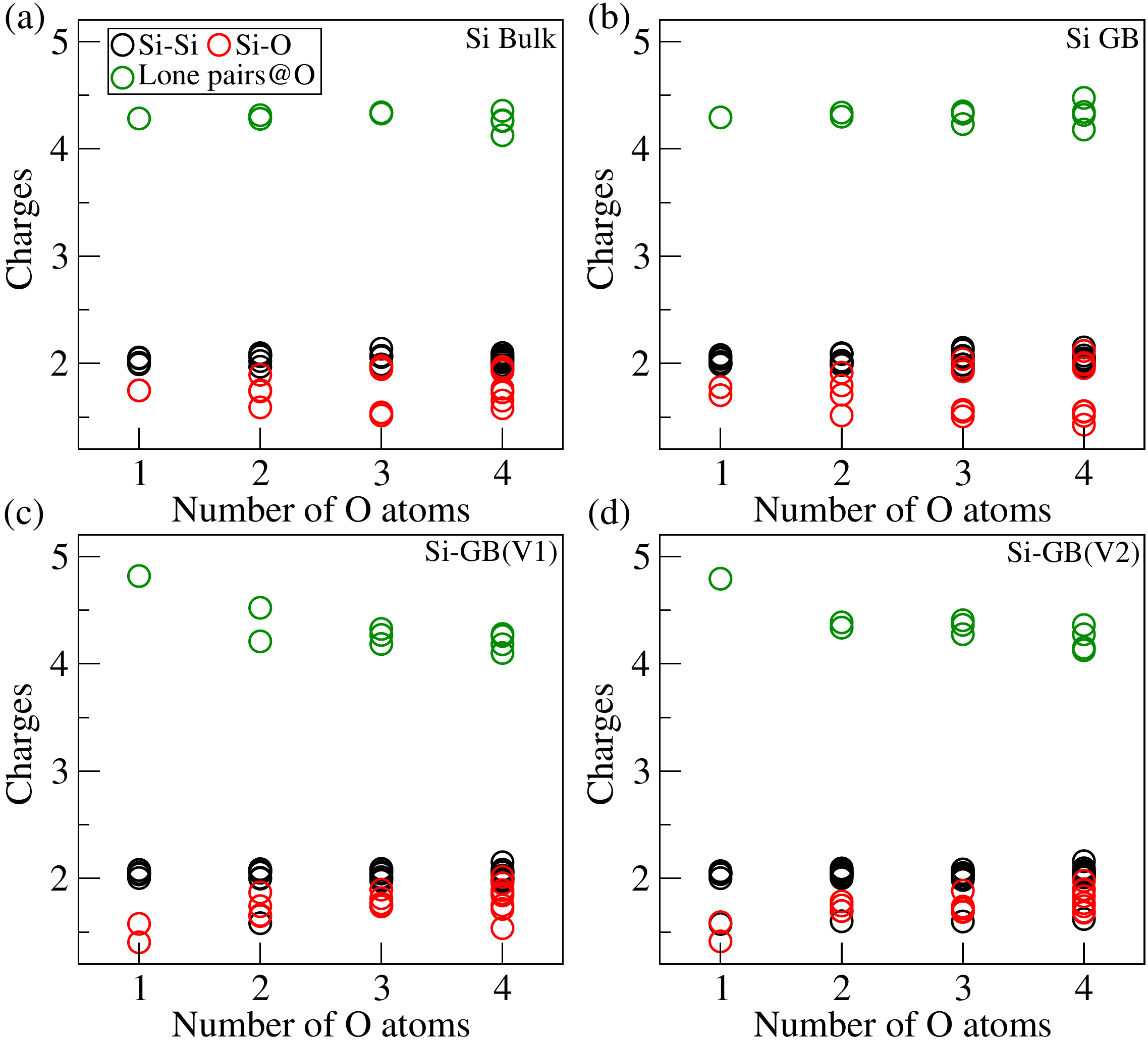}
    \caption{Bond charges (Si-Si and Si-O) and lone pairs at O (LP@O) variation  with the inclusion of different numbers of oxygen atoms in a)bulk Si b) GB c) V1 d) V2.}
    \label{fig:olp}
\end{figure}

When we introduce oxygen as an interstitial, patterns are very regular. Oxygen atoms always appear in between two Si atoms, O-O bonds are never observed. Overall, this leads to structures reminiscent of SiO$_2$, with Si-O bonds holding ca. 1.6-1.7 e and lone pairs on the oxygen atoms. Here the lone pairs holding ca. 4.4-4.8 e, total charge close to 8 electrons (see Fig. \ref{fig:olp}) for oxygen atom.

\begin{figure}[b]    
    \centering
    \includegraphics[scale=0.218]{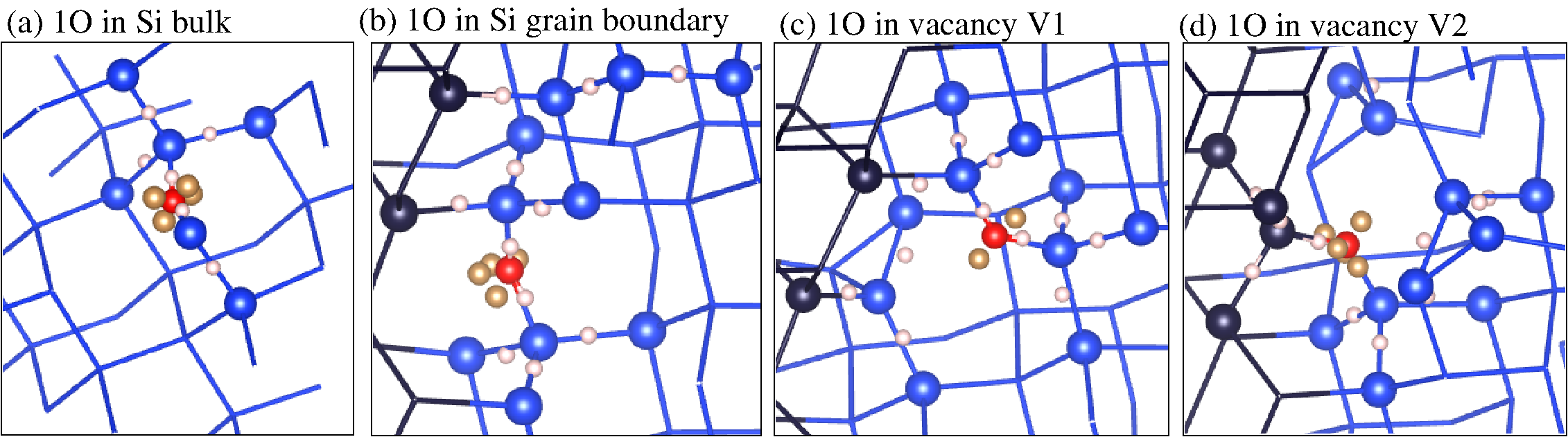}
    \includegraphics[scale=0.2]{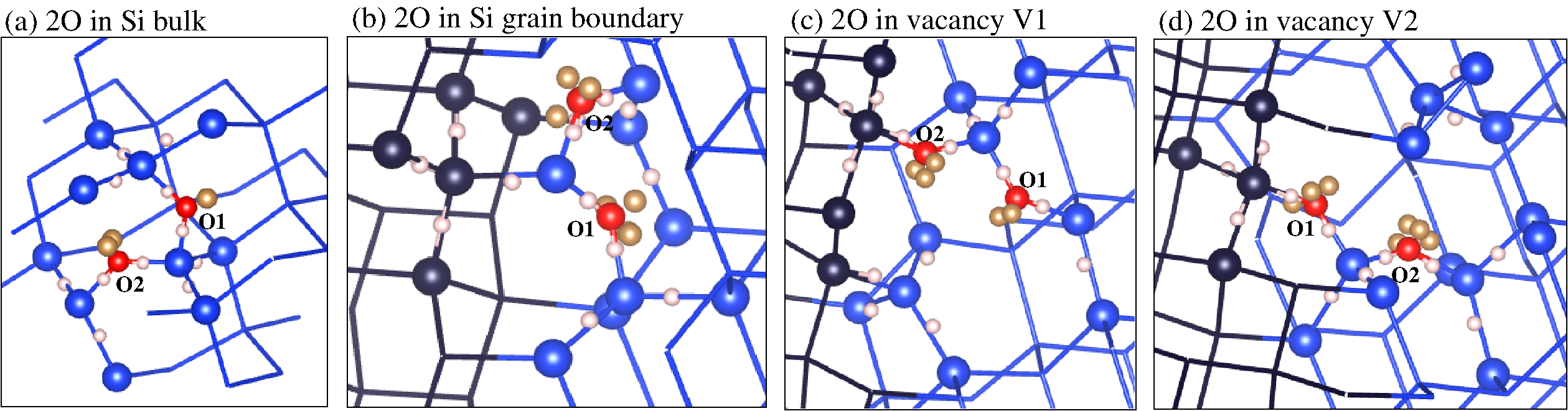}
    \caption{Structures with the inclusion of one oxygen atom (top) and two oxygen atoms (bottom) in a) bulk Si b) GB c) V1 d) V2. Silicon atoms in blue. Bonds and lone pairs are highlighted with small white spheres and brown spheres respectively}
    \label{fig:1Obond}
\end{figure}
In all cases, the insertion of oxygen atoms leads to Si-O bonds that have a slightly lower population than in bulk SiO$_2$ bulk. In some cases, this can go down to 1.5 e, essentially related to elongated Si-O bonds. The Si structure balances this effect, slightly increasing the Si-Si charges (Fig. \ref{fig:olp}).

\begin{figure}[h!]    
    \centering
    \includegraphics[scale=0.218]{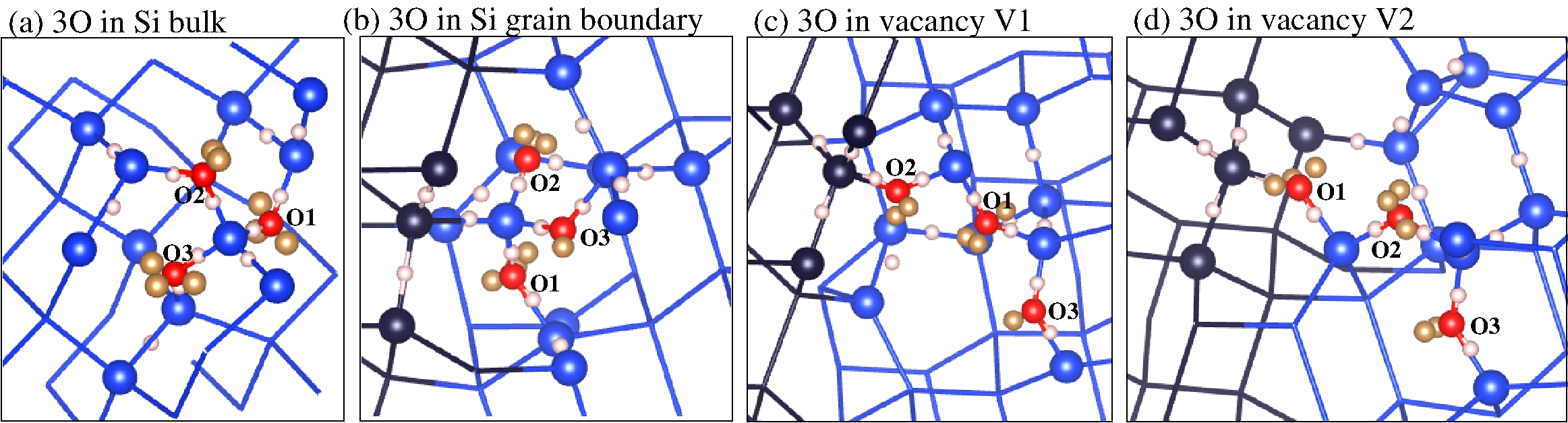}
    \includegraphics[scale=0.2]{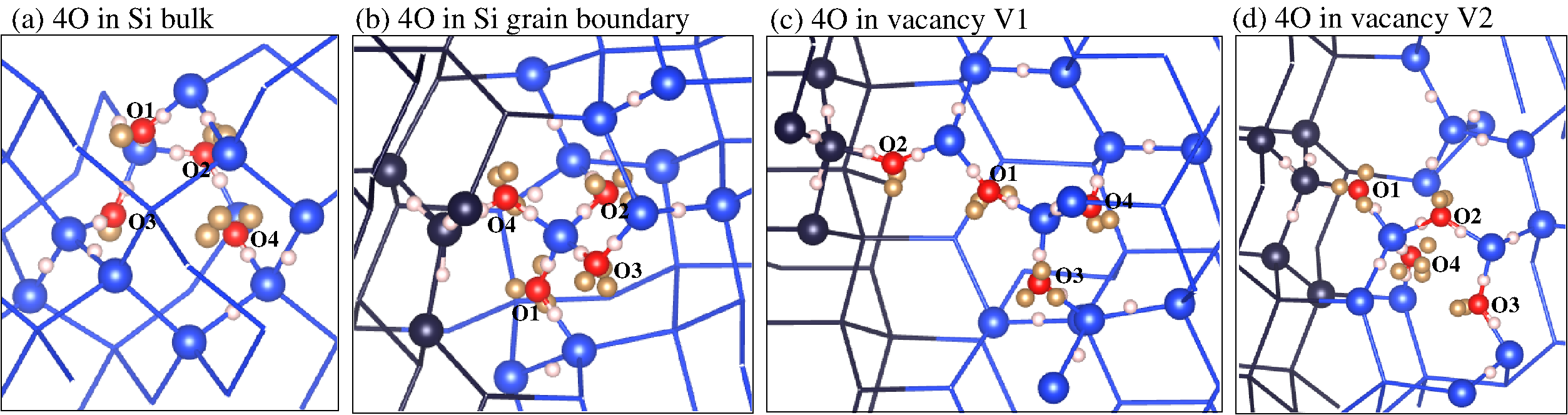}
\caption{Structures with the inclusion of three oxygen atoms (top) and four oxygen atoms (bottom) in a) bulk Si b) GB c) V1 d) V2. Silicon atoms in blue. Bonds and lone pairs are highlighted with small white spheres and brown spheres respectively}
    \label{fig:3Obond}
\end{figure}

Whereas two oxygen atoms always arrange linearly (Si-O-Si-O-Si), three and four oxygen atoms lead to two different arrangement types: linear (e.g. 4O in Si bulk and 3O in V1) or dendritic (e.g. 3O in Si bulk) (Fig. \ref{fig:3Obond}). The latter usually lead to a greater Si-O charge dispersion (Fig. \ref{fig:olp}).

Fig. \ref{fig:olp} allows to see the overall charge distribution in the different systems. Whereas the LP charges are rather stable both in the bulk and GB systems, so that the changes in Si-O bond charges are slightly compensated by the bulk, in the vacancy systems, the charge redistribution is mainly buffered within the interstitial: Si-O changes are balanced out by the LPs charge. This means that the charge on the oxygen suffers more important changes in the vacancies, with the total oxygen charge decreasing with the number of interstitials towards the bulk value.
The pattern repetition leads to very linear energetic behaviors (Fig. \ref{fig:en-n} top).

\section{Summary and conclusions}
\label{sec:conclusions}
\begin{figure}[t]
    \centering
    \includegraphics[scale=0.5]{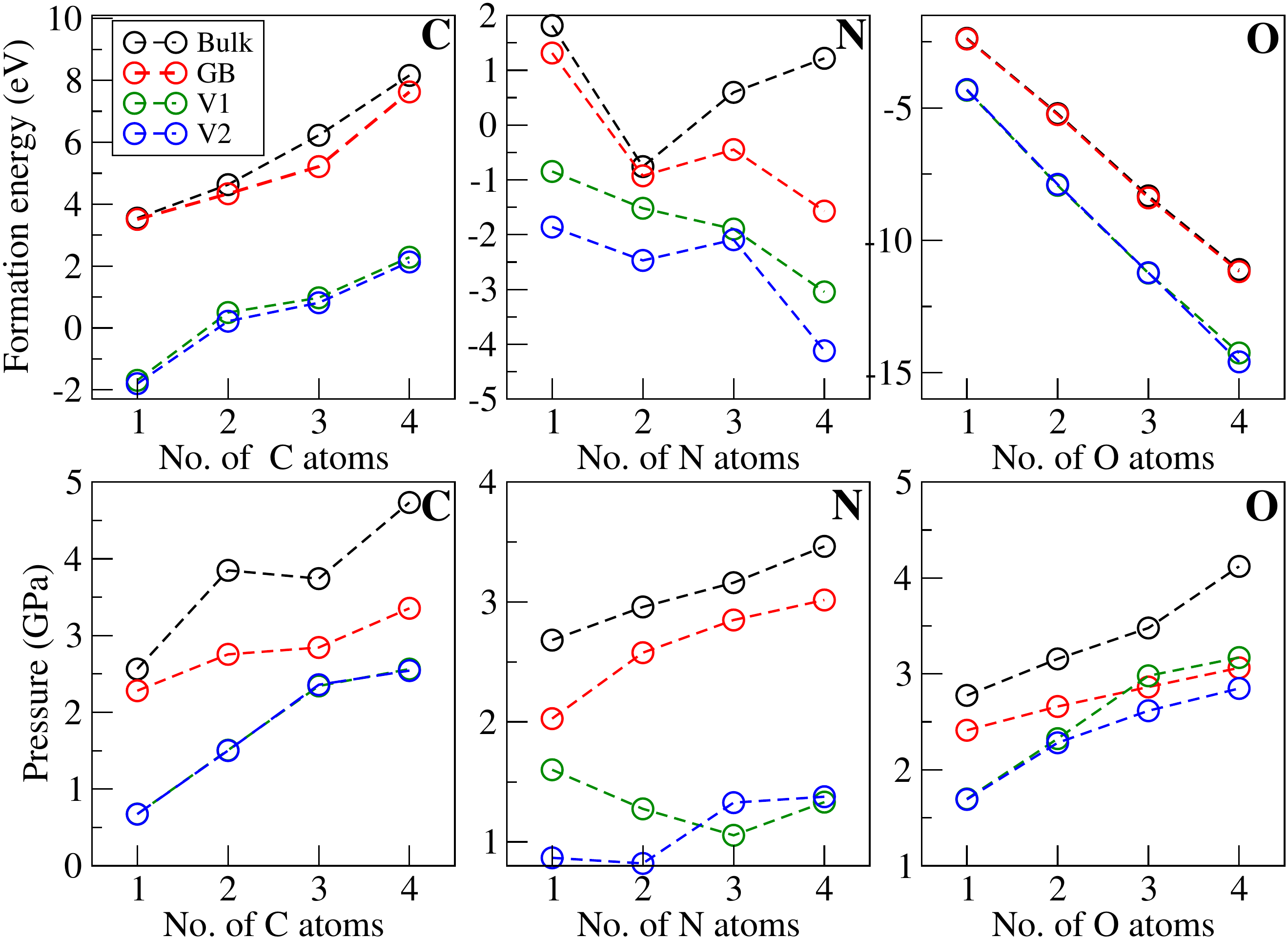} 
    \caption{Formation energy variation (top) and system pressure (bottom) with the inclusion of different numbers of C/N/O impurity atoms (1 to 4) in  bulk Si, GB, V1 and V2 as mentioned in legends}.
    \label{fig:en-n}
\end{figure}
Overall, we see that the system tries to keep as much as possible the stable tetrahedral electron pair organization when defects are introduced. Grain boundaries easily stabilize forming new bonds at the interface. Vacancies lead to missing coordinations that are compensated by confronted lone pairs that take the position and the total charge of the missing bond. Where possible, new bonds are also formed. In this case, planar geometries and weak bonds are formed. In both cases, local deformations are needed to promote these arrangements, leading to strained bonds.

When C atoms are added as interstitials, due to the strong C-C bonds, C atoms tend to form C-based arrangements reminiscent of C-based molecules. The presence of defects lowers the formation energy, allowing for the ``molecules'' to be included.
When there is not enough space, the C-based molecules will fit in by completing their coordination with bridges. Nevertheless, this leads to higher perturbations in the Si matrix. 

The formation energy for one carbon atom changes enormously depending on the presence or not of vacancies. Indeed, ca. 3.5 eV are obtained for bulk and GB. Instead it is favorable to introduce one carbon atom in V1 and V2 ($E_\text{f}^\text{XB}=-1.69$ eV and -1.80 eV, respectively). 
This is easy to understand if we compare these systems before and after the introduction of the interstitial.

Nitrogen patterns are complex: dendrites, bridges and linear arrangements are found, generally formed by Si-N bonds. N-N bonds are scarcely found. 
From the  formation energy and pressure trends point of view (Fig. \ref{fig:en-n}), only perceptible trends are observed for the bulk and GB pressure. This can be explained by the fact that these structures grow around the Si$_2$N$_2$ structure. Instead, vacancy structures tend to yield more dendritic organizations which, due to the sharing of bridges, induce different charge reorganizations for each new nitrogen atom.


Contrary to nitrogen, oxygen inclusions are extremely regular: Si-O very polar bonds holding 1.7-1.8 e in average and lone pairs on the Oxygen of about 4.2e (like chains of “water”). Si-O-Si arrangements are found in dendritic and linear arrangements. Overall, this pattern is rather additive, as directly reflected on the linear formation energies and pressures (Fig. \ref{fig:en-n}).

\section*{Acknowledgements}
We would like to thank the University of Modena and Reggio Emilia for the financial support (FAR2020 and FAR2021) and Centro Interdipartimentale En$\&$Tech, as well as the CINECA HPC facility for the approved ISCRA C projects SiGB-NMI (IsC86\_SiGB-NMI), GBimp-TS (IsC95\_GBimp-TS) and Interpol (IsC90\_Interpol).



\pagebreak
\newpage
\hspace{-11.0 in}\section*{{Supporting Information}}

\subsection*{Bulk systems}

\begin{table*}[h!]
\begin{center}
\begin{tabular}{ |c |c |c |c | c |c| } 
\hline
 & {\text{Si}} & \text{SiC} & \text{SiO$_2$} & {\text{Si$_3$N$_4$}} &\text{Diamond} \\ 
\hline
Bond Charges(BC) &1.999  &2.000  &1.908, 1.957  & 2.313&1.999 \\
\hline
Bond length(BL)\AA{} &2.351  & 1.896 & 1.625, 1.628 &1.747 &1.547 \\
\hline
Lone Pair(LP)        &       &       & 4.1340   & N2:(0.5289 each) 1.0578      & \\
\hline
\end{tabular}
\end{center}
\caption{References from bulk systems} 
\label{Bulk_ref}
\end{table*}

\subsection*{Molecular systems}
\begin{table*}[h]
\begin{center}
\begin{tabular}{ |c |c |c |c | c | c|} 
\hline
 & {\text{Ethane}} & \text{Ethylene} & \text{Cyclopropane} &\text{Cyclopropene} & {\text{Bicyclobutane}}  \\ 
\hline
Bond Charges(BC) & 1.81 &3.42  &1.70 & 1.80, 3.20 & 1.87, 1.72 \\
\hline
Bond length(BL)\AA{} & 1.53 &1.33  &1.52 &1.53, 1.30  &1.46, 1.55  \\
\hline
\end{tabular}
\end{center}
\caption{References from molecular systems}
\label{mol_ref}
\end{table*}


\begin{table*}[h]
\begin{center}
\begin{tabular}{ |c |c |c |c | } 
\hline
& Bond Charges(BC) & Bond length(BL)\AA{}& \text{LP} \\
\hline      
\text{Si-GB} & 2.00329  & 2.35084& \\ 
             & 2.00096  & 2.35122& \\
             & 1.98795  & 2.36642& \\
\hline
\text{Si-GB: V1}& 2.0697    & 2.32564 & Si1: 0.99\\
                & 2.0138    & 2.37601 & Si2: 1.00\\
& 2.0138    & 2.37601 &\\
& 2.0192    & 2.36607 &\\
& 2.0192    & 2.36607 &\\
& 2.0712    & 2.31773 &\\
& 2.0712    & 2.31773 &\\
& 1.9712    & 2.42741 &\\
& 2.0282    & 2.47125 &\\
& 2.0062    & 2.50193 &\\
& 2.0280    & 2.31424 &\\
& 2.0280    & 2.31424 &\\
& 1.5429    & 2.45097 &\\
\hline
\text{Si-GB: V2}  &2.0172    & 2.37177  &Si1: 0.97\\
 &2.0172    & 2.37177  &   Si2: 1.03\\
 &2.0609    & 2.32423  &\\
 &2.0609    & 2.32423  &\\
 &2.0676    & 2.32906  &\\
 &2.0676    & 2.32906  &\\
 &1.9926    & 2.35628  &\\
 &1.9782    & 2.43172  &\\
 &2.0207    & 2.43038  &\\
 &2.0207    & 2.43038  &\\
 &2.0273    & 2.48162  &\\
 &2.0251    & 2.31610  &\\
 &1.5522    & 2.44485  &\\
\hline
\end{tabular}
\end{center}
\caption{References from GB}
\label{Bulkdef_ref}
\end{table*}
\begin{table*}[h!]
\begin{center}
\begin{tabular}{ |c |c |c |c | c | } 
\hline
X &$E^{\text{XB}}_{\text{f}}$ & $E^{\text{XGB}}_{\text{f}}$ & $E^{\text{XV1GB}}_{\text{f}}$ & $E^{\text{XV2GB}}_{\text{f}}$ \\ 
\hline
\hline 
1C & 3.5467  & 3.5072 & -1.6935  & -1.8003 \\ 
\hline
2C & 4.6315  & 4.3359 & 0.4985  & 0.2165 \\
\hline
3C & 6.2242  & 5.2122 &  0.9758 & 0.8144 \\
\hline
4C & 8.1596  & 7.6275 &  2.2861 & 2.1306 \\
\hline
\end{tabular}
\end{center}
\caption{Formation energy (eV) of Si bulk and $\Sigma$3\{111\} Si GB without/with vacancies V1 and V2 with interstitial C atoms from 1 to 4. The values are reported for the lowest energy structures.}
\label{Ef_C}
\end{table*}
\begin{table*}[h!]
\begin{center}
\begin{tabular}{ |c |c |c |c | c | } 
\hline
X &$E^{\text{Si-bulk}}_{\text{g}}$ & $E^{\text{GB}}_{\text{g}}$ & $E^{\text{V1GB}}_{\text{g}}$ & $E^{\text{V2GB}}_{\text{g}}$ \\ 
\hline
\hline 
1C & 0  & 0.0477 &  0.3587 & 0.4014 \\ 
\hline
2C & 0.4291  & 0.5235 &  0 & 0 \\
\hline
3C & 0.6944  & 0.4729 &  0.6066 & 0.5578 \\
\hline
4C & 0.4549  & 0.1319 &  0.6502 & 0.5978 \\
\hline
\end{tabular}
\end{center}
\caption{Energy gaps (eV) of Si bulk and $\Sigma$3\{111\} Si GB without/with vacancies V1 and V2 with interstitial C atoms from 1 to 4. The values are reported for the lowest energy structures.}
\label{Eg_C}
\end{table*}

\begin{table*}[h!]
\begin{center}
\begin{tabular}{ |c |c |c |c | c | } 
\hline
X &$P{\text{(X+B)}}$ & $P{\text{(X+GB)}}$ & $P{\text{(X+V1GB)}}$ & $P{\text{(X+V2GB)}}$ \\ 
\hline
\hline 
1C & 2.564  &2.281   & 0.674  & 0.673 \\ 
\hline
2C &  3.850 & 2.756  & 1.507  & 1.502 \\
\hline
3C & 3.742  & 2.845   & 2.343  & 2.359  \\
\hline
4C &  4.734 & 3.356  & 2.563  &  2.543 \\
\hline
\end{tabular}
\end{center}
\caption{Hydrostatic pressure (GPa) of Si bulk and $\Sigma$3\{111\} Si GB without/with vacancies V1 and V2 with interstitial C atoms from 1 to 4. The values are reported for the lowest energy structures.}
\label{Pressure_C}
\end{table*}


\begin{table*}[h!]
\begin{center}
\begin{tabular}{ |c |c |c |c | c | } 
\hline
X &$E^{\text{XB}}_{\text{f}}$ & $E^{\text{XGB}}_{\text{f}}$ & $E^{\text{XV1GB}}_{\text{f}}$ & $E^{\text{XV2GB}}_{\text{f}}$ \\ 
\hline
\hline 
1N & 1.8155  &  1.3127 & -0.8477  & -1.8618 \\ 
\hline
2N & -0.7577  & -0.9272 & -1.5182  & -2.4698  \\
\hline
3N & 0.5968  &-0.4488  & -1.8968 & -2.0912 \\
\hline
4N & 1.2163  & -1.5700 & -3.0445  & -4.1177 \\
\hline
\end{tabular}
\end{center}
\caption{Formation energy (eV) of Si bulk and $\Sigma$3\{111\} Si GB without/with vacancies V1 and V2 with interstitial N atoms from 1 to 4. The values are reported for the lowest energy structures.}
\label{Ef_N}
\end{table*}
\begin{table*}[h!]
\begin{center}
\begin{tabular}{ |c |c |c |c | c | } 
\hline
X &$E^{\text{Si-bulk}}_{\text{g}}$ & $E^{\text{GB}}_{\text{g}}$ & $E^{\text{V1GB}}_{\text{g}}$ & $E^{\text{V2GB}}_{\text{g}}$ \\ 
\hline
\hline 
1N & 0  & 0 &  0 & 0 \\ 
\hline
2N & 0.5427  & 0.5444 &  0.2439 & 0.6191 \\
\hline
3N & 0  & 0 &  0 & 0 \\
\hline
4N & 0.3464  & 0.6274 &  0.3966 & 0.6397 \\
\hline
\end{tabular}
\end{center}
\caption{ Energy gaps (eV) of Si bulk and $\Sigma$3\{111\} Si GB without/with vacancies V1 and V2 with interstitial N atoms from 1 to 4. The values are reported for the lowest energy structures.}
\label{Eg_N}
\end{table*}
\begin{table*}[h!]
\begin{center}
\begin{tabular}{ |c |c |c |c | c | } 
\hline
X &$P{\text{(X+B)}}$ & $P{\text{(X+GB)}}$ & $P{\text{(X+V1GB)}}$ & $P{\text{(X+V2GB)}}$ \\ 
\hline
\hline 
1N & 2.682  &  2.027 & 1.601  & 0.867 \\ 
\hline
2N &  2.959 & 2.577  &  1.275 & 0.820 \\
\hline
3N & 3.161  & 2.850   & 1.054  & 1.327 \\
\hline
4N & 3.464  & 3.017  & 1.331   &  1.376 \\
\hline
\end{tabular}
\end{center}
\caption{Hydrostatic pressure (GPa) of Si bulk and $\Sigma$3\{111\} Si GB without/with vacancies V1 and V2 with interstitial N atoms from 1 to 4. The values are reported for the lowest energy structures.}
\label{Pressure_N}
\end{table*}
\begin{table*}[h!]
\begin{center}
\begin{tabular}{ |c |c |c |c | c | } 
\hline
X &$E^{\text{XB}}_{\text{f}}$ & $E^{\text{XGB}}_{\text{f}}$ & $E^{\text{XV1GB}}_{\text{f}}$ & $E^{\text{XV2GB}}_{\text{f}}$ \\ 
\hline
\hline 
1O & -2.3649  & -2.3950 & -4.3281  & -4.3036 \\ 
\hline
2O &-5.2126  & -5.2515  & -7.9224  & -7.8855  \\
\hline
3O & -8.3195  & -8.3974 & -11.2368  & -11.2213 \\
\hline
4O & -11.0994  & -11.1843 & -14.2579  & -14.5924 \\
\hline
\end{tabular}
\end{center}
\caption{Formation energy (eV) of Si bulk and $\Sigma$3\{111\} Si GB without/with vacancies V1 and V2 with interstitial O atoms from 1 to 4. The values are reported for the lowest energy structures.}
\label{Ef_O}
\end{table*}
\begin{table*}[h!]
\begin{center}
\begin{tabular}{ |c |c |c |c | c | } 
\hline
X &$E^{\text{Si-bulk}}_{\text{g}}$ & $E^{\text{GB}}_{\text{g}}$ & $E^{\text{V1GB}}_{\text{g}}$ & $E^{\text{V2GB}}_{\text{g}}$ \\ 
\hline
\hline 
1O & 0.6086  & 0.5506 &  0.6274 & 0.6283 \\ 
\hline
2O & 0.5913  & 0.6073 &  0.6353 & 0.6292 \\
\hline
3O & 0.7285  & 0.6185 &  0.6306 & 0.6319 \\
\hline
4O & 0.7430  & 0.5569 &  0.6850 & 0.6989 \\
\hline
\end{tabular}
\end{center}
\caption{ Energy gaps (eV) of Si bulk and $\Sigma$3\{111\} Si GB without/with vacancies V1 and V2 with interstitial O atoms from 1 to 4. The values are reported for the lowest energy structures.}
\label{Eg_O}
\end{table*}
\begin{table*}[h!]
\begin{center}
\begin{tabular}{ |c |c |c |c | c | } 
\hline
X &$P{\text{(X+B)}}$ & $P{\text{(X+GB)}}$ & $P{\text{(X+V1GB)}}$ & $P{\text{(X+V2GB)}}$ \\ 
\hline
\hline 
1O & 2.776   & 2.413  & 1.697  & 1.692 \\ 
\hline
2O & 3.157  & 2.660  & 2.325  & 2.282 \\
\hline
3O & 3.479  & 2.864  & 2.981  & 2.616 \\
\hline
4O & 4.121   & 3.064  & 3.171  & 2.849 \\
\hline
\end{tabular}
\end{center}
\caption{Hydrostatic pressure (GPa) of Si bulk and $\Sigma$3\{111\} Si GB without/with vacancies V1 and V2 with interstitial O atoms from 1 to 4. The values are reported for the lowest energy structures.}
\label{Pressure_O}
\end{table*}


\begin{table*}[h!]
\begin{center}
\begin{tabular}{ |c |c |c |c | c | c|c |c|} 
\hline
 Si-Bulk & {\text{BC(Si-Si)}} & \text{BL(Si-Si)\AA{}} & \text{BC(Si-C)} & {\text{BL(Si-C)\AA{}}}&\text{BC(C-C)}&\text{BL(C-C)\AA{}} & \text{LP} \\ 
\hline
1C & 1.9917  & 2.31736  & 2.4598 &1.81289  & & & 0.14269  \\
   & 1.9916  & 2.31759  & 2.5371 &1.81233  & & &\\
   & 1.9950  & 2.31807  & 2.3758 &1.74694  & & &\\
   & 1.9948  & 2.31842  &  &  &  & &\\
   & 2.2328  & 2.28068 &  &  &  & &\\
   & 2.2350  & 2.28173 &  &  &  &&\\
   & 1.9995  & 2.44675   &  &  &  &&\\
\hline
2C &1.9779  &2.31447 & 2.4845 &1.78563  & 3.0651&1.31413 &\\
   &1.9781  &2.31436 &2.4843  & 1.78585 & & &\\
   &1.9779  &2.31428 &1.7554  & 1.84197 & & &\\
   &1.9782  &2.31450 &1.7557  & 1.84192 & & &\\
   &2.1463  &2.25425 &  &  & & &\\
   &2.1460  &2.25412 &  &  & & &\\
   &1.9930  &2.30624 &  &  & & &\\
\hline
3C & 2.0201 &2.31590  &2.3825 &1.81499  &1.5792  &1.53572 &\\
   & 2.0046 &2.30954  &2.3820 &1.81535  &1.5777  &1.53568 &\\
   & 2.0047 &2.30951  &1.9941 &1.87728  &1.41599 &1.50008 &\\
   & 2.0202 &2.31584  &1.9021 &1.87268 & &&\\
   & 2.0693 &2.34550  &1.9028 &1.87244  & &&\\ 
   & 2.0693 &2.28606  &  &  & &&\\ 
   & 2.1675 &2.20866  &  &  & &&\\ 
   & 1.9581 &2.34297  &  &  & &&\\ 
   & 1.9577 &2.34290 &  &  & &&\\ 
\hline
4C & 1.9954 & 2.36665  & 2.2361 &1.84433  &2.1508 &1.41648&C3: 0.01604\\
   & 2.0178 & 2.32830  &2.3127  &1.80898  &3.0399 &1.34593&\\
   & 2.0250 & 2.38233  &2.2938  &1.85218  &1.7943 &1.46732&\\
   & 2.0650 & 2.26366  &2.3754  &1.81756  &1.0551 &1.66790&\\
   & 2.0407 & 2.28754  &1.9845  &1.86359  & &&\\
   & 1.9946  & 2.30892 &2.1936  &1.87557  & &&\\
   & 1.9880  & 2.31720 &  &  & &&\\
   & 1.9545  & 2.34714 &  &  & &&\\
   & 1.9410  & 2.35165 &  &  & &&\\
\hline
\end{tabular}
\end{center}
\caption{Bond charge(BC), bond length (BL) and lone pair(LP) data for Si-Si, Si-C and C-C bonds with the inclusion of n numbers of C atoms (n=1,2,3 \& 4) in Si-bulk.}
\label{C@Si-Bulk}
\end{table*}

\begin{table*}[h!]
\begin{center}
\begin{tabular}{ |c |c |c |c | c | c|c |c|} 
\hline
 Si-GB & {\text{BC(Si-Si)}} & \text{BL(Si-Si)\AA{}} & \text{BC(Si-C)} & {\text{BL(Si-C)\AA{}}}&\text{BC(C-C)}&\text{BL(C-C)\AA{}}& \text{LP} \\ 
\hline
1C & 1.9903 & 2.31786  &2.2747  &1.81425 & &&0.5948\\
   & 2.0164 & 2.31981  & 2.2497 &1.81470 & &&\\
   & 1.9814 & 2.31807  & 2.4034 &1.74762 & &&\\
   & 2.2083 & 2.29283 &  & & &&\\
   & 2.0041 & 2.33252 &  & & &&\\
   & 2.2183 & 2.29363 &  & & &&\\
   & 2.0015 & 2.45067 &  & & &&\\
\hline
2C & 2.0507 & 2.37415 & 2.3492 & 1.83840 &2.8306 &1.37792 &\\
   & 2.0595 & 2.36930 & 2.3430 & 1.84499 & &&\\
   & 2.0793 & 2.27725 & 2.1074 & 1.86866 & &&\\
   & 2.0129 & 2.29788 & 2.1091 & 1.86785 & &&\\
   & 2.0158 & 2.29966 &  &  & &&\\
   & 1.9780 & 2.34655 &  &  & &&\\
   & 1.9662 & 2.35109 &  &  & &&\\
   & 2.0841 & 2.26822 &  &  & &&\\
   & 2.0117 & 2.34787 &  &  & &&\\
   & 1.9379 & 2.31357 &  &  & &&\\
\hline
3C& 2.0549 & 2.42702  & 2.2420  & 1.85243 &1.5274 &1.57712 & C3: 0.001514\\
  & 2.0493 & 2.30263  & 2.2603  & 1.85057 & &&\\
  & 2.0167 & 2.33603  & 2.2152  & 1.84016 & &&\\
  & 2.0128 & 2.33161  & 2.1312  & 1.85078 & &&\\
  & 2.0341 & 2.35625  & 2.1595  & 1.84555 & &&\\
  & 2.0395 & 2.32409  & 1.8675  & 1.93941 & &&\\
  & 2.0554 & 2.44381  & 1.8601  & 1.93139 & &&\\
  & 1.9637 & 2.32060  & 1.8795  & 1.88733 & &&\\
  & 2.2222 & 2.21525  & 1.6191  & 1.88109 & &&\\
  &             &     & 1.5615  & 1.89062 & &&\\
\hline
4C& 2.0053  & 2.31422 & 2.3129 & 1.83722 &2.5353 &1.39499 &\\
  & 2.0744  & 2.32324 & 2.2215 & 1.87488 & &&\\
  & 2.0044  & 2.33814 & 2.1408 & 1.89536 & &&\\
  & 2.0281  & 2.36433 & 2.4743 & 1.78531 & &&\\
  & 1.9727  & 2.34566 & 2.4729 & 1.78473 & &&\\
  & 1.9471  & 2.32925 & 2.4820 & 1.80381 & &&\\
  & 1.8315  & 2.48968 & 2.5122 & 1.77339 & &&\\
  & 1.8336  & 2.44290 & 2.1517 & 1.79929 & &&\\
  & 1.1289  & 2.57283 & 2.7127 & 1.75854 & &&\\
  &         &         & 1.7269 & 1.95371 & &&\\
  &         &         & 1.1668 & 2.03012 & &&\\
\hline
\end{tabular}
\end{center}
\caption{Bond charge(BC), bond length (BL) and lone pair(LP) data for Si-Si, Si-C and C-C bonds with the inclusion of n numbers of C atoms (n=1,2,3 \& 4) in Si-GB}
\label{C@Si-GB}
\end{table*}

\begin{table*}[h!]
\begin{center}
\begin{tabular}{|c |c |c |c |c | c | c|c |c|} 
\hline
& Si-VGB & {\text{BC(Si-Si)}} & \text{BL(Si-Si)\AA} & \text{BC(Si-C)} & {\text{BL(Si-C)\AA{}}}&\text{BC(C-C)}&\text{BL(C-C)\AA{}} & \text{LP} \\ 
\hline
&1C  & 2.0243  & 2.39351& 1.9629 & 1.99583   &  &  &\\
 &  & 2.0242  & 2.39361       & 1.9526 & 2.00851   &  &  &\\
 &  & 2.0205  & 2.39538       & 1.9201 & 2.03954   &  &  &\\
 &  & 2.0289  & 2.39751       & 1.9199 & 2.03983   &  &  &\\
 &  & 2.0210  & 2.40463       &        &           &  &  &\\
 &  & 2.0211  & 2.40463       &        &           &  &  &\\
 &  & 2.0291  & 2.39740       &        &           &  &  &\\
 &  & 2.0115  & 2.34935       &        &           &  &  &\\
 &  & 2.0237  & 2.40396       &        &           &  &  &\\
 &  & 2.0115  & 2.34945       &        &           &  &  &\\
 &  & 2.0238  & 2.40392       &        &           &  &  &\\
 &  & 1.9847  & 2.34152       &        &           &  &  &\\
 &  & 2.0026  & 2.46212       &        &           &  &  &\\
 &  & 1.9689  & 2.42414       &        &           &  &  &\\
\cline{2-9}
V1& 2C  & 2.0236 & 2.40835   &  2.4013   & 1.88745     & 2.3241  &1.41529 & \\
  &     & 2.0234 & 2.42022   &  2.4174   & 1.88984     &         &        & \\
  &     & 2.0299 & 2.42408   &  2.4146   & 1.88913     &         &        & \\
  &     & 2.0176 & 2.42635   &  2.4217   & 1.88409     &         &        & \\
  &     & 2.0019 & 2.33929   &           &             &         &        & \\
  &     & 2.0023 & 2.34743   &           &             &         &        & \\
  &     & 2.0037 & 2.34256   &           &             &         &        & \\
  &     & 1.9859 & 2.37766   &           &             &         &        & \\
  &     & 1.9560 & 2.37192   &           &             &         &        & \\
  &     & 1.9545 & 2.37375   &           &             &         &        & \\
\cline{2-9}
&3C & 2.0355    &  2.32286   & 2.6560     & 1.80175        & 3.0039   & 1.31195   &\\          
&   & 2.0148    &  2.39841   & 2.7033     & 1.80450        & 1.5660   & 1.51629   & \\            
&   & 2.0329    &  2.29982   & 2.1869     & 1.89111        & 1.5181   & 1.52702   &   \\          
&   & 2.0273    &  2.33162   & 2.1780     & 1.90020        &          &           &  \\
&   & 2.0332    &  2.32608   &            &                &          &           &    \\    
&   & 1.9774    &  2.32650   &            &                &          &           &      \\  
&   & 1.9799    &  2.34479   &            &                &          &           & \\       
&   & 1.9779    &  2.34151   &            &                &          &           & \\       
&   & 2.0099    &  2.42987   &            &                &          &           &  \\      
&   & 1.9848    &  2.31307   &            &                &          &           &  \\      
&   & 1.9863    &  2.37639   &            &                &          &           &  \\      
&   & 1.9979    &  2.37766   &            &                &          &           &  \\      
&   & 1.9875    &  2.37762   &            &                &          &           &  \\      
\cline{2-9}
&4C & 2.0668 &  2.30703  & 2.3035   & 1.85418  & 1.6203   & 1.53713  &C1: 0.010741\\
&  & 2.0139  &  2.31936  & 2.5360   & 1.83508  & 1.6193   & 1.53703  &C3: 0.035845\\
&  & 2.0058  &  2.34112  & 2.5114   & 1.83507  & 1.5869   & 1.55733  &C3: 0.021072\\
&  & 2.0131  &  2.34511  & 2.3171   & 1.85991  & 1.5648   & 1.55739  &\\
&  & 2.0274  &  2.37703  & 2.0619   & 1.84544  & 1.3678   & 1.46903  &\\
&  & 2.0276  &  2.37684  & 2.0494   & 1.90386  &          &          &\\
&  & 2.0429  &  2.35443  &          &          &          &          &\\
&  & 2.0429  &  2.35446  &          &          &          &          &\\
&  & 2.0115  &  2.29645  &          &          &          &          &\\
&  & 2.0414  &  2.36713  &          &          &          &          &\\
\hline
\end{tabular}
\end{center}
\caption{Bond charge(BC), bond length (BL) and lone pair(LP) data for Si-Si, Si-C and C-C bonds with the inclusion of n numbers of C atoms (n=1,2,3 \& 4) in Si-GB with vacancy V1.}
\label{C@Si-V1GB}
\end{table*}

\begin{table*}[h!]
\begin{center}
\begin{tabular}{|c |c |c |c |c | c | c|c |c|} 
\hline
& Si-VGB & {\text{BC(Si-Si)}} & \text{BL(Si-Si)\AA{}} & \text{BC(Si-C)} & {\text{BL(Si-C)\AA{}}}&\text{BC(C-C)}&\text{BL(C-C)\AA{}} & \text{LP} \\ 
\hline
 &1C & 2.0281  &   2.38089  & 1.9623 & 2.00709  &  &   &\\
        &  & 2.0281  &   2.38089  & 1.9271 & 2.01777  &  &   &\\
        &  & 2.0184  &   2.38908  & 1.9333 & 2.03516  &  &   &\\
        &  & 2.0207  &   2.40451  & 1.9333 & 2.03516  &  &   &\\
        &  & 2.0207  &   2.40451  &        &          &  &   &\\
        &  & 2.0272  &   2.39927  &        &          &  &   &\\
        &  & 2.0272  &   2.39927  &        &          &  &   &\\
        &  & 2.0224  &   2.39941  &        &          &  &   &\\
        &  & 2.0224  &   2.39941  &        &          &  &   &\\
        &  & 2.0196  &   2.39996  &        &          &  &   &\\
        &  & 2.0022  &   2.46209  &        &          &  &   &\\
        &  & 2.0022  &   2.46209  &        &          &  &   &\\
\cline{2-9}
V2 &2C & 2.0119  &   2.32572 & 2.3918 & 1.89263  & 2.3152  & 1.41603& \\
& & 2.0235  &   2.40835 & 2.4065 & 1.89224  &         &        &\\
& & 2.0171  &   2.42459 & 2.4949 & 1.87145  &         &        &\\
& & 2.0034  &   2.32550 & 2.4071 & 1.88101  &         &        &\\
& & 1.9984  &   2.38682 &        &          &         &        &  \\                   
& & 1.9612  &   2.35986 &        &          &         &        &\\                     
& & 1.9912  &   2.33270 &        &          &         &        &  \\                   
& & 1.9641  &   2.41901 &        &          &         &        &\\                     
& & 1.9530  &   2.37274 &        &          &         &        &  \\                   
& & 1.9518  &   2.37180 &        &          &         &        &    \\    
\cline{2-9}
&3C & 2.0037 & 2.33228  & 2.6418   & 1.80187  & 2.9989    & 1.31248  &\\
&   & 1.9961 & 2.33392  & 2.7338   & 1.79859  & 1.5669    & 1.51368  &\\
&   & 1.9890 & 2.33664  & 2.1864   & 1.89848  & 1.5185    & 1.52813  &\\
&   & 2.0321 & 2.30156  & 2.1773   & 1.90683  &           &          & \\
&   & 2.0308 & 2.33111  &          &          &           &          & \\
&   & 2.0230 & 2.32828  &          &          &           &          & \\
&   & 2.0121 & 2.41445  &          &          &           &          & \\
&   & 2.0278 & 2.32947  &          &          &           &          & \\
\cline{2-9}
&4C& 2.0689 &   2.31040  & 2.2894  & 1.86170  & 1.6442  &  1.53314  &\\
&  & 2.0040 &   2.33369  & 2.5498  & 1.83354  & 1.6105  &  1.53866  &\\
&  & 1.9870 &   2.32764  & 2.5521  & 1.84761  & 1.5948  &  1.55655  &\\
&  & 2.0023 &   2.33598  & 2.3272  & 1.86247  & 1.5870  &  1.55328  &\\
&  & 2.0042 &   2.33820  & 2.0410  & 1.84755  & 1.3498  &  1.47835  &\\
&  & 2.0345 &   2.35649  & 2.0608  & 1.90940  &         &           &\\       
&  & 2.0533 &   2.31292  &         &          &         &           & \\      
&  & 2.0063 &   2.33383  &         &          &         &           & \\      
&  & 2.0353 &   2.34643  &         &          &         &           & \\      
\hline
\end{tabular}
\end{center}
\caption{Bond charge(BC), bond length (BL) and lone pair(LP) data for Si-Si, Si-C and C-C bonds with the inclusion of n numbers of C atoms (n=1,2,3 \& 4) in Si-GB with vacancy V2.}
\label{C@Si-V2GB}
\end{table*}
\begin{table*}[h!]
\begin{center}
\begin{tabular}{ |c |c |c |c | c | c|c |c|} 
\hline
 Si-Bulk & {\text{BC(Si-Si)}} & \text{BL(Si-Si)\AA{}} & \text{BC(Si-N)} & {\text{BL(Si-N)\AA{}}}&\text{BC(N-N)}&\text{BL(N-N)\AA{}}&\text{LP}  \\ 
\hline
1N & 2.0478   & 2.33613  & 2.4274  & 1.66719 & & &2.4509\\
 & 1.9314   & 2.32035  & 2.4141  & 1.66703  & & &\\
 & 1.9270   & 2.32165  &         &          & & &\\
 & 2.0029   & 2.35253  &         &          & & &\\
 & 2.0013   & 2.35318  &         &          & & &\\
 & 2.0181   & 2.35985  &         &          & & &\\
 & 1.9412   & 2.33676  &         &          & & &\\
 & 1.9427   & 2.33660  &         &          & & &\\
\hline
2N & 2.0849  & 2.30713   & 2.2752  & 1.73464  &  & &N1: 1.4121\\
  & 2.0843  & 2.30644   & 2.2956  & 1.73455  &  & &N2: 1.3700\\
  & 2.0843  & 2.30645   & 2.0008  & 1.73999  &  & &\\
  & 2.0849  & 2.30701   & 2.0007  & 1.73986  &  & &\\
  & 2.0244  & 2.31767   & 2.0293  & 1.76801  &  & &\\
  & 2.0239  & 2.31746   & 2.0508  & 1.76801  &  & &\\
\hline
3N  &2.0352   &2.28999   & 2.2909   &1.73687  & & &N1: 0.7259\\
    &2.0792   &2.30490   & 2.3008   &1.73777  & & &N2: 0.8651\\
    &2.0869   &2.31795   & 1.9742   &1.74611  & & &N3: 1.9743\\
    &2.0827   &2.33198   & 2.0136   &1.74460  & & &\\
    &2.0912   &2.30624   & 1.9209   &1.76536  & & &\\
    &2.0617   &2.28782   & 2.1595   &1.76647  & & &\\
    &2.0273   &2.31692   & 1.8634   &1.73632  & & &\\
    &2.0109   &2.31369   & 2.0166   &1.76668  & & &\\
    &         &          & 1.8249   &1.78669  & & &\\
\hline
4N  &2.0829  &2.33553  & 2.3121 &1.73744  &  &   &N1: 1.2854\\
    &2.0801  &2.32301  & 2.3443 &1.71281  &  &   &N2: 1.3015\\
    &2.0871  &2.29693  & 2.0021 &1.75323  &  &   &N3: 1.5256\\
    &2.1036  &2.30566  & 2.0234 &1.74916  &  &   &N4: 1.9553\\
    &2.1494  &2.25315  & 1.9686 &1.75665  &  &   &\\
    &2.1334  &2.35389  & 1.9389 &1.77627  &  &   &\\
    &1.9851  &2.30198  & 2.2747 &1.72530  &  &   &\\
    &        &         & 1.9527 &1.74615  &  &   &\\
    &        &         & 2.0948 &1.76061  &  &   &\\
    &        &         & 2.0943 &1.77118  &  &   &\\
    &        &         & 1.9726 &1.73255  &  &   &\\
    &        &         & 1.7351 &1.76306  &  &   &\\
\hline
\end{tabular}
\end{center}
\caption{Bond charge(BC), bond length (BL) and lone pair(LP) data for Si-Si, Si-N and N-N bonds with the inclusion of n numbers of N atoms (n=1,2,3 \& 4) in Si-bulk.}
\label{N@Si-Bulk}
\end{table*}

\begin{table*}[h!]
\begin{center}
\begin{tabular}{ |c |c |c |c | c | c|c |c|} 
\hline
 Si-GB & {\text{BC(Si-Si)}} & \text{BL(Si-Si)\AA{}} & \text{BC(Si-N)} & {\text{BL(Si-N)\AA{}}}&\text{BC(N-N)}&\text{BL(N-N)\AA{}}&\text{LP}  \\ 
\hline
1N  &2.0376  &2.32647   &2.0010  &1.76298  &  &  &1.9985\\
   &2.0375  &2.32636   &1.9200  &1.74187  &  &  &\\
   &2.0492  &2.32064   &1.7552  &1.76611  &  &  &\\
   &2.0493  &2.32045   &        &         &  &  &\\
   &2.0331  &2.45711   &        &         &  &  &\\
   &2.1035  &2.31514   &        &         &  &  &\\
   &2.1035  &2.31462   &        &         &  &  &\\
   &2.0271  &2.43737   &        &         &  &  &\\
\hline
2N & 2.0834  & 2.31385    & 2.3180  &1.73044 & & &N1: 1.3607\\
  & 2.0832  & 2.31385    & 2.5108  &1.72975 & & &N2: 1.0621\\
  & 2.0883  & 2.31135    & 2.1484  &1.74457 & & &\\
  & 2.0881  & 2.31135    & 2.0497  &1.74220 & & &\\
  & 2.0104  & 2.31162    & 1.9979  &1.77111 & & &\\
  & 2.0247  & 2.32479    & 1.9923  &1.76485 & & &\\
\hline
3N  &2.1013  &2.29004  & 2.1939  &1.74968  & &  &N1: 1.1192\\
    &2.0927  &2.31424  & 2.2970  &1.72659  & &  &N2: 1.2278\\
    &2.0902  &2.30417  & 2.6832  &1.71560  & &  &N3: 1.1573\\
    &2.0275  &2.32465  & 2.4427  &1.74610  & &  &\\
    &2.0207  &2.33297  & 2.2157  &1.74352  & &  &\\
    &2.0532  &2.35819  & 2.1275  &1.73394  & &  &\\
    &        &         & 2.0487  &1.71729  & &  &\\
    &        &         & 1.8478  &1.75508  & &  &\\
    &        &         & 1.9054  &1.79481  & &  &\\
\hline
4N & 2.1249  &2.27680  & 2.4731  &1.73868  & & &N1: 1.4252\\
   & 2.0931  &2.30091  & 2.1082  &1.76076  & & &N2: 1.0691\\
   & 2.0432  &2.30684  & 2.3015  &1.72478  & & &N3: 1.1897\\
   & 2.0634  &2.30310  & 2.4608  &1.72397  & & &N4: 1.2506\\
   & 2.0821  &2.34419  & 2.3014  &1.74044  & & &\\
   & 2.0068  &2.31122  & 2.0111  &1.75477  & & &\\
   &         &         & 2.1785  &1.74475  & & &\\
   &         &         & 2.2933  &1.78921  & & &\\
   &         &         & 1.9397  &1.77425  & & &\\
   &         &         & 2.0967  &1.73882  & & &\\
   &         &         & 1.9236  &1.77890  & & &\\
   &         &         & 1.9361  &1.77797  & & &\\
\hline
\end{tabular}
\end{center}
\caption{Bond charge(BC), bond length (BL) and lone pair(LP) data for Si-Si, Si-N and N-N bonds with the inclusion of n numbers of N atoms (n=1,2,3 \& 4) in Si-GB.}
\label{N@Si-GB}
\end{table*}

\begin{table*}[h!]
\begin{center}
\begin{tabular}{|c |c |c |c |c | c | c|c |c|} 
\hline
& Si-VGB & {\text{BC(Si-Si)}} & \text{BL(Si-Si)\AA{}} & \text{BC(Si-N)} & {\text{BL(Si-N)\AA{}}}&\text{BC(N-N)}&\text{BL(N-N)\AA{}}&\text{LP}  \\ 
\hline
&1N & 2.0247  &2.41953  & 2.3100  & 1.69449 & &  &2.9121\\
 & & 2.0197  &2.43623    & 2.2639  & 1.69582 & &  &\\
 & & 2.0010  &2.35263    &         &         & &  &\\
 & & 2.0038  &2.34135    &         &         & &  &\\
 & & 2.0009  &2.35266    &         &         & &  &\\
 & & 1.9653  &2.38943    &         &         & &  &\\
 & & 1.9649  &2.38938    &         &         & &  &\\
 & & 2.0462  &2.30787    &         &         & &  &\\
 & & 1.9577  &2.38419    &         &         & &  &\\
 & & 1.9570  &2.38426    &         &         & &  &\\
 & & 2.0887  &2.32091    &         &         & &  &\\
 & & 2.0894  &2.31500    &         &         & &  &\\
 & & 1.9718  &2.30958    &         &         & &  &\\
\cline{2-9}
V1 &2N & 2.0434  &2.34686    &1.7099  &1.80391   &1.0824  &1.45049   & N1: 2.8568\\
 & & 2.0256  &2.41936    &1.6528  &1.80577   &        &          & N2: 3.1440\\
 & & 2.0453  &2.34935    &1.5105  &1.83934   &        &          &\\
 & & 2.0452  &2.34960    &1.5587  &1.83902   &        &          &\\
 & & 2.0501  &2.33689    &        &          &        &          &\\
 & & 2.0484  &2.35730    &        &          &        &          &\\
 & & 2.0241  &2.44403    &        &          &        &          &\\
 & & 2.0094  &2.34903    &        &          &        &          &\\
 & & 1.9838  &2.33999    &        &          &        &          &\\
 & & 1.9960  &2.55257    &        &          &        &          &\\
\cline{2-9}
&3N &2.0773  & 2.41155  &  2.3000   &1.73292  &  & &N1: 2.0548\\
 &  &2.0422   &2.37194   & 1.9799    &1.82098  &  & &N2: 2.7514\\
 &  &2.0225   &2.33420   & 1.9268    &1.78660  &  & &N3: 1.4688\\
 &  &2.0439   &2.40624   & 2.3839    &1.68197  &  & &\\
 &  &1.9813   &2.31695   & 1.9969    &1.79269  &  & &\\
 &  &2.0262   &2.37227   & 2.3275    &1.70177  &  & &\\
 &  &2.0281   &2.42340   & 1.9324    &1.81505  &  & &\\
 &  &2.0372   &2.40966   & 1.7521    &1.84096  &  & &\\
 &  &2.0574   &2.48591   &           &         &  & &\\
\cline{2-9}
&4N  &2.1485  &2.46424  & 2.0854  &1.74349  &  & &N1: 1.8152\\
 &  &2.0467  &2.35171  & 2.0270  &1.73643  &  & &N2: 1.2873\\
 &  &2.0365  &2.42135  & 2.0928  &1.77086  &  & &N3: 1.2731\\
 &  &2.0349  &2.38921  & 2.1196  &1.75251  &  & &N4: 2.2097\\
 &  &2.0348  &2.41911  & 1.9919  &1.77712  &  & &\\
 &  &2.0138  &2.36413  & 2.0611  &1.79525  &  & &\\
 &  &1.9812  &2.32272  & 2.0450  &1.79535  &  & &\\
 &  &2.0344  &2.49119  & 1.8110  &1.76966  &  & &\\
 &  &        &         & 2.2998  &1.80866  &  & &\\
 &  &        &         & 1.8331  &1.84124  &  & &\\
 &  &        &         & 2.3149  &1.77682  &  & &\\
 &  &        &         & 1.6849  &1.80845  &  & &\\
\hline
\end{tabular}
\end{center}
\caption{Bond charge(BC), bond length (BL) and lone pair(LP) data for Si-Si, Si-N and N-N bonds with the inclusion of n numbers of N atoms (n=1,2,3 \& 4) in Si-GB with vacancy V1.}
\label{N@Si-V1GB}
\end{table*}
\begin{table*}[h!]
\begin{center}
\begin{tabular}{|c |c |c |c |c | c | c|c |c|} 
\hline
& Si-VGB & {\text{BC(Si-Si)}} & \text{BL(Si-Si)\AA{}} & \text{BC(Si-N)} & {\text{BL(Si-N)\AA{}}}&\text{BC(N-N)}&\text{BL(N-N)\AA{}}&\text{LP}  \\ 
\hline
&1N  &2.0564    &2.36784   & 1.9553  &1.87247 &  & & N: 1.6904\\
&  &2.0509    &2.40413   & 1.9558  &1.87367 &  & &   Si: 0.4432\\
&  &2.0493    &2.40528   & 1.9353  &1.86030 &  & &\\
&  &2.0511    &2.38624   &         &        &  & &\\
&  &2.0483    &2.38886   &         &        &  & &\\
&  &1.9908    &2.40434   &         &        &  & &\\
&  &1.9654    &2.40723   &         &        &  & &\\
&  &2.1537    &2.33401   &         &        &  & &\\
&  &2.1767    &2.33784   &         &        &  & &\\
&  &2.2444    &2.32922   &         &        &  & &\\
\cline{2-9}
V2  &2N  &2.0061   &2.31947    & 2.0590  &1.76352  & & &N1: 1.8892\\
 &  &2.0063   &2.31967    & 2.0490  &1.76331  & & &N2: 1.9015\\
 &  &2.0476   &2.36605    & 1.9409  &1.80508  & & &\\
 &  &2.0227   &2.33318    & 1.9121  &1.80526  & & &\\
 &  &2.0426   &2.45558    & 1.7975  &1.82537  & & &\\
 &  &2.0550   &2.53332    & 1.7913  &1.82514  & & &\\
 &  &2.0549   &2.53382    &         &         & & &\\
\cline{2-9}
& 3N  &2.0252  &2.36235   & 2.4391  &1.75975  &  &  &N1: 1.2519\\
&   &2.0156  &2.33604   & 1.9526  &1.77458  &  &  &N2: 2.1413\\
&   &2.0422  &2.35106   & 2.7003  &1.65230  &  &  &N3: 2.1548\\
&   &2.0451  &2.35443   & 1.9947  &1.77620  &  &  &\\
&   &2.0241  &2.38609   & 2.0178  &1.78340  &  &  &\\
&   &2.0507  &2.43521   & 2.5160  &1.66644  &  &  &\\
&   &2.0473  &2.41717   & 1.8130  &1.79287  &  &  &\\
&   &2.0488  &2.51035   & 1.7839  &1.81480  &  &  &\\
&   &2.0415  &2.51595   &         &         &  &  &\\
&   &1.7991  &2.35569   &         &         &  &  &\\
\cline{2-9}
& 4N  &2.0319  &2.28568   & 2.2734   &1.73678 &   &   &N1: 1.8719\\
&   &2.1039  &2.42680   & 2.1041   &1.75573 &   &   &N2: 1.3477\\
&   &2.0438  &2.26015   & 2.1349   &1.76866 &   &   &N3: 1.2853\\
&   &2.0431  &2.35302   & 2.0410   &1.74235 &   &   &N4: 2.0208\\
&   &2.0439  &2.35439   & 2.4054   &1.75972 &   &   &\\
&   &2.0269  &2.37705   & 1.9884   &1.77301 &   &   &\\
&   &1.9962  &2.33643   & 2.0737   &1.76081 &   &   &\\
&   &2.0385  &2.42899   & 1.9729   &1.78870 &   &   &\\
&   &2.0439  &2.42410   & 1.8965   &1.77478 &   &   &\\
&   &2.0702  &2.47255   & 1.9491   &1.77956 &   &   &\\
&   &2.0709  &2.51903   & 1.8610   &1.79077 &   &   &\\
&   &2.0000  &2.46747   & 1.7530   &1.80498 &   &   &\\
\hline
\end{tabular}
\end{center}
\caption{Bond charge(BC), bond length (BL) and lone pair(LP) data for Si-Si, Si-N and N-N bonds with the inclusion of n numbers of N atoms (n=1,2,3 \& 4) in Si-GB with vacancy V2.}
\label{N@Si-V2GB}
\end{table*}
\begin{table*}[h!]
\begin{center}
\begin{tabular}{ |c |c |c |c | c | c|c |c|} 
\hline
 Si-Bulk & {\text{BC(Si-Si)}} & \text{BL(Si-Si)\AA{}} & \text{BC(Si-O)} &{\text{BL(Si-O)\AA{}}}&\text{LP} \\ 
\hline
1O & 1.9918 & 2.32115 & 1.7489  &1.63550 &O1: 4.283\\
   & 2.0069 & 2.32194 & 1.7479 &1.63423 &\\
   &2.0467  & 2.32461 &  & &\\
   & 2.0555 & 2.32523 &  & &\\
\hline
2O & 2.0789   &2.30065    & 1.8939  &1.64273 &O1: 4.2801\\
 & 2.0795   &2.30074    & 1.7391  &1.66098   &O2: 4.3145\\
 & 2.0221   &2.32101    & 1.7493  &1.63986   &\\
 & 1.9685   &2.31220    & 1.5902  &1.67594   &\\
 & 2.0680   &2.33876    &         &          &\\
 & 2.0681   &2.33886    &         &          &\\
 & 2.0950   &2.33589    &         &          &\\
\hline
3O &2.1373  &2.24814  &1.9606  &1.62698  &O1: 4.3299\\
&1.9904  &2.30827  &1.9473  &1.62684  &O2: 4.3404\\
&2.0606  &2.33173  &1.9752  &1.62684  &O3: 4.3299\\
&2.0602  &2.33169  &1.5148  &1.68729  &\\
&2.0598  &2.33175  &1.5429  &1.68740  &\\
&2.0714  &2.35013  &1.5191  &1.68737  &\\
&2.0710  &2.34967  &        &         &\\
\hline
4O &2.0971  &2.22921   & 1.9675  &1.63190 &O1: 4.2585\\
  & 2.0108  &2.32069   & 1.9450  &1.62957 &O2: 4.1269\\
  & 2.0322  &2.32228   & 1.9281  &1.62211 &O3: 4.2659\\
  & 1.9837  &2.30422   & 1.7651  &1.66014 &O4: 4.3555\\
  & 2.0497  &2.33004   & 1.7301  &1.63470 &\\
  & 2.0734  &2.34109   & 1.7331  &1.65317 &\\
  &         &          & 1.6537  &1.66647 &\\
  &         &          & 1.5824  &1.68762 &\\
\hline
\end{tabular}
\end{center}
\caption{Bond charge(BC), bond length (BL) and lone pair(LP) data for Si-Si and Si-O bonds with the inclusion of n numbers of O atoms (n=1,2,3 \& 4) in Si-bulk.}
\label{O@Si-Bulk}
\end{table*}

\begin{table*}[h!]
\begin{center}
\begin{tabular}{ |c |c |c |c | c |c|} 
\hline
 Si-GB & {\text{BC(Si-Si)}} & \text{BL(Si-Si)\AA{}} & \text{BC(Si-O)} & {\text{BL(Si-O)\AA{}}}&\text{LP}  \\ 
\hline
1O& 2.0093  & 2.32532  & 1.7792 & 1.63913 & 4.2934\\
& 1.9902  & 2.32383  & 1.7035 & 1.63789 &\\
& 2.0134  & 2.33911  &        &         &\\
& 2.0425  & 2.32990  &        &         &\\
& 2.0564  & 2.32704  &        &         &\\
& 2.0781  & 2.32612  &        &         &\\
& 1.9997  & 2.32640  &        &         &\\
& 1.9931  & 2.34696  &        &         &\\
& 1.9959  & 2.36746  &        &         &\\
\hline
2O&2.0872  &2.31206  & 1.9142    & 1.64514     &O1: 4.3003\\
  &2.0250  &2.32108  & 1.7951    & 1.64091     &O2: 4.3380\\
  &2.0197  &2.32159  & 1.7078    & 1.66111     &\\
  &2.0920  &2.31034  & 1.5149    & 1.67498     &\\
  &2.0950  &2.33289  &           &             &\\
  &2.0132  &2.35452  &           &             &\\
  &2.0226  &2.32704  &           &             &\\
  &1.9877  &2.36383  &           &             &\\
  &1.9847  &2.32704  &           &             &\\
  &1.9957  &2.35741  &           &             &\\
\hline
3O&2.1448   &2.25427   &1.9600  &1.62934  &O1: 4.2309\\
  &2.1231   &2.28737   &2.0395  &1.62583  &O2: 4.3284\\
  &2.0699   &2.34079   &1.9255  &1.63628  &O3: 4.3491\\
  &2.0687   &2.36253   &1.5670  &1.68212  &\\
  &2.0445   &2.33292   &1.5497  &1.68240  &\\
  &2.0624   &2.34653   &1.5057  &1.69960  &\\
  &1.9502   &2.32814   &        &         &\\
  &1.9914   &2.33156   &        &         &\\
  &1.9873   &2.33906   &        &         &\\
\hline
4O&1.9968   &2.30698   &2.1162   &1.60775   & O1: 4.1798\\
  &2.0109   &2.31411   &1.9557   &1.60825   & O2: 4.3162\\
  &2.0089   &2.29839   &2.0040   &1.60321   & O3: 4.4751\\
  &2.0324   &2.32370   &1.9907   &1.61242   & O4: 4.3444\\
  &2.0459   &2.36708   &1.5109   &1.69476   &\\
  &2.0352   &2.32384   &1.5546   &1.69474   &\\
  &2.0739   &2.37431   &1.5502   &1.69645   &\\
  &2.1489   &2.29031   &1.4290   &1.71093   &\\
  &1.9766   &2.33417   &         &          &\\
\hline
\end{tabular}
\end{center}
\caption{Bond charge(BC), bond length (BL) and lone pair(LP) data for Si-Si and Si-O bonds with the inclusion of n numbers of O atoms (n=1,2,3 \& 4) in Si-GB}
\label{O@Si-GB}
\end{table*}

\begin{table*}[h!]
\begin{center}
\begin{tabular}{|c |c |c |c |c | c |c| } 
\hline
& Si-VGB & {\text{BC(Si-Si)}} & \text{BL(Si-Si)\AA{}} & \text{BC(Si-O)} & {\text{BL(Si-O)\AA{}}}&\text{LP} \\ 
\hline
&1O& 2.0433   &2.35933   &1.5763   &1.70276   & 4.8177\\
 & & 2.0432   &2.35930   &1.4038   &1.70352   &\\
 & & 2.0504   &2.34937   &         &          &\\
 & & 2.0503   &2.34933   &         &          &\\
 & & 2.0493   &2.41995   &         &          &\\
 & & 2.0374   &2.44706   &         &          &\\
 & & 2.0781   &2.32314   &         &          &\\
 & & 2.0788   &2.31593   &         &          &\\
 & & 2.0029   &2.31171   &         &          &\\
\cline{2-7}
V1&2O & 2.0854   &2.34609   &1.8687   &1.63189  &O1: 4.5218\\
 &   & 2.0853   &2.34610   &1.6492   &1.68177   &O2: 4.2086\\
 &   & 1.9984   &2.31011   &1.7409   &1.66199   &\\
 &   & 2.0531   &2.38759   &1.6583   &1.66506   &\\
 &   & 2.0612   &2.35561   &         &          &\\
 &   & 2.0613   &2.35561   &         &          &\\
 &   & 2.0745   &2.32627   &         &          &\\
 &   & 2.0759   &2.32289   &         &          &\\
 &   & 2.0046   &2.31500   &         &          &\\
 &   & 1.5825   &2.45085   &         &          &\\
\cline{2-7}
&3O  &2.0851  &2.31788    & 1.8948  &1.62656   & O1: 4.2695\\
&    &2.0886  &2.32236    & 1.8150  &1.63978   & O2: 4.1855\\
&    &2.0016  &2.31408    & 1.7386  &1.62749   & O3: 4.3248\\
&    &2.0592  &2.34710    & 1.7672  &1.63987   &\\
&    &2.0592  &2.34711    & 1.7558  &1.65652   &\\
&    &2.0446  &2.31293    & 1.7408  &1.65403   &\\
&    &2.0107  &2.32353    &         &          &\\
&    &1.9913  &2.34144    &         &          &\\
&    &1.9769  &2.33572    &         &          &\\
&    &2.0050  &2.31155    &         &          &\\
\cline{2-7}
 &4O  &2.1503   & 2.27352   & 2.0133  &1.62843   & O1: 4.1015\\
 &  &2.0809   & 2.31442   & 1.8618  &1.62652   & O2: 4.1822\\
 &  &2.0001   & 2.31435   & 1.8409  &1.64666   & O3: 4.2785\\
 &  &1.9944   & 2.30186   & 1.7146  &1.66121   & O4: 4.2594\\
 &  &2.0569   & 2.32987   & 1.9526  &1.62095   &\\
 &  &2.0262   & 2.30338   & 1.8908  &1.62452   &\\
 &  &2.0582   & 2.34486   & 1.7483  &1.65339   &\\
 &  &2.0608   & 2.34711   & 1.5363  &1.69373   &\\
 &  &2.0576   & 2.38358   &         &          &\\
 &  &2.0007   & 2.33742   &         &          &\\
 &  &1.9774   & 2.34089   &         &          &\\
\hline
\end{tabular}
\end{center}
\caption{Bond charge(BC), bond length (BL) and lone pair(LP) data for Si-Si and Si-O bonds with the inclusion of n numbers of O atoms (n=1,2,3 \& 4) in Si-GB with vacancy V1.}
\label{O@Si-V1GB}
\end{table*}
\begin{table*}[h!]
\begin{center}
\begin{tabular}{|c |c |c |c |c | c |c| } 
\hline
& Si-VGB & {\text{BC(Si-Si)}} & \text{BL(Si-Si)\AA{}} & \text{BC(Si-O)} & {\text{BL(Si-O)\AA{}}}&\text{LP} \\ 
\hline
&1O &2.0490  &2.35202  &1.5907  &1.70536   &4.7927\\
&   &2.0490  &2.35202  &1.4133  &1.70324   &\\
&   &2.0389  &2.34423  &        &          &\\
&   &2.0479  &2.39899  &        &          &\\
&   &2.0479  &2.39899  &        &          &\\
&   &2.0459  &2.42851  &        &          &\\
&   &2.0673  &2.32292  &        &          &\\
&   &2.0017  &2.31335  &        &          &\\
&   &1.5752  &2.43705  &        &          &\\
\cline{2-7}
V2 &2O &2.0916  &2.34935   &1.7851  &1.63329 &O1: 4.3862\\
&   &2.0916  &2.34935   &1.6953  &1.66673   &O2: 4.3375\\
&   &2.0257  &2.32115   &1.7470  &1.68194   &\\
&   &2.0257  &2.32115   &1.6959  &1.66639   &\\
&   &2.0504  &2.36999   &        &          &\\
&   &2.0504  &2.36999   &        &          &\\
&   &2.0360  &2.33241   &        &          &\\
&   &2.0735  &2.37747   &        &          &\\
&   &2.0611  &2.32703   &        &          &\\
&   &2.0018  &2.31163   &        &          &\\
&   &1.6003  &2.43229   &        &          &\\
\cline{2-7}
& 3O &2.0823  &2.30486   &1.8844  &1.65237   &O1: 4.4063\\
&   &2.0826  &2.30432   &1.7179  &1.64360   &O2: 4.2758\\
&   &2.0494  &2.35862   &1.6933  &1.64503   &O3: 4.3626\\
&   &2.0494  &2.35862   &1.7372  &1.66772   &\\
&   &2.0304  &2.32693   &1.7276  &1.66736   &\\
&   &2.0309  &2.32429   &1.6871  &1.65621   &\\
&   &2.0363  &2.30348   &        &          &\\
&   &1.9855  &2.32768   &        &          &\\
&   &2.0048  &2.31078   &        &          &\\
&   &1.5979  &2.42764   &        &          &\\
\cline{2-7}
&4O &2.1604  &2.33661    & 1.8891  &1.63356  & O1: 4.2751\\
&   &2.0939  &2.32015    & 1.9768  &1.61832  & O2: 4.1492\\
&   &2.0371  &2.34213    & 1.8987  &1.64051  & O3: 4.3613\\
&   &2.0210  &2.31449    & 1.8329  &1.66554  & O4: 4.1261\\
&   &2.0529  &2.36183    & 1.7697  &1.64925  &\\
&   &2.0454  &2.35171    & 1.7404  &1.65316  &\\
&   &2.0487  &2.34450    & 1.6899  &1.67462  &\\
&   &2.0529  &2.32527    & 1.6799  &1.67034  &\\
&   &2.0686  &2.32756    &         &         &\\
&   &2.0095  &2.32386    &         &         &\\
&   &1.6214  &2.41121    &         &         &\\
\hline
\end{tabular}
\end{center}
\caption{Bond charge(BC), bond length (BL) and lone pair(LP) data for Si-Si and Si-O bonds for the inclusion of n numbers of O atoms (n=1,2,3 \& 4) in Si-GB with vacancy V2.}
\label{O@Si-V2GB}
\end{table*}


\pagebreak
\newpage

\begin{figure}[h]    
    \centering
    \includegraphics[width=1.0\textwidth]{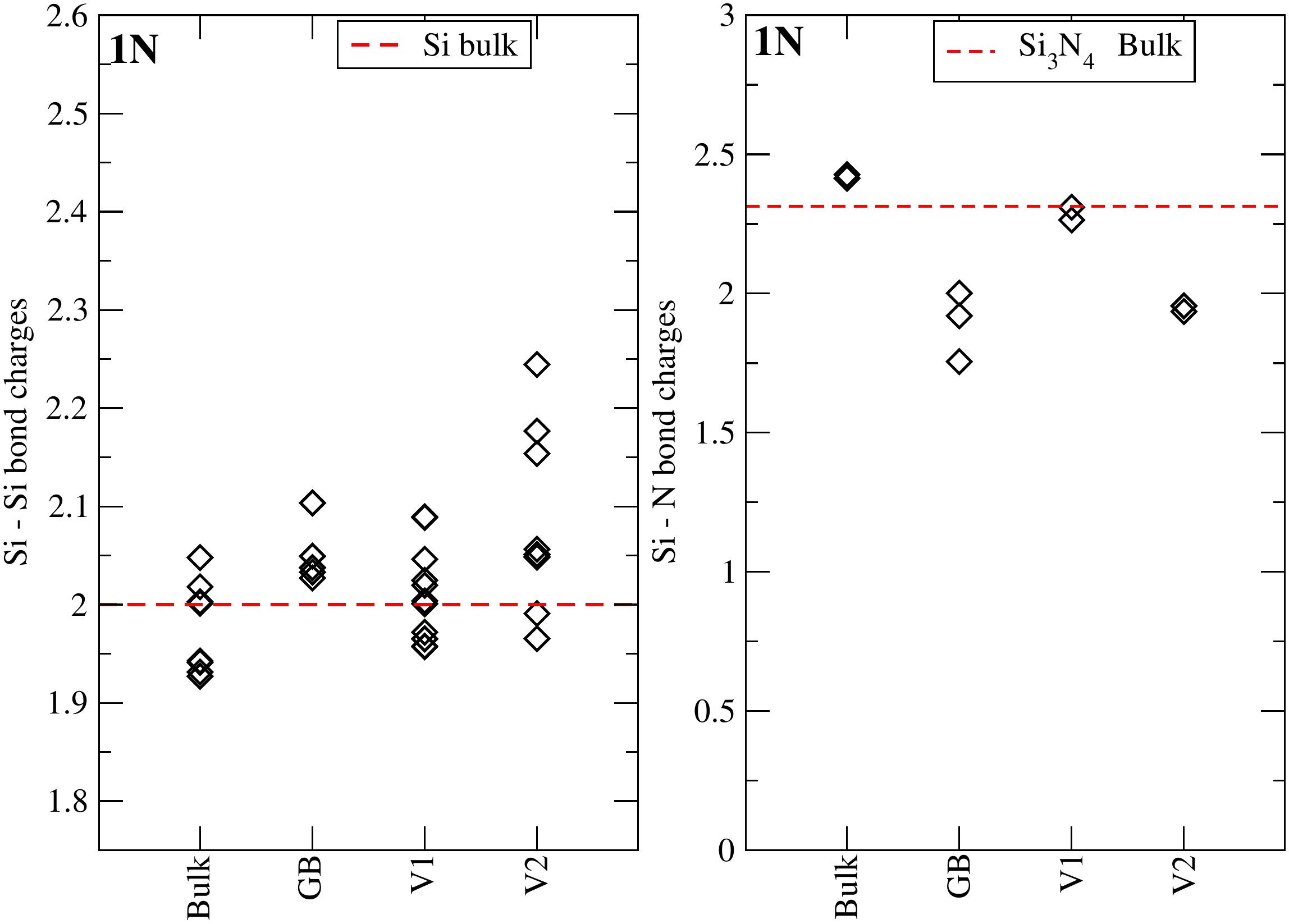}
    \caption{Bond charge variation of Si-Si and Si-N with the inclusion of one nitrogen atom in Si-bulk, Si-GB and in presence of vacancy V1 and V2. As a reference, Si-Si/Si-N bond charges from bulk phases of Si and Si$_3$N$_4$ are marked in the inset of respective plots.}
    \label{fig:1Ncharges}
\end{figure}
\begin{figure}[h]  
    \centering
    \includegraphics[width=1.0\textwidth]{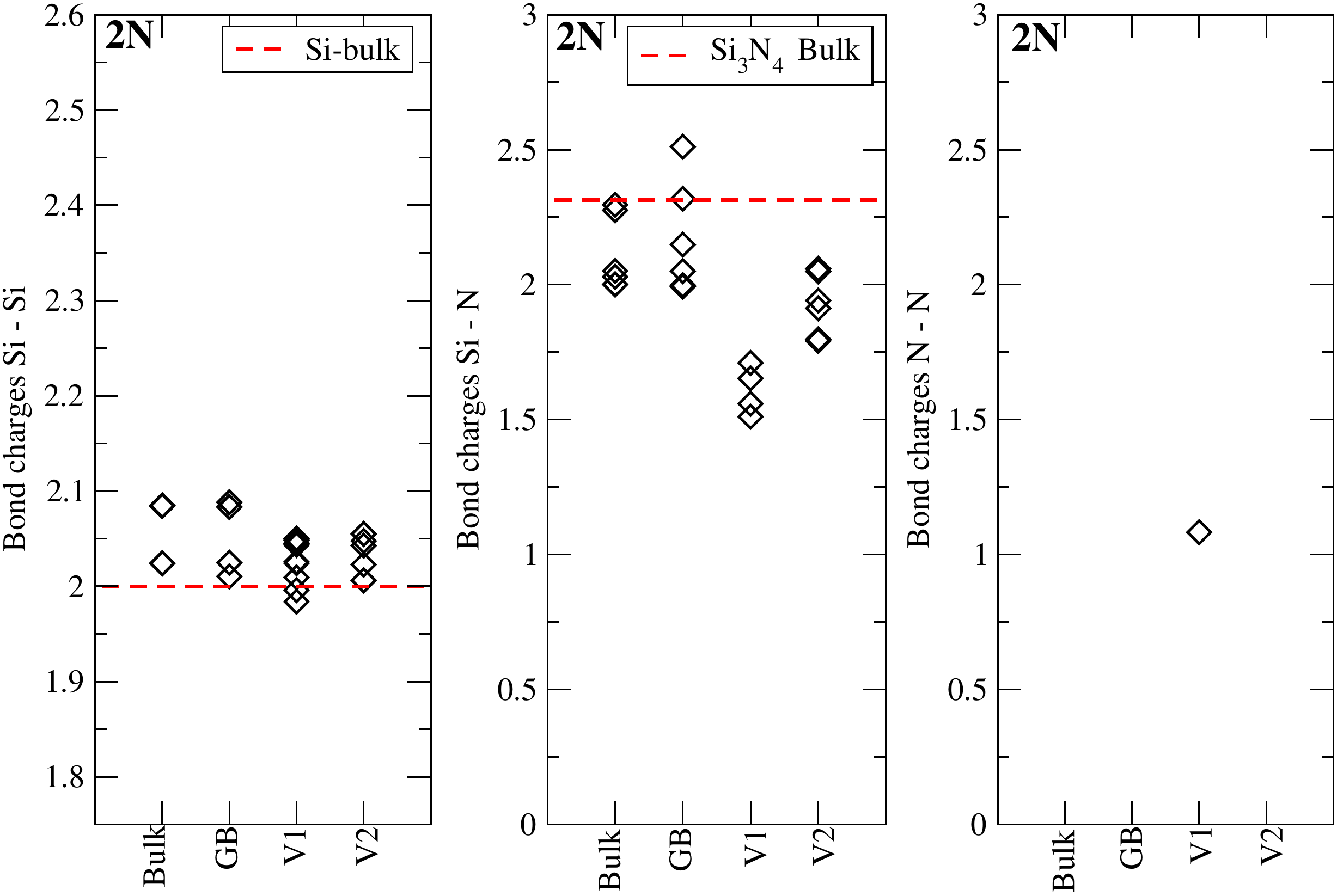}
    \caption{Bond charge variation of Si-Si, Si-N and N-N for the inclusion of two nitrogen atoms in Si-bulk, Si-GB and in presence of vacancy V1 and V2. As a reference, Si-Si/Si-N bond charges from bulk phases of Si and Si$_3$N$_4$ are marked in the inset of respective plots.}
    \label{fig:2Ncharges}
\end{figure}
\begin{figure}[h]    
    \centering
    \includegraphics[width=1.0\textwidth]{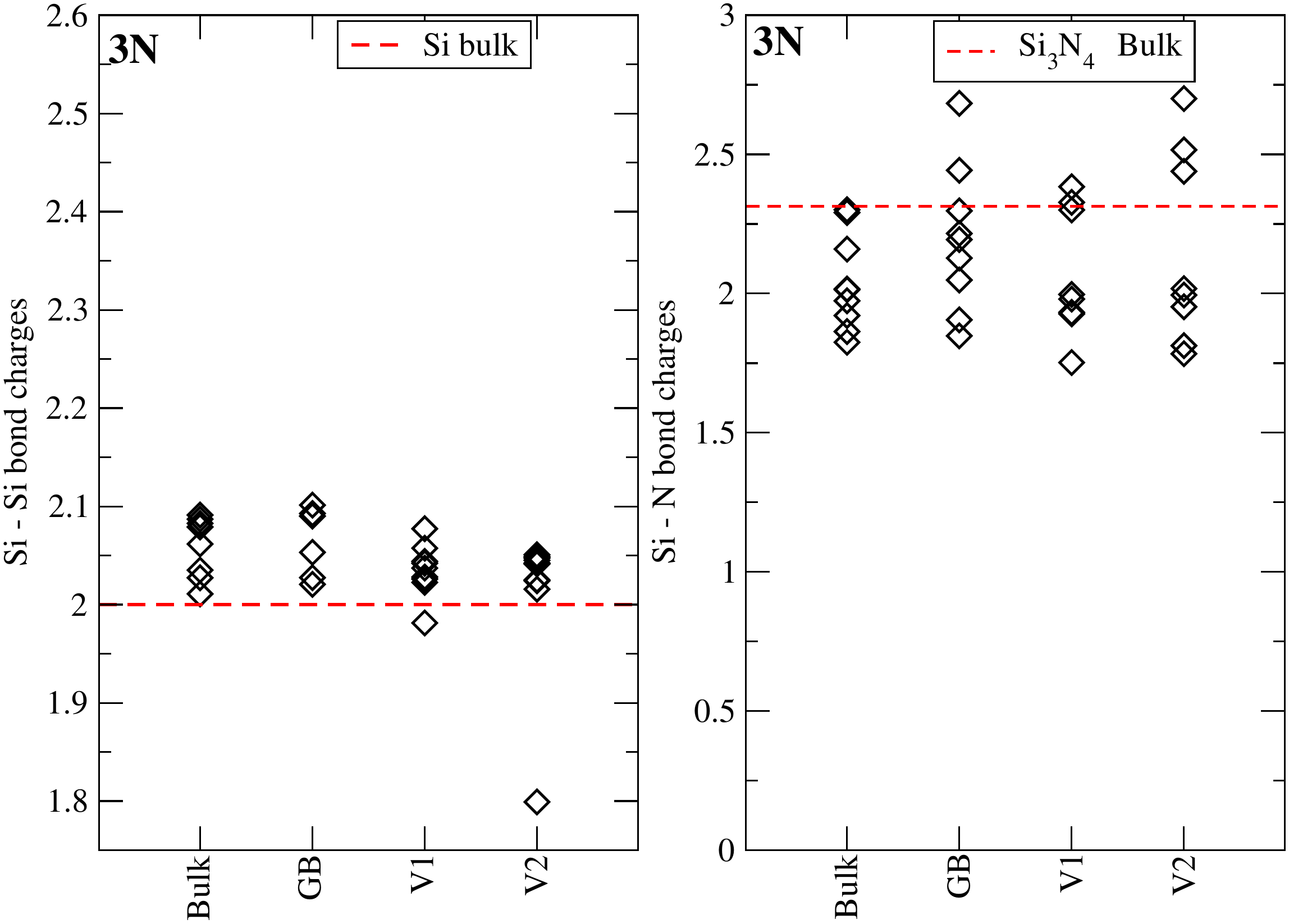}
    \caption{Bond charge variation of Si-Si and Si-N for the inclusion of three nitrogen atoms in Si-bulk, Si-GB and in presence of vacancy V1 and V2. As a reference, Si-Si/Si-N bond charges from bulk phases of Si and Si$_3$N$_4$ are marked in the inset of respective plots.}
    \label{fig:3Ncharges}
\end{figure}
\begin{figure} [h]   
    \centering
    \includegraphics[width=1.0\textwidth]{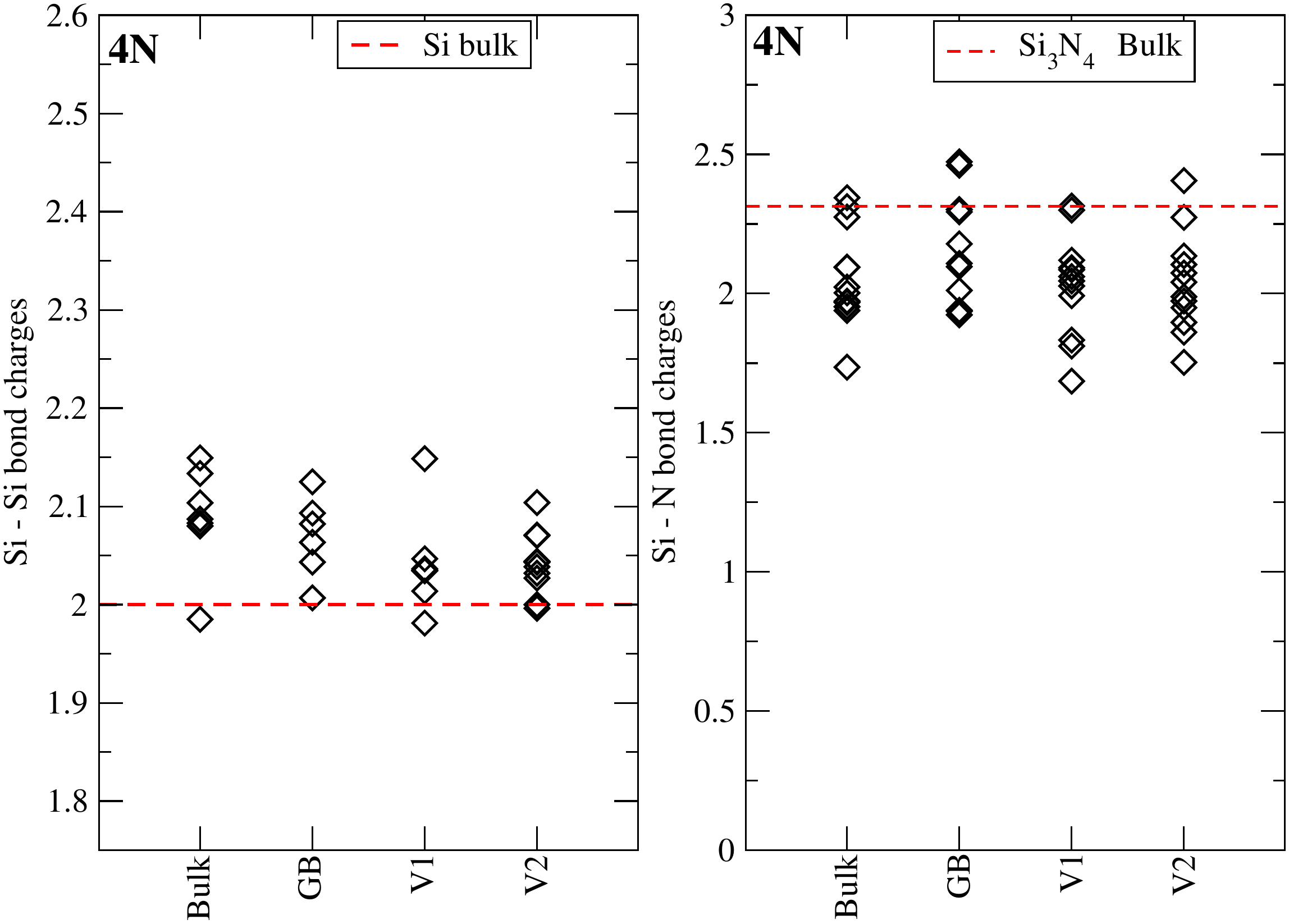}
    \caption{Bond charge variation of Si-Si and Si-N for the inclusion of four nitrogen atoms in Si-bulk, Si-GB and in presence of vacancy V1 and V2. As a reference, Si-Si/Si-N bond charges from bulk phases of Si and Si$_3$N$_4$ are marked in the inset of respective plots.}
    \label{fig:4Ncharges}
\end{figure}
\begin{figure}[h]    
    \centering
    \includegraphics[width=1.0\textwidth]{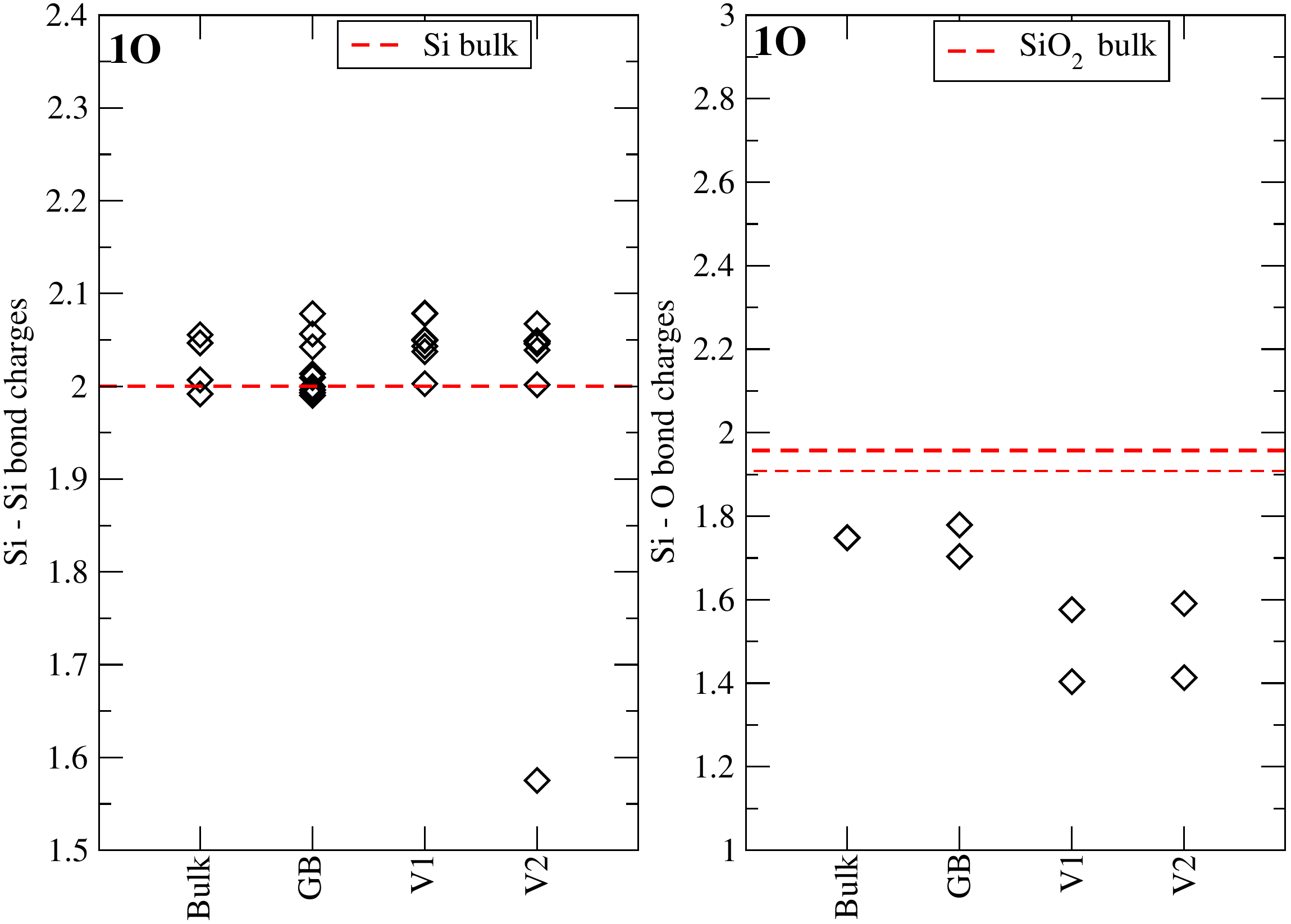}
    \caption{Bond charge variation of Si-Si and Si-O for the inclusion of one oxygen atom in Si-bulk, Si-GB and in presence of vacancy V1 and V2. As a reference, Si-Si/Si-O bond charges from bulk phases of Si and SiO$_2$ are marked in the inset of respective plots.}
    \label{fig:1obonds}
\end{figure}
\begin{figure}[h]    
    \centering
    \includegraphics[width=1.0\textwidth]{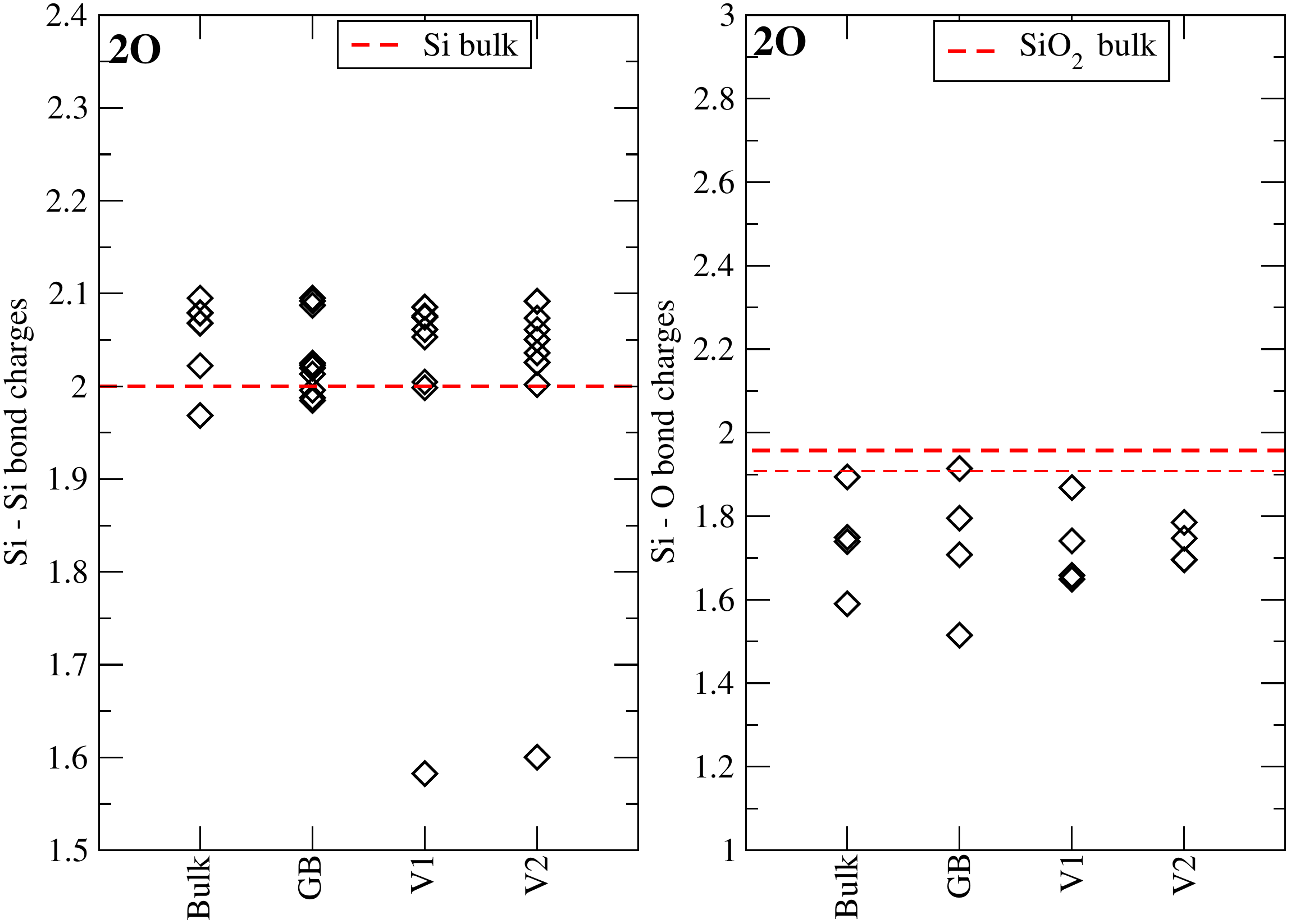}
    \caption{Bond charge variation of Si-Si and Si-O for the inclusion of two oxygen atoms in Si-bulk, Si-GB and in presence of vacancy V1 and V2. As a reference, Si-Si/Si-O bond charges from bulk phases of Si and SiO$_2$ are marked in the inset of respective plots.}
    \label{fig:2obonds}
\end{figure}
\begin{figure}[h]    
    \centering
    \includegraphics[width=1.0\textwidth]{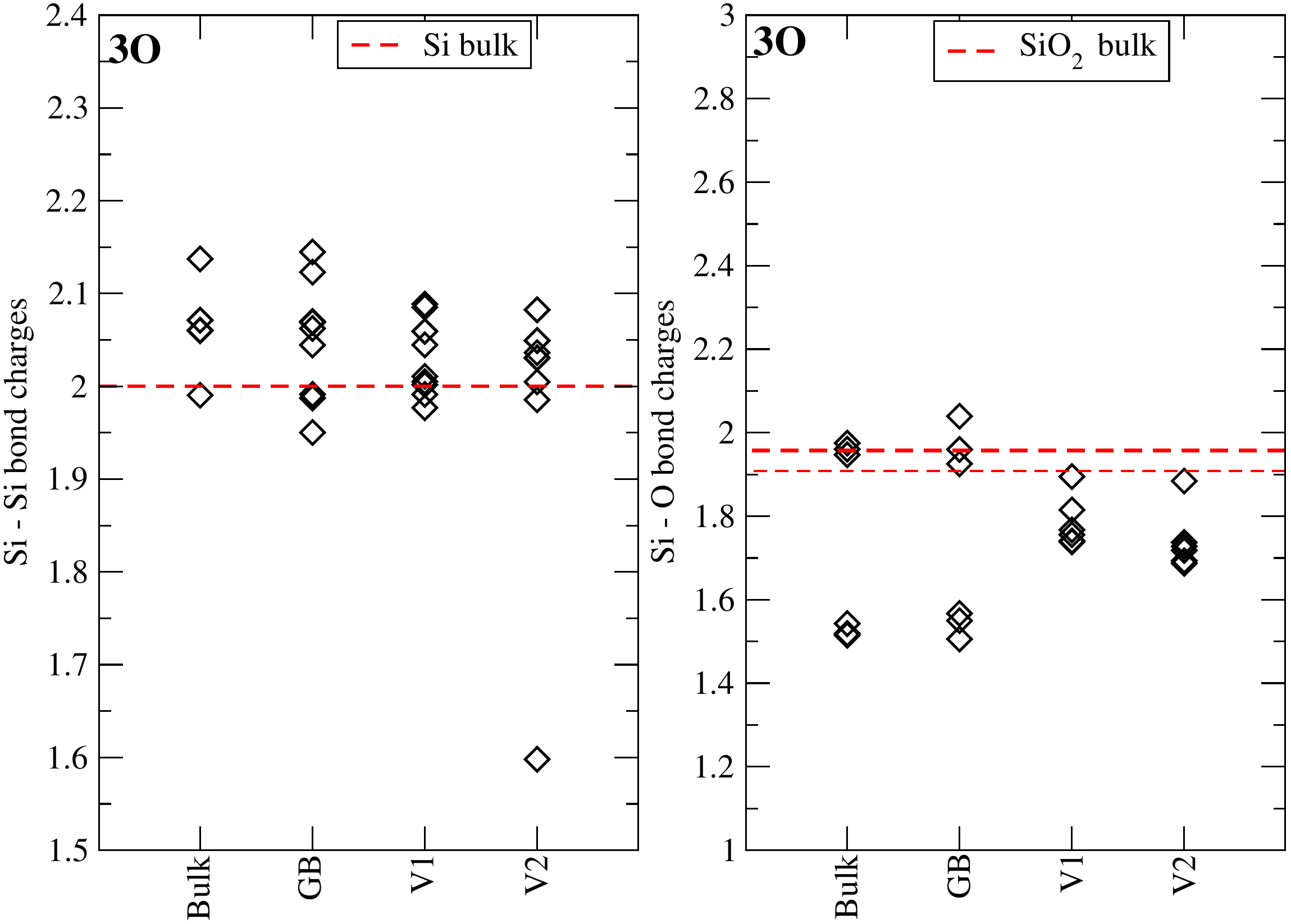}
    \caption{Bond charge variation of Si-Si and Si-O for the inclusion of three oxygen atoms in Si-bulk, Si-GB and in presence of vacancy V1 and V2. As a reference, Si-Si/Si-O bond charges from bulk phases of Si and SiO$_2$ are marked in the inset of respective plots.}
    \label{fig:3obonds}
\end{figure}
\begin{figure} [h]   
    \centering
    \includegraphics[width=1.0\textwidth]{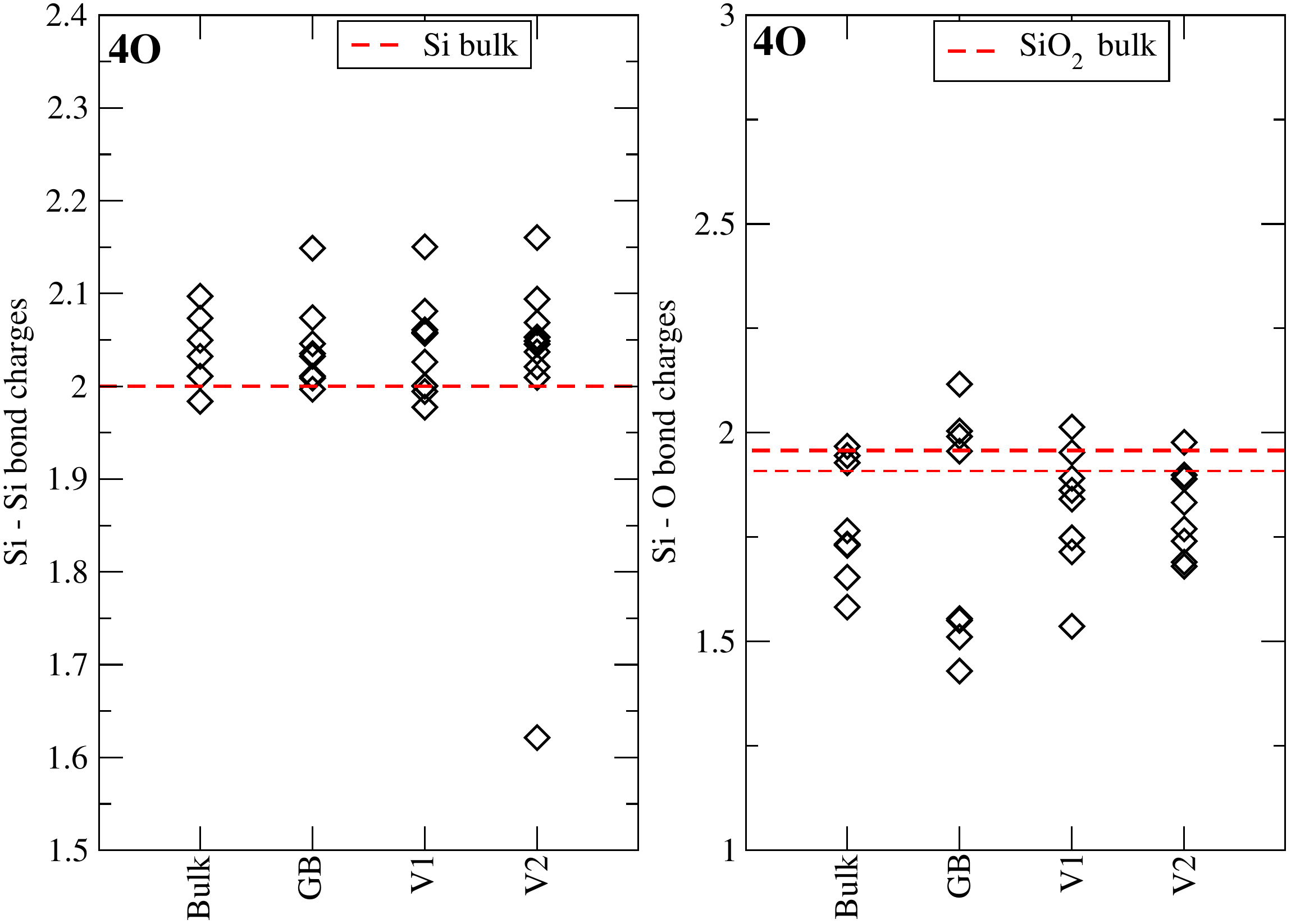}
    \caption{Bond charge variation of Si-Si and Si-O for the inclusion of four oxygen atoms in Si-bulk, Si-GB and in presence of vacancy V1 and V2. As a reference, Si-Si/Si-O bond charges from bulk phases of Si and SiO$_2$ are marked in the inset of respective plots.}
    \label{fig:4obonds}
\end{figure}

\pagebreak
\newpage

\begin{figure} [h!]   
    \centering
  \includegraphics[width=1.0\textwidth]{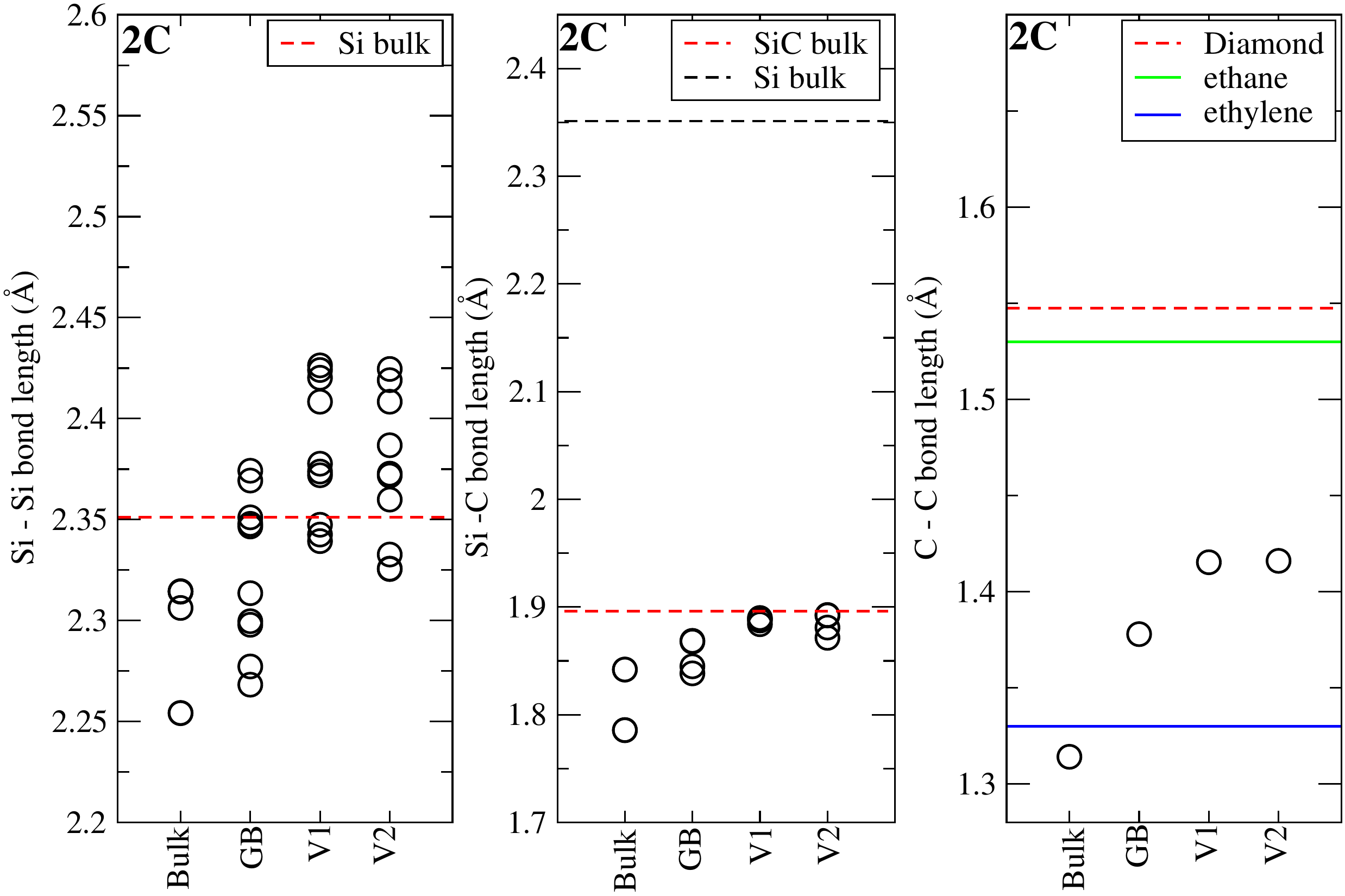}
    \caption{Bond length variation of Si-Si, Si-C and C-C for the inclusion of two carbon atoms in Si-bulk, Si-GB and in presence of vacancy V1 and V2. As a reference, Si-Si/Si-C/C-C bond length from bulk phases of Si, SiC, diamond and molecular systems ethylene and ethane are marked in the inset of respective plots.}
    \label{fig:2cbondsl}
\end{figure}
\begin{figure}[h]   
    \centering
  \includegraphics[width=1.0\textwidth]{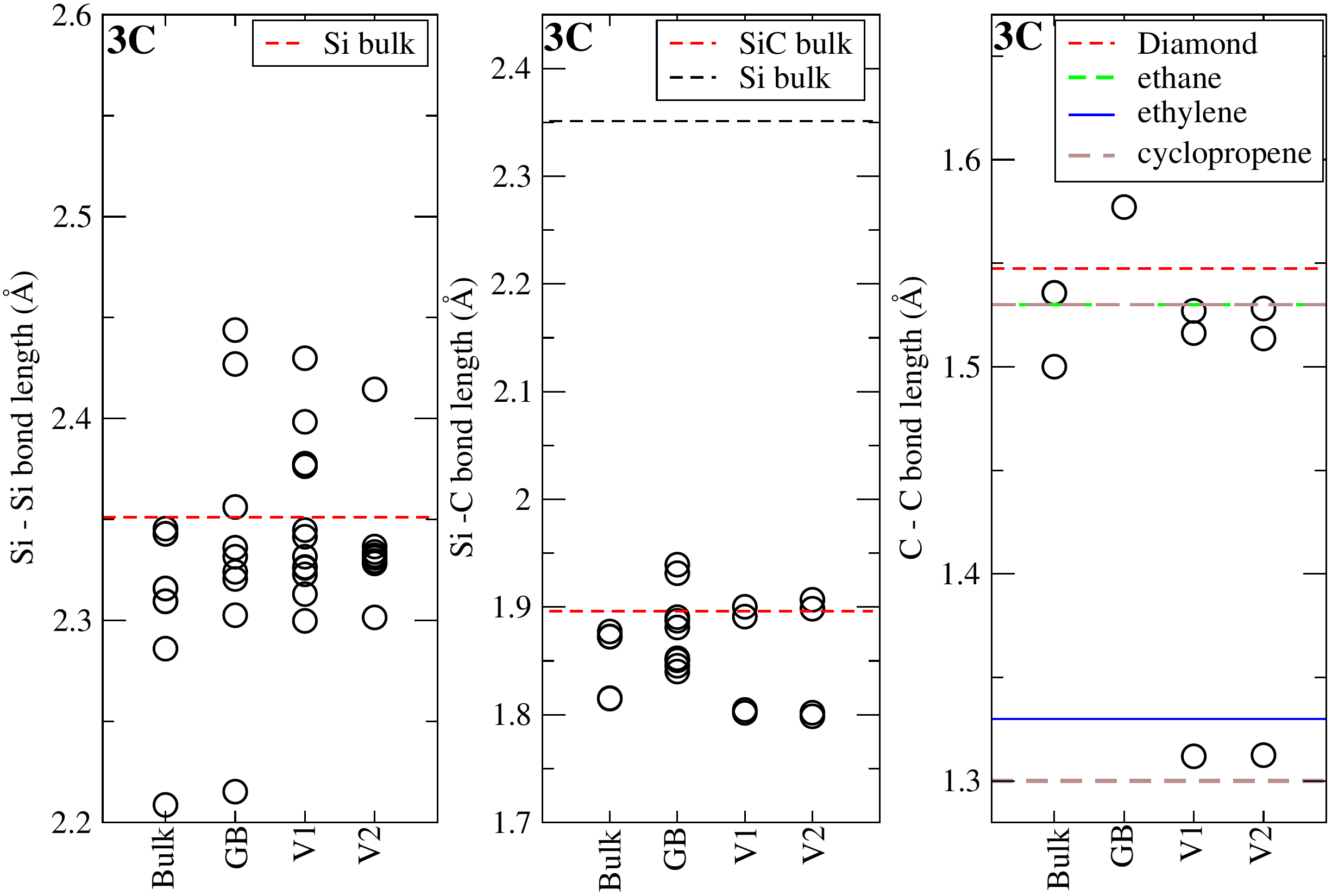}
    \caption{Bond length variation of Si-Si, Si-C and C-C for the inclusion of three carbon atoms in Si-bulk, Si-GB and in presence of vacancy V1 and V2. As a reference, Si-Si/Si-C/C-C bond length from bulk phases of Si, SiC, diamond and molecular systems ethylene, ethane, and cyclopropen are marked in the inset of respective plots}
    \label{fig:3cbondsl}
\end{figure}

\begin{figure}[h!]  
    \centering
    \includegraphics[width=1.0\textwidth]{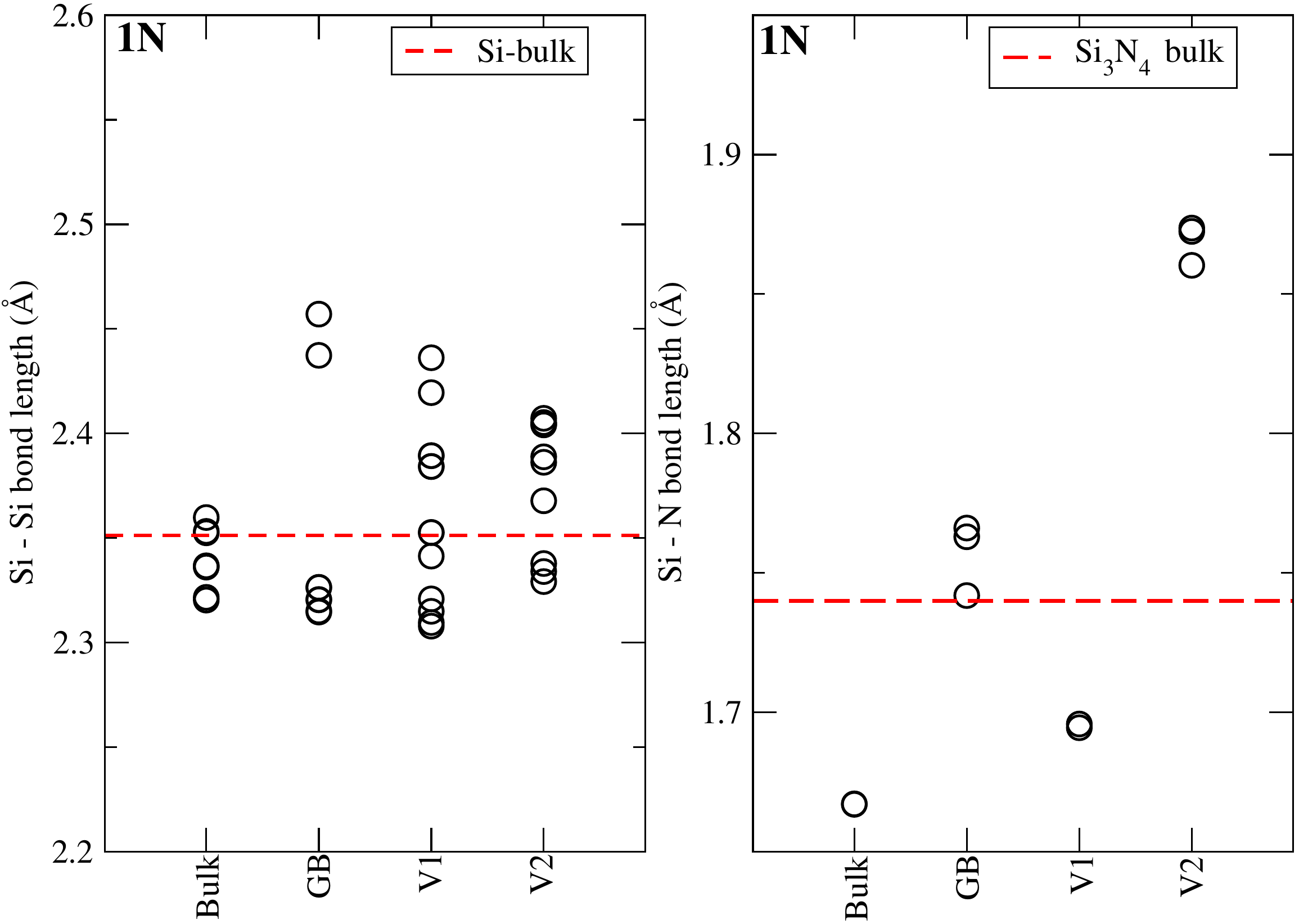}
    \caption{Bond length variation of Si-Si and Si-N for the inclusion of one nitrogen atom in Si-bulk, Si-GB and in presence of vacancy V1 and V2. As a reference, Si-Si/Si-N bond length from bulk phases of Si, Si$_3$N$_4$ are marked in the inset of respective plots.}
    \label{fig:1nbondsl}
\end{figure}
\begin{figure}[h]    
    \centering
    \includegraphics[width=1.0\textwidth]{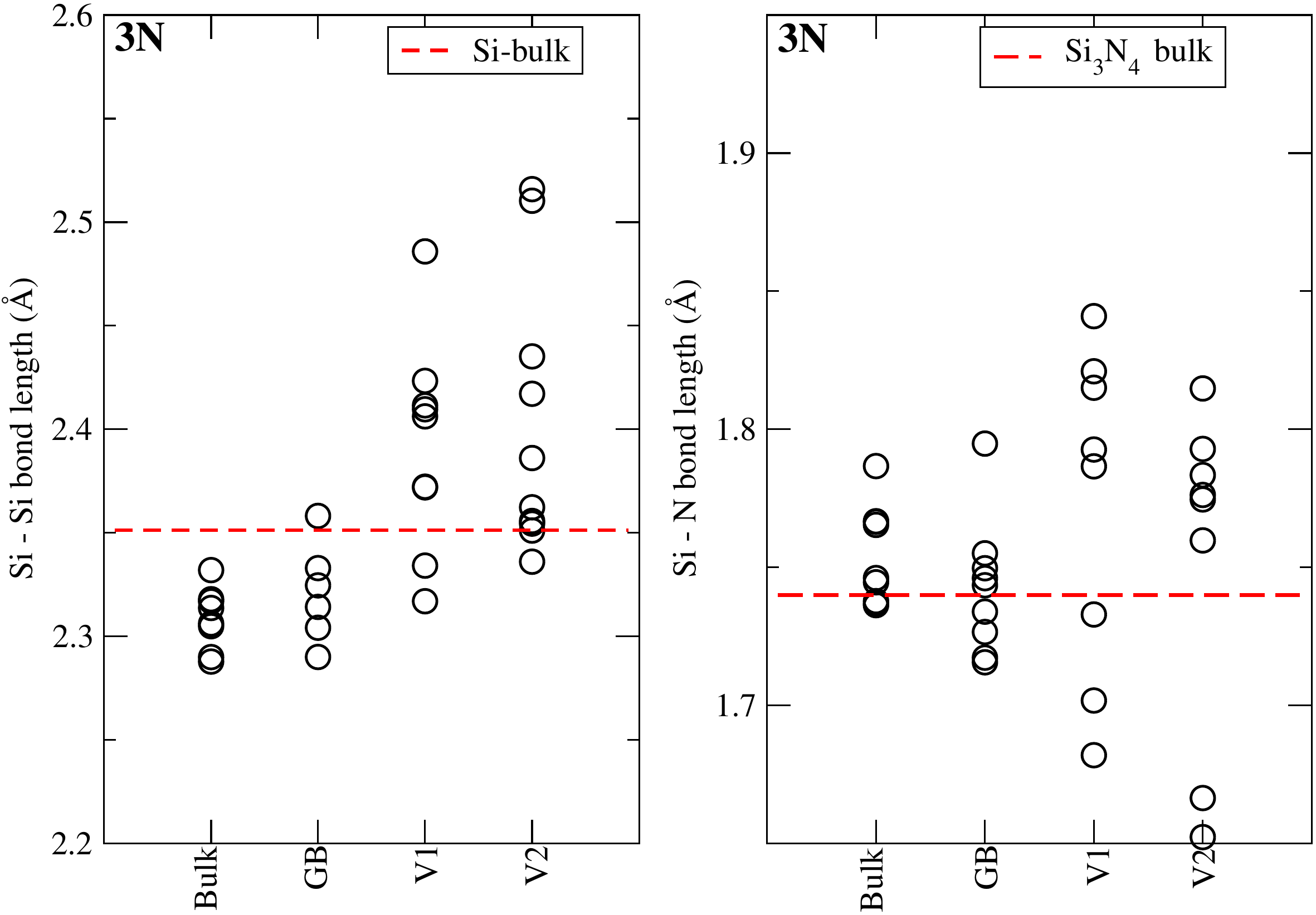}
    \caption{Bond length variation of Si-Si and Si-N for the inclusion of three nitrogen atom in Si-bulk, Si-GB and in presence of vacancy V1 and V2. As a reference, Si-Si/Si-N bond length from bulk phases of Si, Si$_3$N$_4$ are marked in the inset of respective plots.}
    \label{fig:3nbondsl}
\end{figure}
\begin{figure}[h]    
    \centering
    \includegraphics[width=1.0\textwidth]{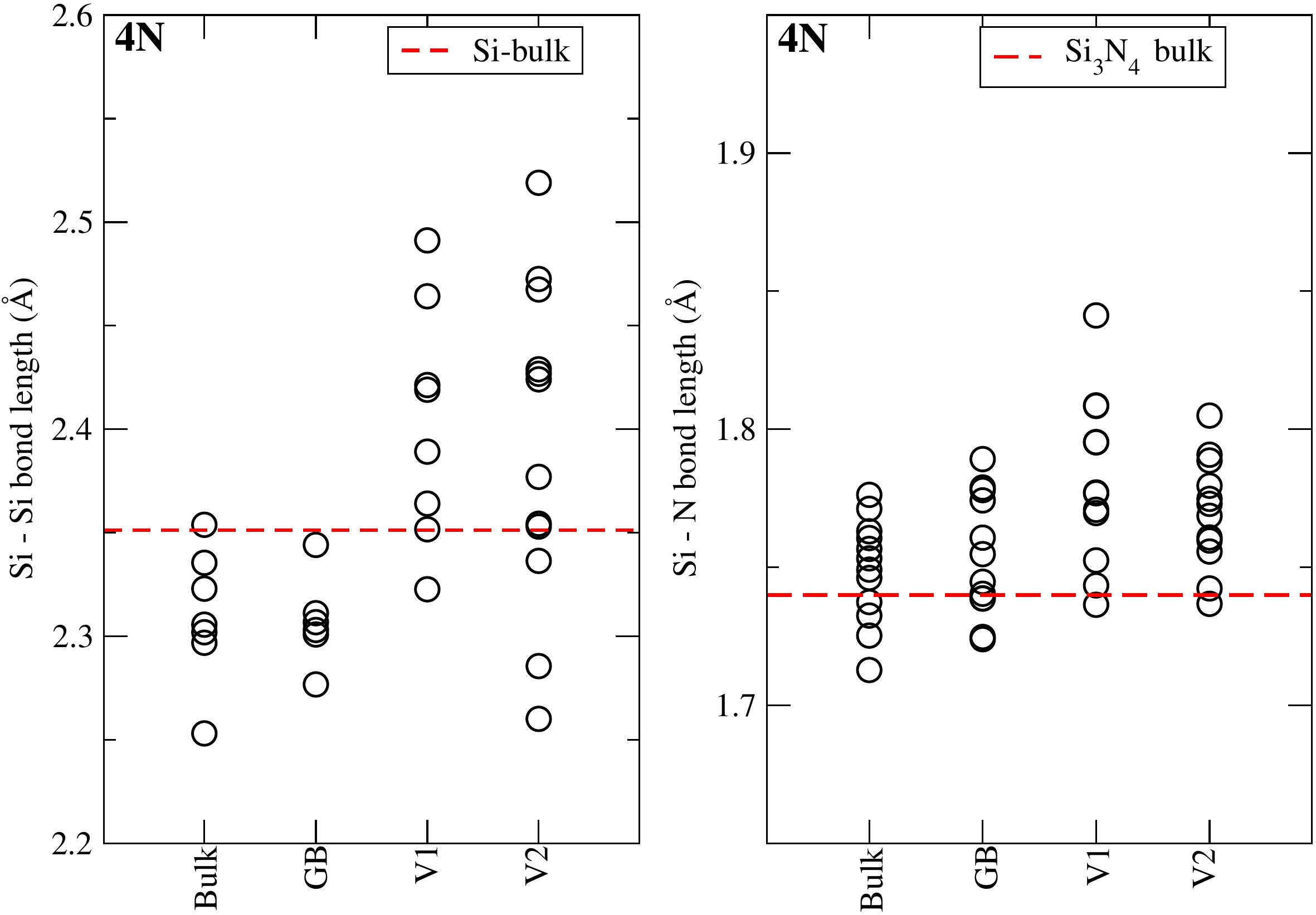}
    \caption{Bond length variation of Si-Si and Si-N for the inclusion of four nitrogen atom in Si-bulk, Si-GB and in presence of vacancy V1 and V2. As a reference, Si-Si/Si-N bond length from bulk phases of Si, Si$_3$N$_4$ are marked in the inset of respective plots.}
    \label{fig:4nbondsl}
\end{figure}

\begin{figure} [h!]   
    \centering
    \includegraphics[width=1.0\textwidth]{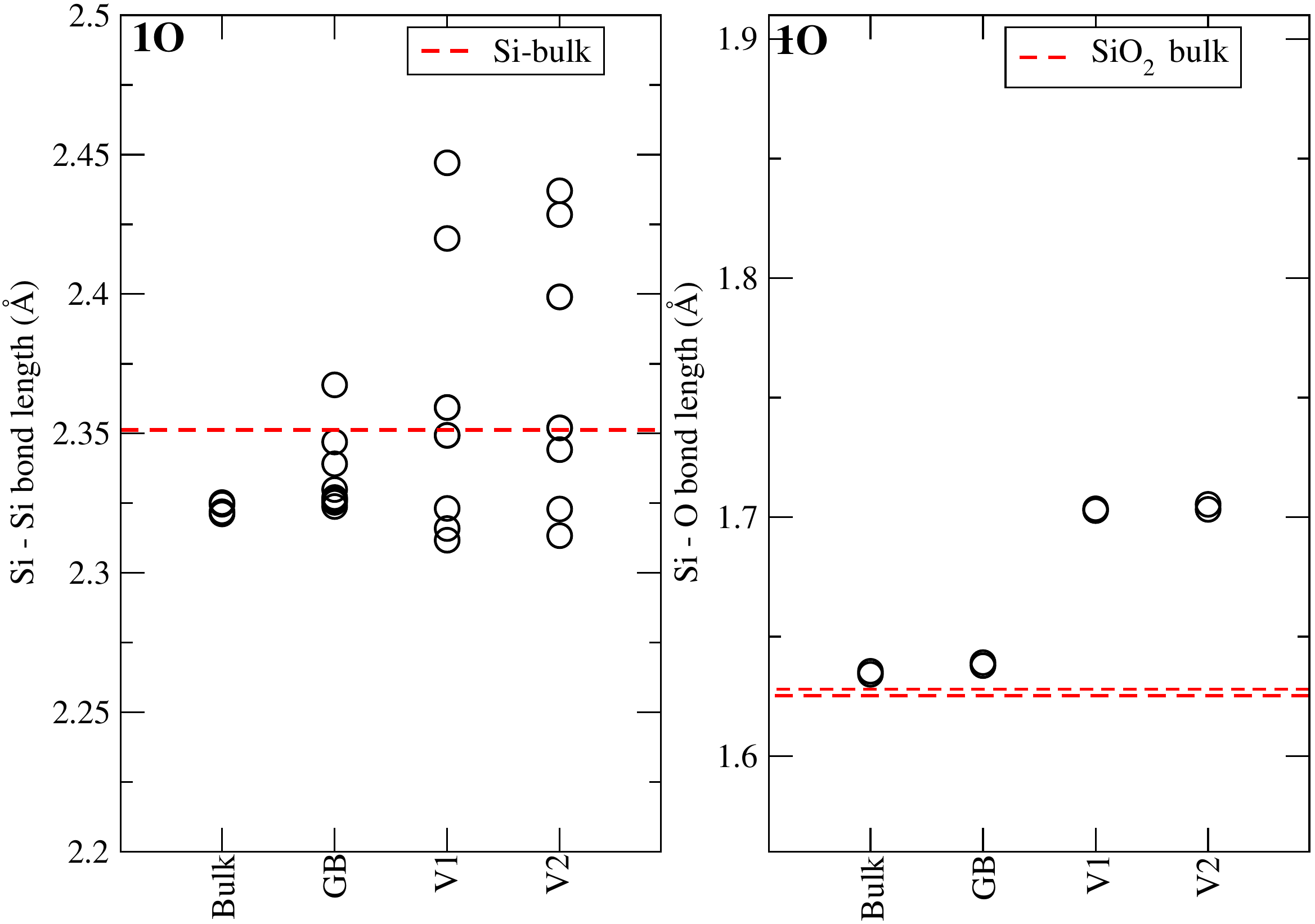}
    \caption{Bond length variation of Si-Si and Si-O for the inclusion of one oxygen atom in Si-bulk, Si-GB and in presence of vacancy V1 and V2. As a reference, Si-Si/Si-O bond length from bulk phases of Si, SiO$_2$ are marked in the inset of respective plots.}
    \label{fig:1obondsl}
\end{figure}
\begin{figure}[h]    
    \centering
    \includegraphics[width=1.0\textwidth]{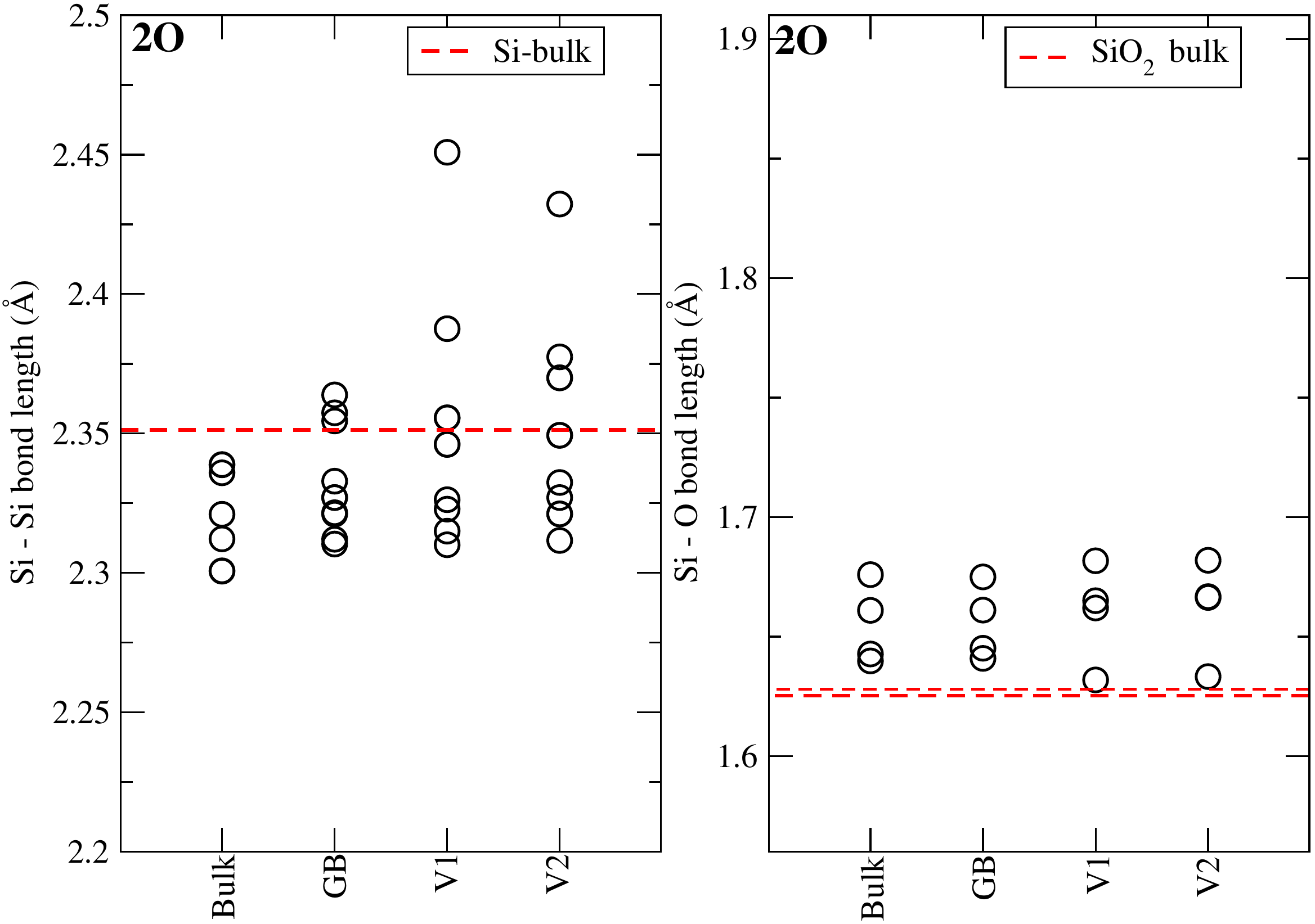}
    \caption{Bond length variation of Si-Si and Si-O for the inclusion of two oxygen atoms in Si-bulk, Si-GB and in presence of vacancy V1 and V2. As a reference, Si-Si/Si-O bond length from bulk phases of Si, SiO$_2$ are marked in the inset of respective plots.}
    \label{fig:2obondsl}
\end{figure}
\begin{figure}[h]    
    \centering
    \includegraphics[width=1.0\textwidth]{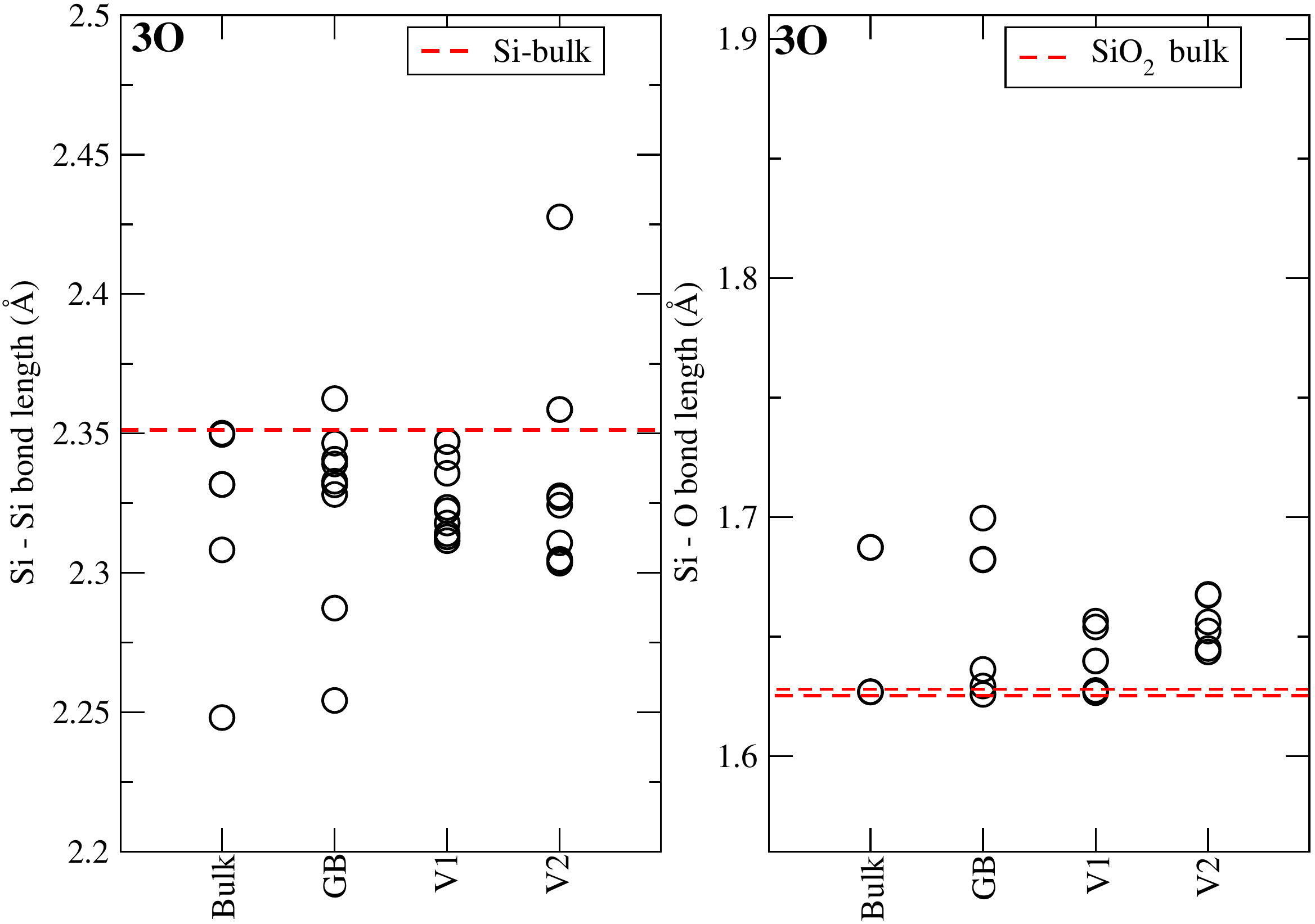}
    \caption{Bond length variation of Si-Si and Si-O for the inclusion of three oxygen atoms in Si-bulk, Si-GB and in presence of vacancy V1 and V2. As a reference, Si-Si/Si-O bond length from bulk phases of Si, SiO$_2$ are marked in the inset of respective plots.}
    \label{fig:3obondsl}
\end{figure}
\begin{figure}[h]    
    \centering
    \includegraphics[width=1.0\textwidth]{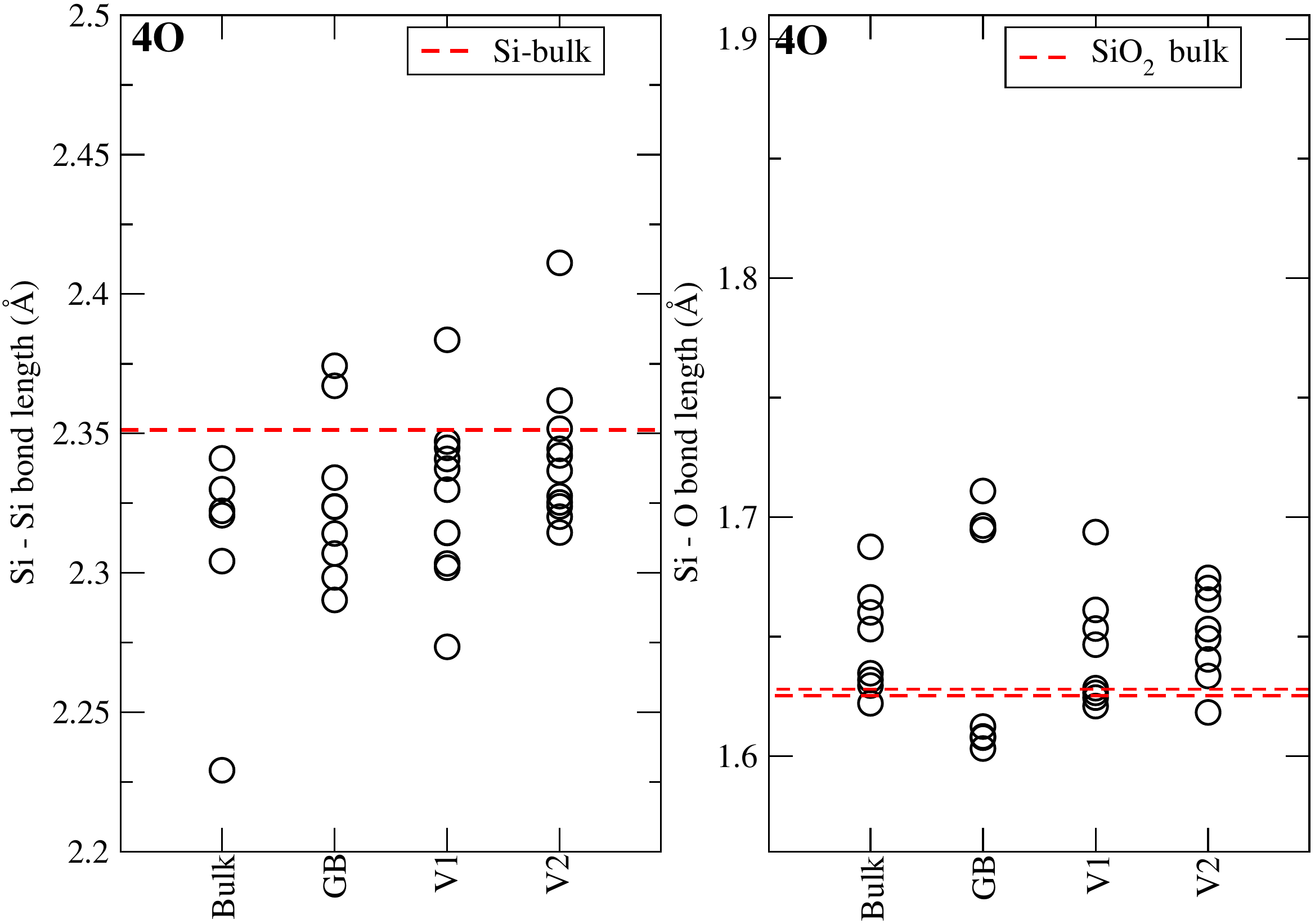}
    \caption{Bond length variation of Si-Si and Si-O for the inclusion of four oxygen atoms in Si-bulk, Si-GB and in presence of vacancy V1 and V2. As a reference, Si-Si/Si-O bond length from bulk phases of Si, SiO$_2$ are marked in the inset of respective plots.}
    \label{fig:4obondsl}
\end{figure}

\end{document}